\documentclass[12pt, preprint]{aastex}
\newcommand{\novalue}{\multicolumn{2}{c}{--------}}
\begin{document}

\title{The Complete Spectral Catalog of Bright BATSE Gamma-Ray~Bursts}

\author{Yuki~Kaneko\altaffilmark{1},
         Robert~D.~Preece\altaffilmark{2},
         Michael~S.~Briggs\altaffilmark{2},
         William~S.~Paciesas\altaffilmark{2},
         Charles~A.~Meegan\altaffilmark{3},
         David~L.~Band\altaffilmark{4}
}

\altaffiltext{1}{Universities Space Research Association,
   NSSTC, XD-12, 320 Sparkman Drive, Huntsville, AL 35805,
{\tt Yuki.Kaneko@nsstc.nasa.gov}}
\altaffiltext{2}{Department of Physics, University of Alabama in Huntsville,
   NSSTC, XD-12, 320 Sparkman Drive, Huntsville, AL 35805,
   {\tt Rob.Preece@nsstc.nasa.gov, Michael.Briggs@nsstc.nasa.gov, 
   Bill.Paciesas@nsstc.nasa.gov}}
\altaffiltext{3}{NASA/Marshall Space Flight Center,
   NSSTC, XD-12, 320 Sparkman Drive, Huntsville, AL 35805,
   {\tt Chip.Meegan@nasa.gov}}
\altaffiltext{4}{NASA/Goddard Space Flight Center, Greenbelt, MD 20771 
   \& JCA/University of Maryland, Baltimore County, 
   Baltimore, MD 21250, {\tt dband@milkyway.gsfc.nasa.gov}}

\begin{abstract}
We present a systematic spectral analysis of 350 bright Gamma-Ray Bursts (GRBs)
observed with the Burst and Transient Source Experiment (BATSE; $\sim$ 30 keV -- 
2 MeV) with high temporal and spectral resolution. 
Our sample was selected from the complete set of 2704 BATSE GRBs based on their 
energy fluence or peak photon flux values to assure good statistics, and
included 17 short GRBs. 
To obtain well-constrained spectral parameters, several photon models 
were used to fit each spectrum.
We compared spectral parameters resulting from the fits using different models, 
and the spectral parameters that best represent each spectrum were statistically 
determined, taking into account the parameterization differences among the 
models.
A thorough analysis was performed on 350 time-integrated and 
8459 time-resolved burst spectra, and the effects of integration times in 
determining the spectral parameters were explored.
Using the results, we studied correlations among spectral parameters and 
their evolution pattern within each burst.
The resulting spectral catalog is the most comprehensive study of spectral
properties of GRB prompt emission to date, and is available electronically 
from the High-Energy Astrophysics Science Archive Research Center (HEASARC).
The catalog provides reliable constraints on particle acceleration and 
emission mechanisms in GRBs.
\end{abstract}


\section{Introduction}
In recent years, multi-wavelength observations of afterglow emission of 
Gamma-Ray Bursts (GRBs) have provided great advancement in our knowledge of 
GRB progenitor, afterglow emission mechanism, and their environment.
Nonetheless, the physical mechanism that creates the prompt gamma-ray 
emission with extremely short variability is still not resolved, thus 
understanding GRB prompt emission spectra remains crucial to revealing their 
true nature. 
GRB spectral analysis attempts to empirically characterize the spectra, 
which are generally well described, in the energy range of $\sim$ 10 keV to 
a few MeV, by two power laws joined smoothly at certain energies \citep{ban93}.
The spectral parameters (the power-law indices and the peak energy in the power 
density spectrum) are then used to infer the GRB emission and particle 
acceleration mechanisms.  

Currently, the most favored GRB emission mechanism is the simple
emission scenario of optically-thin synchrotron radiation by shock-accelerated
electrons \citep[``synchrotron shock model";][]{tav96b}.
This simple theoretical model, however, has previously been challenged by the 
observations in the context of GRB prompt emission
\citep{pre98a, pre98b, pre00, ghi02, llo02}.
While the synchrotron shock model can account for many of the observed spectra, 
a considerable number of spectra exhibit behavior inconsistent with this 
theoretical model.
Meanwhile, it is also true that many observed spectra could be fitted with 
various photon models statistically as well as each other, due to the limited 
spectral resolution of available data and detector sensitivity.
Since the photon models usually used in GRB spectral analysis are parameterized 
differently, the resulting spectral parameters are found to be highly dependent 
on photon model choices \citep{pre02, ghi02}.
Additionally, to deduce the emission mechanism from observations, spectra with 
fine time resolution are necessary because of the short timescales involved 
in typical emission processes (i.e., the radiative cooling time, dynamical time, 
or acceleration time).
This is also indicated by the extremely short variability observed in GRB 
lightcurves \citep[e.g.,][]{bha92}, although the detectors' finest time 
resolution is still longer than the shortest physical timescales involved 
to produce GRBs. 
The integration times of spectra certainly depend on the capabilities of the
detector systems as well as the brightness of events and photon flux evolution.
GRB spectral analyses, therefore, have been performed on various timescales,
yet a comprehensive study of the relations between time-averaged and 
time-resolved spectra, and the effects of various integration times on spectral
properties has not been done.
Thus, in order for the spectral parameters to meaningfully constrain the 
physical mechanisms, a comprehensive spectral study with finest 
possible spectral and temporal resolution, using various photon models, 
should be carried out with a sufficiently large database. 

Among all the gamma-ray experiments that have detected GRBs, the Burst and 
Transient Source Experiment \citep[BATSE;][]{fis89}, aboard the 
{\it Compton Gamma-Ray Observatory} \citep[{\it CGRO};][]{geh94}, 
provided the largest GRB database from a single experiment, consisting of 
observational data for 2704 GRBs (M.S. Briggs in preparation). 
For many of the BATSE GRBs, high time and energy resolution data are 
available. 
BATSE also provided wider energy coverage than current GRB missions
such as {\it HETE-2} and {\it Swift}.
The BATSE data, therefore, are the most suitable for detailed
spectral studies of GRB prompt emission, both in quantity and quality.
The previous BATSE GRB spectral catalog \citep[SP1 hereafter]{pre00} 
consisted of 5500 time-resolved spectra from 156 bright GRBs that occurred before
October 1998.
The SP1 burst sample was selected from a set of 1771 GRBs (from the BATSE 4B
catalog \citep{pac99} and the ``current'' catalog available online\footnotemark[1]),
nearly 1000 bursts
less than what is currently available in the complete database. 
The sample was also limited to the bursts that provided more than eight spectra 
and therefore, excluded relatively shorter and weaker bursts.
In addition, a combination of the Large Area Detector (LAD) and Spectroscopy
Detector (SD) data was used, no time-integrated spectral fit (i.e., spectrum
with integration time of the burst duration) results were presented, and 
only one photon model was fitted to each spectrum.
Finally, the mathematical differences in parameterization of each model were not 
taken into account to obtain corrected overall statistics of the analysis results.
\footnotetext[1]{\url http://gammaray.nsstc.nasa.gov/batse/grb/catalog/current/}  

We present in this paper the high-energy resolution spectral analysis of
350 bright BATSE GRBs with fine time resolution.
The main objective of this work is to obtain consistent spectral properties of 
GRB prompt emission with sufficiently good statistics.
This is done by a systematic analysis of the large sample of GRB spectra, 
using a set of photon models, all fitted to each spectrum.
Our spectral sample includes both time-resolved and time-integrated spectra
for each burst: the time-resolved spectra within each burst has the best 
possible time resolution for the high-energy resolution data types used, 
and the time-integrated spectra are the sum of the resolved spectra within 
the bursts, covering entire duration of the bursts. 
We obtain well-constrained spectral properties by studying characteristics of 
each photon model, taking into account the parameterization differences, and
statistically determining the best-fit model for each spectrum.
The analysis performed here is much more comprehensive and consistent than that 
of SP1 in the sense that (i) only LAD data are used, (ii) the burst sample
selections are more objective, (iii) five different photon models are fitted to 
each spectrum, both time-integrated and time-resolved, and 
(iv) the best-fit parameters of each spectrum are statistically determined.
The use of various models allows us to compare the behavior of different 
models as well as to obtain unbiased statistics for the spectral parameters. 
We also note that the BATSE data and the Detector Response Matrices (DRMs) 
used here have been regenerated since the publication of SP1, 
with a refined detector calibration database. 
This provided more precise LAD energy edges, thus assuring the improved accuracy
of the spectral analysis.

Since our sample only includes bright bursts, there may be some bias in our
analysis results.  As reported previously in literature, there is a tendency
of bright bursts being spectrally harder than dimmer bursts \citep[e.g.,][]{nem94, 
mal95, dez97}.
Therefore, it is likely that our sample of bright bursts belongs to the harder side
of the overall BATSE GRB sample, on average \citep[e.g.,][]{llo+00}.
It is particularly important to keep this in mind when studying global correlations
with our spectral analysis results.
Our selection criteria also excludes many bright short bursts with duration much 
less than a second.  
As a result, only $\sim$ 5\% of our sample are short bursts (duration $<$ 2 s) 
while short bursts comprises $\sim$ 19\% of the entire BATSE GRB sample.

The paper is organized in the following manner. 
We first describe the burst sample and analysis interval selection methodology in 
\S \ref{sec:selection}.  The details of the spectral analysis methods are then 
discussed in \S \ref{sec:analysis}, including descriptions of the
photon models used.
We also discuss our simulation results in \S \ref{sec:simulation}, which were
performed to assist us in interpreting the analysis results correctly.
Finally, the analysis results are presented in \S \ref{sec:results}
and summarized in \S \ref{sec:summary}. 
We note that there are two ways of identifying a BATSE burst; by the GRB name 
(``GRB {\it yymmdd}") and by the BATSE trigger number.  
We use both the names and trigger numbers to refer to individual GRBs 
throughout this paper\footnotemark[2]. \\
\footnotetext[2]{Tables \ref{tab:grblist} and \ref{tab:results} list both the
GRB names and the BATSE trigger numbers for the GRBs in our sample.  For other
BATSE-triggered events, a lookup table to convert between them is available at
\url http://gammaray.nsstc.nasa.gov/batse/data/trigger\_lookup.html}
\section{Selection Methodology}\label{sec:selection}
During its nine-year lifetime that ended in June 2000, BATSE triggered on 
a total of 8021 gamma-ray transient events, of which 2704
events were identified as GRBs.
The BATSE detectors were sensitive enough to detect relatively weak GRBs down 
to a peak photon flux in 256 ms of $\sim$ 0.3~s$^{-1}$~cm$^{-2}$ (50 -- 300 keV)
and a total energy fluence in 25 -- 2000~keV of $\sim 10^{-8}$~ergs~cm$^{-2}$.  
Unfortunately, many dim bursts do not provide enough signal above background
for high energy resolution spectral analysis, particularly for time-resolved
spectroscopy.
Therefore, we need to select and limit our analysis to GRBs with 
sufficient signal.
In addition, the available data types and data ranges vary for each burst.
In this section, we describe the methodology employed for the burst selection, 
the data type selection, and time and energy interval selections.

\subsection{Burst Sample Selection}\label{sec:burst_sel}

The primary selection was made based on the peak photon flux and the 
total energy fluence in the BATSE 4B catalog \citep{pac99}, as well as the
current BATSE GRB catalog\footnotemark[1] for the post-4B bursts.  
The catalogs list a total of 2702~GRBs, of which two were later identified 
as non-GRBs (trigger numbers 5458 and 7523).  
Additionally, there are four triggered events that were later identified as 
GRBs (trigger numbers 1505, 2580, 3452, and 3934), bringing the total number
of BATSE GRBs to 2704. The number is consistent with the final BATSE 5B catalog
(M.S. Briggs et~al. in preparation). 
Our sample was selected from these 2704~bursts.
The burst selection criteria are a peak photon flux in 256~ms (50 -- 300 keV)
greater than 10~photons~s$^{-1}$~cm$^{-2}$ or a total energy fluence in the 
summed energy range ($\sim$ 20 -- 2000 keV) larger than $2.0 \times 10^{-5}$
ergs~s$^{-1}$.
Having the criteria in either the peak photon flux or total energy fluence 
allows inclusion of short bright bursts as well as long bursts with low peak
flux.
The peak photon flux criterion remains the same as SP1, although it was 
misstated in SP1 as a 1024-ms integration time.
The energy fluence criterion has been lowered from the value used in SP1 so as 
to include more bursts, while still securing sufficiently good statistics.  
A total of 298 GRBs satisfied these criteria.

In addition, 573 GRBs out of 2704 do not have flux/fluence values published in 
any BATSE GRB catalog, mainly due to gaps in the four-channel discriminator 
data that were used to obtain the flux/fluence values.  
Nonetheless, for many of these bursts, finer energy-resolution data are still 
available for spectral analysis, and some are bright enough to be included in 
this work.  
Therefore, we estimated the photon flux and the energy fluence values for
the bursts with no published flux/fluence values, using available data.
To estimate the peak photon flux and energy fluence for such bursts, 
we used 16-channel MER data (see \S \ref{sec:data} below for the description 
of the data type) binned to a 256-ms integration time 
and fitted with a smoothly-broken power-law model (described later in 
\S\ref{sec:sbpl_model}).
The peak photon flux values (in 256 ms, $\sim$ 50 -- 300 keV) and the total 
energy fluence values (in $\sim$ 30 -- 1900 keV) were calculated from the fitted 
spectra.
We note that using other photon models barely changed the outcome.
There were some cases where MER data were not available but either HERB or 
CONT data existed.  
Most of these cases, however, turned out to be noticeably very weak, or else 
all of the available data were not complete, and therefore did not qualify for 
inclusion.
In this way, 55 GRBs yielded peak photon flux or total energy fluence 
values well above our threshold criteria, and were therefore included in this 
work.

Out of these selected bright bursts, we found two cases (trigger numbers 3366 
and 7835) in which the published flux/fluence values were incorrect and the 
actual flux/fluence values were much lower than the criteria used here.  
Consequently, these two bursts were excluded.
We also found one case (trigger number 5614, peak photon flux $=$ 182 photons
s$^{-1}$ cm$^{-2}$) in which the burst was so bright that the indication of
possible pulse pile-up was seen in the energy spectra of all available 
detectors and the data were not usable.  
Thus, this burst was also excluded from this work.
The resulting total number of GRBs included in this spectral analysis was 350.
The GRBs are listed in Table \ref{tab:grblist}, along with the data types,
time and energy intervals, and numbers of time-resolved spectra contained.
\placetable{tab:grblist}

\subsection{Detector Selection}
BATSE was specifically designed to detect GRBs and to study their temporal and 
spectral characteristics in great resolution.
In order to increase the GRB detection probability, BATSE consisted of eight
modules that were located at the corners of the {\it CGRO} spacecraft, 
so as to cover the entire 4$\pi$ steradian of sky. 
Each module comprised two types of detectors; LAD and SD.
They were both NaI(Tl) scintillation detectors of different dimensions 
designed to achieve different scientific goals.
The LADs were 50.8 cm in diameter and 1.27 cm in thickness, and provided
burst triggering and burst localizations with high sensitivity. 
They were gain-stabilized to cover the energy range of $\sim$ 30 keV -- 2 MeV.
The SDs were 12.7 cm in diameter and 7.62 cm in thickness, and provided 
a higher energy resolution in $\sim$ 10 keV -- 10 MeV, depending on the gain 
of each detector at the time of a burst trigger.

In this spectral analysis, only the LAD data are used, mainly to take 
advantage of its larger effective area \citep{fis89}.
Another reason that only the LAD data are used is due to a problem recently 
identified in the SD DRMs at energies above $\sim$ 3 MeV \citep{kan05}.
These problems could give rise to uncertainties in the SD DRMs and make SD 
data unreliable for spectral analysis at high energies.  
Moreover, limiting to one type of detector eliminates systematic uncertainties
arising from different detector characteristics, and thus keeps the analysis 
more uniform.
For each burst, the data from the LAD with the highest counts are used.
 
\subsection{Data Type Selection}\label{sec:data}
\begin{figure}
\centerline{
\plotone{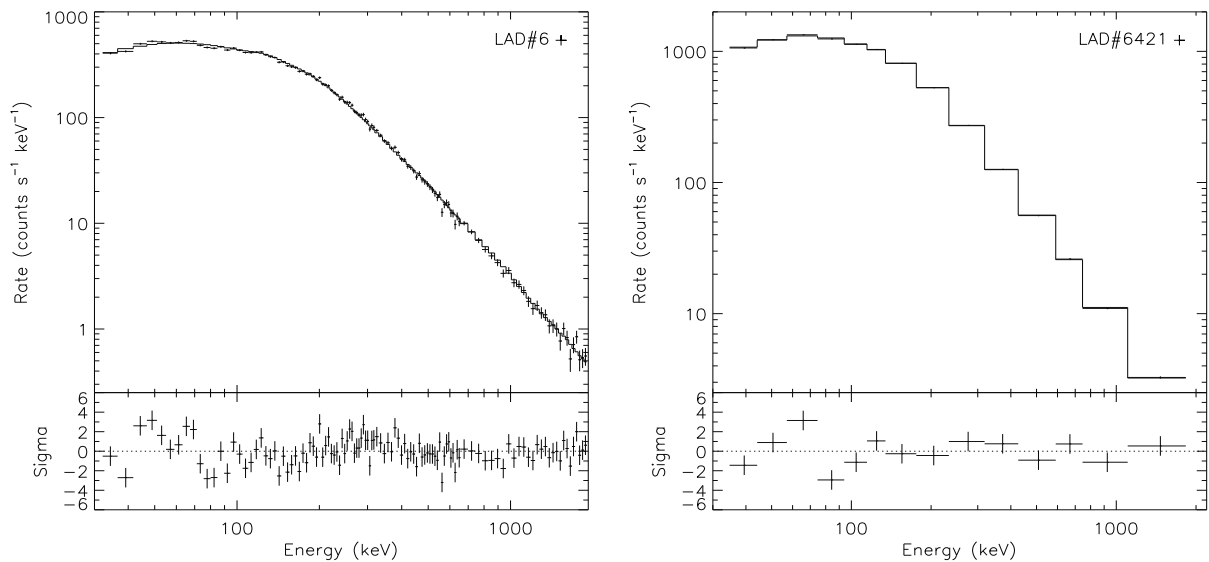}}
\caption{Example count spectra of a bright burst that shows the systematic 
deviations both in the HERB ({\it left}) and MER ({\it right}) data.  
The deviations between the data (crosses) 
and the model (solid lines) are evident below $\sim$ 100 keV.}
\label{fig:system}
\end{figure}

The LADs provided various types of data products to be used 
for various purposes. 
The data were collected in either burst mode or non-burst mode.
The burst mode data accumulation started when a burst was triggered, 
whereas the non-burst mode data were usually continuous, except for possible 
telemetry gaps.  
The burst mode data generally provided higher resolution either in energy 
or in time than the non-burst mode data.
The three LAD data types used in this work, in order of priority, are High 
Energy Resolution Burst data (HERB), Medium Energy Resolution data (MER), and 
Continuous data (CONT).
The characteristics of each data type are listed in Table \ref{tab:datatypes}
(see SP1 for more complete list of BATSE burst data types).  
\placetable{tab:datatypes}
The HERB and MER data are burst-mode data, for which the acquisition began
at the time of a burst trigger, whereas CONT was continuous,
non-burst mode data.  
The accumulation time of the HERB data is rate dependent, with minimum 
time-to-spill of 128 ms with a 64-ms increment.  
HERB provides the highest energy resolution consisting of $\sim 120$ usable 
energy channels in energy range of $\sim$ 30 -- 2000 keV with modest sub-second 
time resolution, and thus is used as the primary data type.
For each burst, the HERB data for the brightest LAD provides the finest time 
resolution, and was always selected for the analysis. 
The coarser-time resolution ($\sim 300$ s) High Energy Resolution data (HER)
are used as background data for the HERB, covering several thousand seconds 
before the trigger and after the HERB accumulation is finished.
However, especially for long, bright bursts, the HERB data can often be 
incomplete since HERB had a fixed memory space that could fill up before the 
burst was over.
In this work, we consider the HERB data incomplete when the data do not cover
the burst duration ($T_{90}$) or when the data do not include
the main peak episodes (this could occur if a burst data accumulation was 
triggered by weak precursor).
In such cases, or if HERB was not available, 16-energy-channel MER data were 
used instead.  
Although MER provides medium energy resolution, it has much finer time 
resolution (16 and 64 ms) than HERB, making it possible to re-create the time 
resolution of the missing HERB.  
In the MER data, the CONT data are used as background; therefore, the spectra 
accumulated before the trigger time and after 163.8 seconds (the MER
accumulation time) are identical to those of CONT, with 2.048-s integration 
times.
The downside of using MER is that the data are summed over multiple sets of
detectors (usually two to four), and therefore, systematic errors tend to 
dominate over statistic errors, especially for bright bursts.
Systematic errors cannot be modeled into the analysis and can contribute to 
large $\chi^2$ especially at lower energies due to the high counts.
This systematic effects can also be visible in the single-detector data (HERB
and CONT) but are much more often found in the MER data of bright bursts.
Examples of HERB and MER spectra of the same bright burst (GRB 910503, trigger 
number 143) that show such systematic deviations are seen in Figure 
\ref{fig:system}.
\placefigure{fig:system}
Possible contributions to the systematics include the uncertainties from the LAD 
calibration \citep{pre98a} and the DRM \citep{pen95, har02}.

Lastly, in the cases where neither HERB nor MER were available for the analysis, 
16-energy-channel CONT data with 2.048-s time resolution from the brightest LAD 
were used.
Despite the lack of sub-second time resolution, the advantage of CONT is that 
the data are from a single detector and continuous.  
Thus, significant precursor activities may be included in the analysis using 
the CONT data.
This was not possible for HERB, because the data accumulations always started 
at triggers, and the background data (HER) did not provide a sufficient time
resolution for pre-trigger data.
For each of 350 bursts, the data type used is listed in column 4 of Table 
\ref{tab:grblist}, and the total number of bursts using each data type is 270 
(HERB), 52 (MER), and 28 (CONT).

\subsection{Time Interval Selection}

We binned the data in time until each spectrum has a large enough 
signal-to-noise ratio (S/N) in the entire LAD energy band to ensure acceptable 
statistics for the time-resolved spectral analysis.  
The S/N is calculated based upon the background model of each burst. 
The background model is determined by fitting (each energy channel separately)
a low-order ($\leq$ 4) polynomial function to spectra that cover time intervals 
before and after each burst, for at least a few hundred seconds.
In cases where the background data are not available for a sufficiently long 
period, the longest available time intervals were used, and the background model 
was checked against those determined using other LAD data available for the same 
burst.  The burst start time is usually the trigger time. 
In the cases where CONT or MER were used, bright pre-trigger activity 
(containing significant amount of emission) was also included in the analysis, 
and thus the start time can be negative relative to the trigger time.

\begin{figure}
\epsscale{0.6}
\centerline{
\plotone{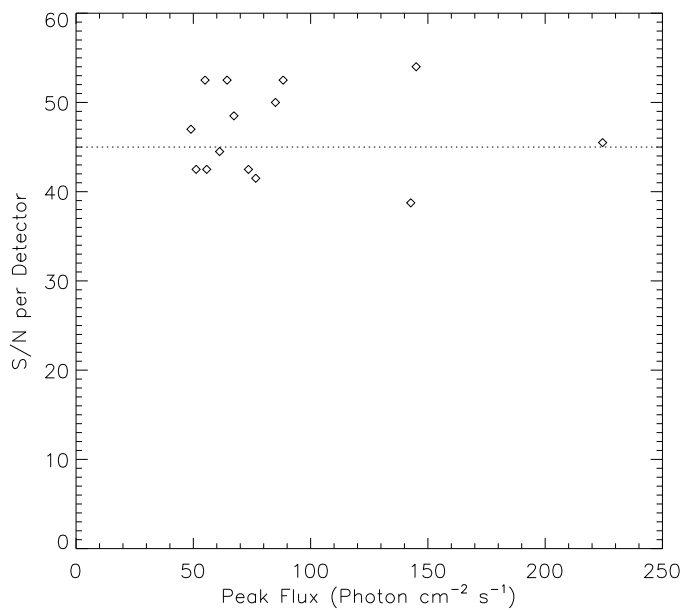}}
\caption{
S/N per detector for 15 MER bursts that produced the equivalent number 
of spectra to HERB binned by S/N $\geq 45$ case as a function of the peak photon
flux (128 ms, 30 -- 2000 keV). The dotted line marks 45$\sigma$.}
\label{fig:snr}
\end{figure}

In SP1, a minimum S/N of 45 was used for the time binning of spectra regardless
of the data types used.
The S/N level was chosen so that each spectral resolution element 
(non-overlapping full-width at half-maximum resolution elements, of which there 
are about 16 -- 20 in the LAD energy range) has approximately 10$\sigma$ of the 
signal on average, assuming a flat spectrum.
The noise (or $\sigma$) defined here is the Poisson error of the observed 
total counts, including the background counts.
Binning by a constant S/N, however, usually yields on average a few times larger 
number of time-resolved spectra when MER data are used, than the HERB and CONT 
cases.
This is because the MER data are summed over two to four detectors, while HERB 
and CONT are single-detector data, and thus the MER data have higher S/N to 
begin with, on top of having a finer time resolution.
As a result, bursts with MER data were overrepresented in the time-resolved
spectral sample of SP1 due to the choice of data type, which in turn
slightly biased the spectral parameter and model statistics presented there.
The MER bursts do provide more time-resolved spectra on average since we use the
MER data for long, bright GRBs; however, this is caused by the nature of the 
bursts and should not be caused by the choice of data type.
To minimize the over-representation caused by the data type selection, we 
investigated which S/N value for MER would yield comparable numbers of 
time-resolved spectra compared with HERB binned by S/N $\geq 45$, 
for the same bursts.
Since the MER data are mainly used for bright, long bursts, and also the 
oversampling problems are more likely to occur when the photon flux is high,
we selected 15 bursts with high peak photon flux ($\gtrsim 50$ photons 
s$^{-1}$ cm$^{-2}$ in 128 ms, $\sim$ 30 -- 2000 keV, determined using HERB data), 
for which both MER and HERB data were available.
We re-binned the HERB data with minimum S/N of 45, and reproduced the same 
number of spectra for the same time intervals with MER data, by increasing the 
minimum S/N in steps.
We found that a minimum S/N of 45~{\it per~detector} could roughly accomplish 
this, regardless of the brightness, as shown in Figure \ref{fig:snr}.
\placefigure{fig:snr}
Therefore, the minimum S/N used for the time binning is 45 $\times$ number of 
detectors, with the exception of 10 MER bursts.
These 10 bursts were mostly with three or four detectors and the minimum S/N 
was found to be too high, for various reasons, when compared with the available 
portion of single-detector data (i.e., HERB or CONT) for the same burst.
Consequently, for these bursts, the minimum S/N was reduced by steps of 45 until 
the binning comparable to that of HERB was achieved.

After time-binning by the minimum S/N, the last time interval, with a S/N less 
than the minimum value, was dropped.
Although the last time bin may constitute a significant tail portion of the 
burst, we found that this exclusion of the last time bin does not greatly affect 
the time-integrated spectral fits.
This is true even when the resulting time interval that is fitted is much shorter 
than the $T_{90}$ of the burst (e.g., a burst with a very long tail).
Unlike for SP1 where bursts with less than eight spectra were dropped, no bursts 
that satisfy the burst selection criteria described above are excluded, 
regardless of the resulting number of spectra after binning in each burst.  
This allowed the inclusion of 17 short GRBs ($T_{90} < 2.0$ s) as well as 
several dimmer bursts in this work.
Note that a set of time-resolved spectra comprises the time-integrated spectrum
of each burst; therefore, the time interval of the integrated spectrum is the
sum of the intervals of all the resolved spectra within the burst.
Most integrated spectra cover the $T_{90}$ duration.
Columns 6 and 7 of Table \ref{tab:grblist} list the time intervals used
for each burst.
There are 11 bursts (BATSE trigger numbers 298, 444, 1525, 2514, 2679, 3087, 
3736, 6240, 6293, 7457, 7610) that provided only one spectrum as a result of 
the time binning.
In addition, there are six weak bursts (2112, 3044, 3410, 3412, 3917, and 6668) 
for which the detection of the entire event was $< 45\sigma$: they provided 
only one spectrum of 28$\sigma$, 15$\sigma$, 26$\sigma$, 28$\sigma$, 
40$\sigma$, and 28$\sigma$, respectively.  
These six spectra were still included in the sample for completeness.
Four of these six bursts are short GRBs.
For these bursts that provided only one spectrum, the same spectra are 
considered as both the time-integrated and time-resolved spectra, and they are 
indicated by the prefix ``W" in column 1 of Table \ref{tab:grblist}.
It must also be noted that there are eight bursts (indicated by the prefix ``C" 
in column 1 of Table \ref{tab:grblist}) whose time-integrated spectra were used 
for calibration of the eight LADs \citep{pre98a}.
Since the fits to a two-component empirical GRB model (\S\ref{sec:models}, 
Equation \ref{band_model})
was used for the calibration, the time-integrated spectra of these calibration 
bursts are, by default, expected to give small $\chi^2$ values when fitted with a
two-component model (Equations \ref{band_model} \& \ref{sbpl_model}), although the 
time intervals used for the calibration and for this work are different.

\subsection{Energy Interval Selection}\label{sec:energy_selection}

All LADs were gain-stabilized; therefore, the usable energy range for spectral
analyses is $\sim$ 30 keV -- 2 MeV for all bursts.
The lowest seven channels of HERB and two channels of MER and CONT are 
usually below the electronic lower-energy cutoff and were excluded.
Likewise, the highest few channels of HERB and normally the very highest channel 
of MER and CONT are unbounded energy overflow channels and also not usable.  
The actual energy range used in the analysis for each burst is shown in columns 
8 and 9 of Table \ref{tab:grblist}. \\

\section{Spectral Analysis}\label{sec:analysis}
Our sample consisted of 350 GRBs, providing 350 time-integrated spectra 
and 8459 time-resolved spectra.
We analyzed both time-integrated and time-resolved spectra, each fitted by a set 
of photon models that are commonly used to fit GRB spectra.
Each of the photon models used consists of a different number of free parameters 
and thus, provides different degrees-of-freedom (dof) for each fit.
This allows statistical comparisons among the model fits.
The fitting procedures and the photon models are discussed in this section.

\subsection{Spectral Fitting Software}\label{sec:fit}
For the spectral analysis presented herein, we used the spectral analysis software 
RMFIT, which was specifically developed for burst data analysis by the BATSE team 
\citep{rmfit}.
It incorporates a fitting algorithm MFIT that employs the forward-folding method 
\citep{bri96a}, and the goodness of fit is determined by $\chi^2$ minimization.
One advantage of MFIT is that it utilizes model variances instead of data 
variances, which enables more accurate fitting even for 
\hbox{low-count} data \citep{for95}.
We analyzed both time-integrated and time-resolved spectra for each burst,
using a set of photon models described below.

\subsection{Photon Models}\label{sec:models}
We have selected five spectral models of interest to fit the BATSE GRB spectra, 
three of which (BAND, COMP, and SBPL) were also employed in SP1.  
Having a variety of models in fitting each spectrum eliminates the need for 
manipulating one model, such as the ``constrained" Band function introduced by 
\citet{sak04a}, which requires some presumptions of the form for the original 
photon spectrum.
GRB spectra are usually well-represented by a broken power law in the BATSE 
energy band.
However, it is possible that the break energy lies outside the energy range, or 
that the spectrum is very soft or dim and the high-energy component is not 
detected.
Therefore, we use a single power-law and a power-law with exponential cutoff model
that may accommodate such spectra, in addition to the more commonly-fit broken 
power-law models.  We review each model used in the analysis below.
All models are functions of energy $E$, measured in keV.

\subsubsection{Power Law Model (PWRL)} \label{sec:pwrl}

The first model is a single power law with two free parameters,
\begin{equation}\label{pwrl_model}
f_{\rm PWRL}(E) = A \left(\frac{\textstyle E}{\textstyle E_{\rm piv}}\right)
   ^{\lambda} ,
\end{equation}
where $A$ is the amplitude in photons~s$^{-1}$~cm$^{-2}$~keV$^{-1}$,
$\lambda$ is a spectral index, and the pivot energy $E_{\rm piv}$ was kept
constant at 100 keV for this work.
The use of this model was motivated by the fact that the break energy of a
broken power law spectrum could lie well outside the LAD passband.
There may also be a case where the signal is weak and the break energy 
cannot be adequately determined.
In such cases, the two-parameter single power law should be able to fit 
the spectra better than the other models with more parameters.

\subsubsection{The GRB Model (BAND)} \label{sec:band}

The next model is the empirical model most widely used to fit GRB spectra
\citep{ban93}:
\begin{eqnarray}\label{band_model}
f_{\rm BAND}(E) = 
   A \left( \frac{\textstyle E}{\textstyle 100}\right)^{\scriptstyle \alpha} 
            \exp{\left(
            -\frac{\textstyle E(2+\alpha)}{\textstyle E_{\rm peak}}
            \right)} &
   \rm{if~~} E < E_{\rm c} 
   \\ \nonumber
f_{\rm BAND}(E) = 
   A  \left[
      \frac{\textstyle (\alpha-\beta)E_{\rm peak}}{\textstyle 100(2+\alpha)}
      \right]^{\alpha-\beta}
            \exp{(\beta-\alpha)} 
            \left(\frac{\textstyle E}{\textstyle 100}\right)^{\beta} &
   \rm{if~~} E \geq E_{\rm c},
\end{eqnarray}
where
\begin{displaymath}
E_{\rm c} = (\alpha-\beta) \frac{\textstyle E_{\rm peak}}{\textstyle 2+\alpha}
\equiv (\alpha - \beta) E_{\rm 0}.
\end{displaymath}
The model consists of four parameters:  
the amplitude $A$ in photons~s$^{-1}$~cm$^{-2}$~keV$^{-1}$,
a low-energy spectral index $\alpha$, a high-energy spectral index $\beta$,
and a $\nu F_{\nu}$ peak energy $E_{\rm peak}$ in keV, which is related to
the $e$-folding energy, $E_{\rm 0}$.
$\nu F_{\nu}$ spectrum represents the total energy flux per energy band
(i.e., power density spectrum, $E^2 f(E)$).
We stress that the $\alpha$ index characterizes an asymptotic power-law (i.e., 
the tangential slope determined at $E = 0$ in a logarithmic scale).
This may not characterize the actual logarithmic tangential slope determined 
within the data energy range, when the $e$-folding energy $E_{\rm 0}$ 
approaches the lower energy bound.
It has now become common practice to use this empirical model to fit 
both time-integrated and -resolved GRB spectra.
Very frequently, however, we find some time-resolved spectra cannot be
adequately fitted with this model.
By fitting this model to both time-integrated and time-resolved spectra, 
we test the validity of the scope of this model.

\subsubsection{The GRB Model with fixed $\beta$ (BETA)}\label{sec:beta}
The BETA model is a variation of the BAND, with a fixed high-energy index 
$\beta$, and is only used to fit the time-resolved spectral fits.
The fixed value of $\beta$ is determined from the time-integrated spectral fit 
using the regular BAND model.
Since $\beta$ is fixed in the fit, this is essentially a 3-parameter model 
with $A$, $\alpha$, and $E_{\rm peak}$.
Our motivation for using this model is to test the hypothesis that the energy 
distribution of the shock-accelerated electron remains constant throughout a 
burst. 
The post-shock electron distribution should be thermal at low energies and 
non-thermal (power law) at high energies \citep{tav96b, bar04}.
The synchrotron spectrum emitted by electrons in a power-law distribution,
$N(\gamma)d\gamma = \gamma^{-p}d\gamma$, is a power-law of index $-(p+1)/2$
\citep{ryb79}, which should correspond to the high-energy index, $\beta$.
It has been found that for Fermi-type acceleration, the accelerated 
particles have a power-law distribution with index $p \sim 2.2 - 2.3$ that is 
constant in time \citep{gal02}. 
Therefore, if $p$ remains constant throughout a burst, $\beta$ should also 
remain constant in the context of the synchrotron shock model.
In fact, it has been found with a smaller sample that the majority of GRBs do 
not exhibit strong evolution in $\beta$ \citep{pre98a}, so we examine this here 
with a larger sample.

\subsubsection{Comptonized Model (COMP)} \label{sec:comp}

The next model considered is a low-energy power law with an exponential  
high-energy cutoff.
It is equivalent to the BAND model without a high energy power law, namely 
$\beta \rightarrow -\infty$, and has the form
\begin{equation}\label{comp_model}
f_{\rm COMP}(E) = A \left(\frac{\textstyle E}{\textstyle E_{\rm piv}}\right)
   ^{\alpha} 
   \exp{\left(-\frac{\textstyle E(2+\alpha)}{\textstyle E_{\rm peak}}\right)}.
\end{equation}
Like the PWRL case, $E_{\rm piv}$ was always fixed at 100 keV in this work; 
therefore, the model consists of three parameters: $A$, $\alpha$, and 
$E_{\rm peak}$.
There are many BATSE GRB spectra that lack high-energy photons \citep{pen97},
and these no-high-energy spectra are usually fitted well with this model.
Another case where this model would be a good fit is when the $e$-folding
energy ($E_{\rm 0} \equiv E_{\rm peak}/(2 + \alpha)$) approaches $\sim$ 1 MeV, 
and the high-energy index of the BAND model cannot be determined by the data.
The model is so named because in the special case of $\alpha = -1$, it represents
the Comptonized spectrum from a thermal medium; however, $\alpha$ is kept as
a free parameter here.
Note that when $\alpha < -2$, the exponential term in the model diverges (i.e., 
$E_{\rm 0}< 0$), and the resulting spectrum has a concave-up shape.

\subsubsection{Smoothly-Broken Power Law (SBPL)}\label{sec:sbpl_model}

The last model we have selected is a broken power law with flexible curvature 
at the break energy, and thus the model can accommodate spectra with very sharp 
breaks, as well as ones with very smooth curvature.  
This SBPL model is expressed by 
\begin{equation}\label{sbpl_model}
f_{\rm SBPL}(E) = A \left(\frac{\textstyle E}{\textstyle E_{\rm piv}}\right)^b 
   10^{(a - a_{\rm piv})}, 
\end{equation}
where
\begin{eqnarray*}
a = m \Lambda \ln{\left(
   \frac{\textstyle e^{q} + e^{-q}}{\textstyle 2}\right)},
&a_{\rm piv} &= m \Lambda \ln{\left(
   \frac {\textstyle e^{q_{\rm piv}} + e^{-q_{\rm piv}}}{\textstyle 2}
   \right)}, \\
q = \frac {\textstyle \log {(E/E_{\rm b})}} {\textstyle \Lambda}, 
&q_{\rm piv} &= \frac {\textstyle \log {(E_{\rm piv}/E_{\rm b})}} 
   {\textstyle \Lambda}, \nonumber \\
m = \frac{\textstyle \lambda_2 - \lambda_1}{\textstyle 2}, 
\rm{ and }
&b &= \frac{\textstyle \lambda_1 + \lambda_2}{\textstyle 2}. 
\end{eqnarray*}
The parameters are
the amplitude $A$ in photons~s$^{-1}$~cm$^{-2}$~keV$^{-1}$,
a lower power-law index $\lambda_{1}$, a break energy $E_{\rm b}$ in keV,
a break scale  $\Lambda$, in decades of energy, and
an upper power-law index $\lambda_{2}$.
The amplitude $A$ represents the photon flux at $E_{\rm piv}$.
The model introduces a break scale $\Lambda$ as the fifth parameter; this is 
thus a five-parameter model.
Like the PWRL and COMP models above, the pivot energy $E_{\rm piv}$ is 
always fixed at 100 keV here.  
The amplitude $A$ is determined at this $E_{\rm piv}$, and it represents a 
convenient overall energy scale.  
This model was originally created to be implemented into MFIT, and the full
derivation is found in Appendix \ref{ap:sbpl_model}.
The basic idea in deriving this model was to have the derivative of the photon
flux (in logarithmic scale) to be a continuous function of the hyperbolic
tangent \citep[SP1]{wingspan, ryd99}.
The main difference between this model and the BAND model is that the break 
scale is not coupled to the power laws, and it approaches the asymptotic 
low-energy power law much quicker than the BAND model case.
Therefore, the low-energy spectral index $\lambda_{1}$ could characterize values 
that are closer to the $true$ power law indices indicated by the actual data 
points, than is possible with $\alpha$ of the BAND model. 
Note also that as $\Lambda \rightarrow 0$, the model reduces to a sharply-broken 
power law.

However, introducing a fifth parameter can be a problem in fitting the LAD 
spectra.
Although the HERB data provides 126 energy channels, the energy range 
encompasses only about 20 energy resolution elements, as mentioned earlier.
Fitting a four-parameter model to the HERB data can already cause the covariance 
matrix between parameters [$C$] to be ill-determined, resulting in unconstrained 
parameters.
This is indicated by a condition number for [$C$]$^{-1}$ that is of the order of 
the reciprocal of the machine precision, meaning that the matrix is nearly 
singular \citep[e.g.,][]{pre92}.
Consequently, having an additional free parameter usually results in highly 
cross-correlated, unconstrained parameter determinations, and is not favored.
For this reason, in SP1, $\Lambda$ was fixed for each time-resolved spectral
fit to the value determined by the time-integrated fit for the corresponding 
burst; however, there is no reason to presume that $\Lambda$ remains constant 
throughout a burst, and also it could be problematic if the initial 
time-integrated break scale is unconstrained. 
On the other hand, we may not be able to constrain $\Lambda$ any better than a 
particular value, due to the finite energy resolution of the LADs, even if the 
five-parameter model fit can be done. 
To resolve this issue, we have simulated SBPL spectra with various parameters 
and fitted these spectra with the SBPL model for various values of $\Lambda$.  
The simulation results are discussed in \S \ref{sec:sbpl_sim}.

We emphasize that the break energy $E_{\rm b}$ of the SBPL model should not 
be confused with $E_{\rm peak}$ of the BAND and COMP models.  
The break energy is simply the energy at which the spectrum changes from the 
low-energy to high-energy power law, whereas $E_{\rm peak}$ is the energy at which 
the $\nu F_{\nu}$ spectrum peaks.
The break energy $E_{\rm b}$ is also different from the characteristic energy 
$E_{\rm c}$ in the BAND model (Equation \ref{band_model}), which is the energy 
where the low-energy power law 
with exponential cutoff ends and the pure high-energy power law starts.
However, the $\nu F_{\nu}$ peak energy of the SBPL spectra, as well as the 
spectral break energy of BAND, can be easily derived (see the Appendices
\ref{ap:sbpl_ep} and \ref{ap:band_eb}) for comparison among the various models, 
which we have done here for the first time. \\

\section{Spectral Simulations}\label{sec:simulation}

In order to interpret the quantitative analysis results correctly, we first need 
to understand the general characteristics and behavior of each photon model when 
applied to the BATSE LAD GRB spectra.
Therefore, we have generated a large set of simulated burst spectra with various 
spectral shapes and signal strengths, and subjected them to our analysis regime.
To create a set of simulated count spectra, a source photon model with specific 
parameters and a background count model of an actual (typical) LAD burst are 
folded through the corresponding LAD DRM, and Poisson noise is added.
It should be noted that the simulated spectra do not include any sources of 
systematic effects that are present in the real spectra.
There are two main objectives in simulating data for this study.  
One is to investigate the behavior of the BAND and COMP models in the limit of low 
S/N and the other is to explore the break scale determination of SBPL.

\subsection{BAND vs. COMP}\label{sec:band_sim}

\begin{figure}
\epsscale{0.6}
\centerline{
\plotone{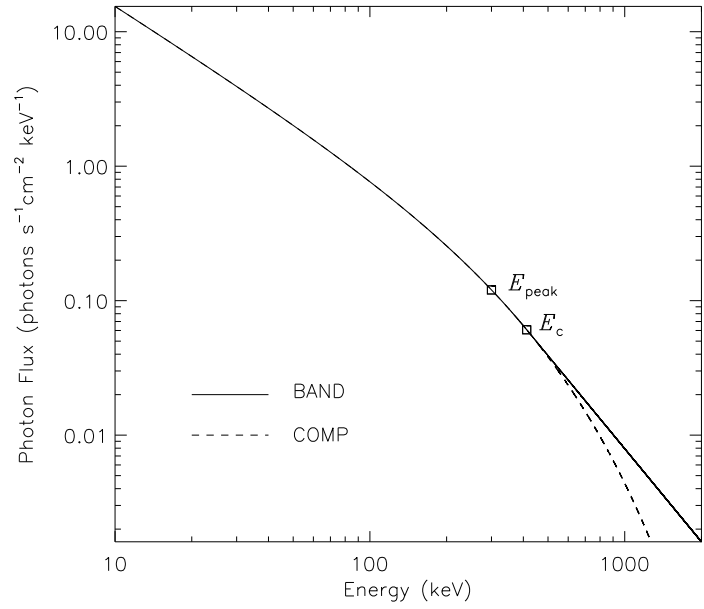}}
\caption{Comparison of BAND and COMP models for $A = 0.1$, $\alpha = -1.2$,
$E_{\rm peak} = 300$ keV, and $\beta = -2.3$. 
$E_{\rm c}$ is where the high-energy power law of BAND begins.}
\label{fig:band_comp}
\end{figure}

The broken-power law nature of the GRB spectra indicates that there typically 
are considerably lower photon fluxes at higher energies.
Because of this,  there is a good chance that the LADs are not sensitive enough 
to detect the non-thermal high-energy power law component of spectra in fainter 
bursts.
In such cases, even if the original source spectra have high-energy components,
our data may not be able to identify this component and therefore, the 
no-high-energy COMP model may statistically fit as well as the BAND model.
As an example, we show in Figure \ref{fig:band_comp} a comparison between the
BAND and the COMP photon spectra with the same $A$, $\alpha$, and 
$E_{\rm peak}$ values.
\placefigure{fig:band_comp}
In fact, \citet{ban93} found that the simulated four-parameter BAND spectra 
with low S/N could be adequately fitted with the three-parameter COMP model, 
although there were some shifts in the COMP-fit parameters.
In order to validate this for our dataset and using our analysis tool (RMFIT, 
\S \ref{sec:fit}), we have further explored these two models by creating sets 
of simulated burst spectra, based on the actual fit parameters of some of 
the observed GRB spectra that clearly have high-energy components.

To start with, we selected a sample of six bright GRB spectra (three each with 
HERB and CONT) to which the BAND model fits substantially better than the COMP 
model, with well-constrained parameters, resulting in large $\chi^2$ 
improvements ($\Delta\chi^2 > 20$) for the additional 1 dof.
This assures that the observed spectra have a high-energy power law component
that is statistically significant.
Based on the spectral parameters provided with the BAND fits to the sample 
spectra (i.e., spectra with high-energy component), sets of 100 simulated 
spectra with various S/N were created.
For the S/N variation, we used the actual fitted amplitude value and the values 
decreased by a step of a factor of 10 until the S/N was a few.
As a result, 
a total of 19 sets provided $ 2 \lesssim \langle {\rm S/N} \rangle \lesssim 200$
in the entire LAD energy range, based on the typical LAD background count model
that was taken as input for the simulation.

The sets of simulated spectra were then fitted with the BAND and the COMP models.
Some example results are presented in Table \ref{tab:comp_band_res}
(upper two tables), where
$\langle x \rangle$ indicates a median value of the parameter $x$ and the 
standard deviation is $\sigma_x = \sqrt{\langle x^2 \rangle - \bar{x}^2}$.
\placetable{tab:comp_band_res}
The fit results indicated that S/N $\gtrsim 80$ is needed for the BAND fits
to be better than the COMP fits, at $\gtrsim $ 99.9\% 
confidence level ($\Delta \chi^2 > 10$ for $\Delta {\rm dof} = 1$).
For spectra with S/N $\sim$ 40, the confidence level of improvements in
BAND over COMP were $<$ 70\%.
Given that the minimum S/N of our time-resolved spectra in this work was set to 
45, for many dimmer spectra we should only be able to determine a better fit 
between the BAND and COMP by only about 1$\sigma$ (68.3\%), although this may 
depend on the spectral parameters in each fit.

In accord with the \citet{ban93} results, the COMP model resulted in higher 
$E_{\rm peak}$ values and steeper values of $\alpha$, due to compensating 
for the lack of high-energy spectral component in the model.  
Consequently, the difference in the COMP $E_{\rm peak}$ and the actual 
$E_{\rm peak}$ value tends to be correlated with $\beta$.
In addition, a strong anti-correlation was always found between $E_{\rm peak}$ 
and $\alpha$ in both BAND and COMP fits, regardless of S/N or values of other 
parameters.  
As a result, the amplitude $A$ is also highly anti-correlated to $E_{\rm peak}$ 
because the parameter $A$ of BAND and COMP is the photon flux at 100 keV 
of the low-energy power-law $without$ the exponential cutoff. 
This is different from $A$ of the SBPL model.
We also fitted the five-parameter SBPL model to the same sets of simulated
BAND spectra, to investigate the possible parameterization differences in the
indices.
We found that the SBPL $\lambda_{1}$ tends to be smaller than $\alpha$
while $\lambda_{2}$ seems to be consistent with $\beta$, as we expected.

\begin{figure}
\epsscale{0.6}
\centerline{
\plotone{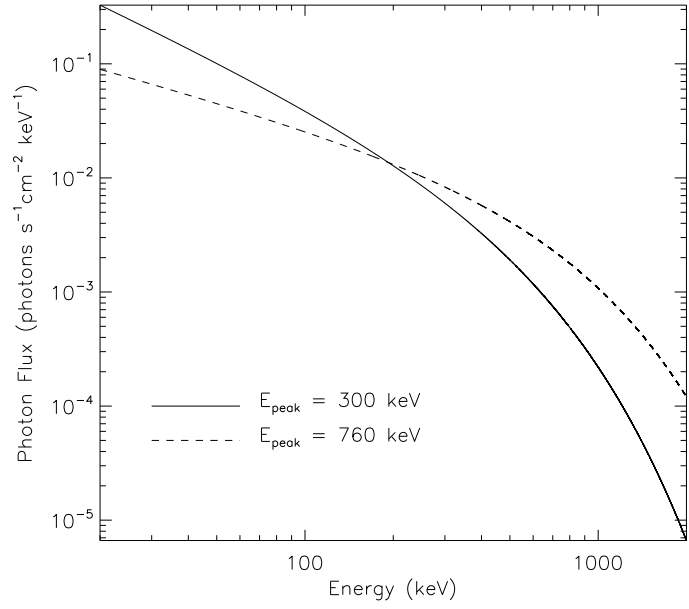}}
\caption{Two COMP spectra that produced simulated spectra with S/N $\sim$ 80.  
Low-$E_{\rm peak}$ spectrum (solid curve) can be fitted with the BAND model
much more frequently than the high-$E_{\rm peak}$ spectrum (dashed curve).
}
\label{fig:comp2}
\end{figure}

Similarly, we have also simulated sets of COMP spectra with low and high 
$E_{\rm peak}$ values (300 and 760 keV, respectively) with different S/N 
($ 80 \lesssim \langle {\rm S/N} \rangle \lesssim 500$), and 
fitted them with the BAND model.
The results are summarized in Table \ref{tab:comp_band_res} (lower two tables). 
In the high $E_{\rm peak}$ case, we found that the BAND fits did not converge 
about a third of the time, regardless of the S/N.
The fitting failure is caused by a very poorly-constrained parameter ($\beta$ in
this case).
On the other hand, in the low $E_{\rm peak}$ case, the number of failed fits
was significantly smaller for the spectra with lower S/N.
In both cases, the $\langle E_{\rm peak} \rangle$ and $\langle \alpha \rangle$ 
values fitted by the BAND model were consistent with the simulated COMP
parameters, while $\langle \beta \rangle$ only gave upper limits in a range
of $\sim -2.5$ to $\sim -4$.
The simulation results suggest that the BAND model fails to converge when
a spectrum has sufficient high-energy photon flux but lacks the high-energy 
power-law component with finite spectral index.
The spectrum in such a case essentially is the COMP model, which is the 
BAND model with $\beta \rightarrow -\infty$ (see \S\ref{sec:comp}).
As an example, Figure \ref{fig:comp2} shows two COMP models that produced 
simulated spectra with S/N $\sim$ 80 but with different $E_{\rm peak}$ values.
\placefigure{fig:comp2}
As mentioned above, the BAND model fails to fit the high-$E_{\rm peak}$
spectrum much more frequently than the low-$E_{\rm peak}$ one.
From Figure \ref{fig:comp2}, it is evident that the high-$E_{\rm peak}$ spectrum 
has much larger photon flux at about 1 MeV although the overall signal
strengths are similar.
Therefore, it is very likely that the spectra that the BAND model fails
to fit lack a high-energy power law component, yet this does not mean
that these are the no-high-energy (NHE) spectra identified by \citet{pen97},
which show no counts above 300 keV. 

\subsection{SBPL Break Scales}\label{sec:sbpl_sim}
\begin{figure}
\epsscale{0.6}
\centerline{
\plotone{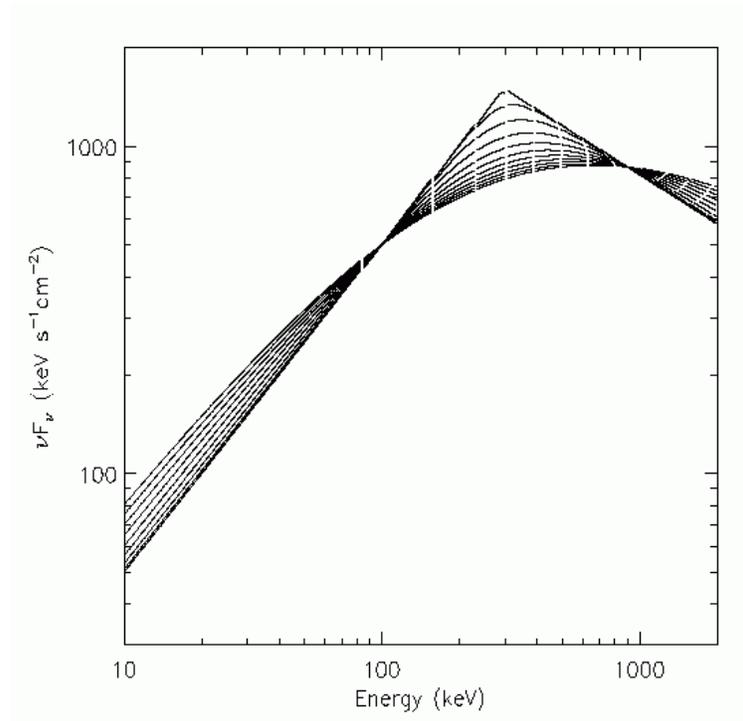}}
\caption{
The simulated SBPL model spectra with $\Lambda = $
0.01, 0.1, 0.2, 0.3, 0.4, 0.5, 0.6, 0.7, 0.8, 0.9, and 1.0, from top to 
bottom at 300 keV.  The other parameters are fixed at $A = 0.05$, 
$E_{\rm b} = 300$ keV, $\lambda_{1} = -1.0$, and $\lambda_{2} = -2.5$.}
\label{fig:sbpl_bsc}
\end{figure}

Another topic that needs to be addressed is the break scale ($\Lambda$) of the 
SBPL model, as mentioned in \S \ref{sec:sbpl_model}.  
The purpose of this simulation is to test the feasibility of performing the 
5-parameter SBPL model fits with $\Lambda$ as a free parameter, as well as to 
examine the capability of the determination of $\Lambda$ by alternatively using 
a set of 4-parameter SBPL models with fixed $\Lambda$.
First, we created sets of 100 simulated SBPL spectra with $\Lambda$ values 0.01 
and between 0.1 and 1.0 with an increment of 0.1 (11 total, in decades of 
energy), while keeping the other parameters fixed at typical fit values of 
$E_{\rm b}$ = 300 keV, $\lambda_{1} = -1.0$, and $\lambda_{2} = -2.5$.  
Figure \ref{fig:sbpl_bsc} shows the 11 simulated spectra in $\nu F_{\nu}$,
with $A$ = 0.05.
\placefigure{fig:sbpl_bsc}
The upper limit of $\Lambda = 1.0$ (in decades of energy) is reasonable, 
considering that the LAD spectra span less than two decades of energy.
The spectrum with $\Lambda = 0.01$ represents a sharply-broken power law.
To provide variations in the signal strength, the amplitude $A$ was set to 
a typical value of 0.05 in one group, and was 0.01 in the other group, 
corresponding to the median S/N of $\sim 100$ and $\sim 30$ per spectrum, 
respectively.
Each of the simulated spectra was then fitted with the full 5-parameter SBPL 
model allowing $\Lambda$ to vary, with a set of 4-parameter SBPL models, each 
with $\Lambda$ fixed to the 11 values mentioned above, as well as with the BAND 
and the COMP models for comparison.

\subsubsection{Finding the Break Scale} \label{sec:bscale}
\begin{figure}
\epsscale{1.0}
\centerline{
\plotone{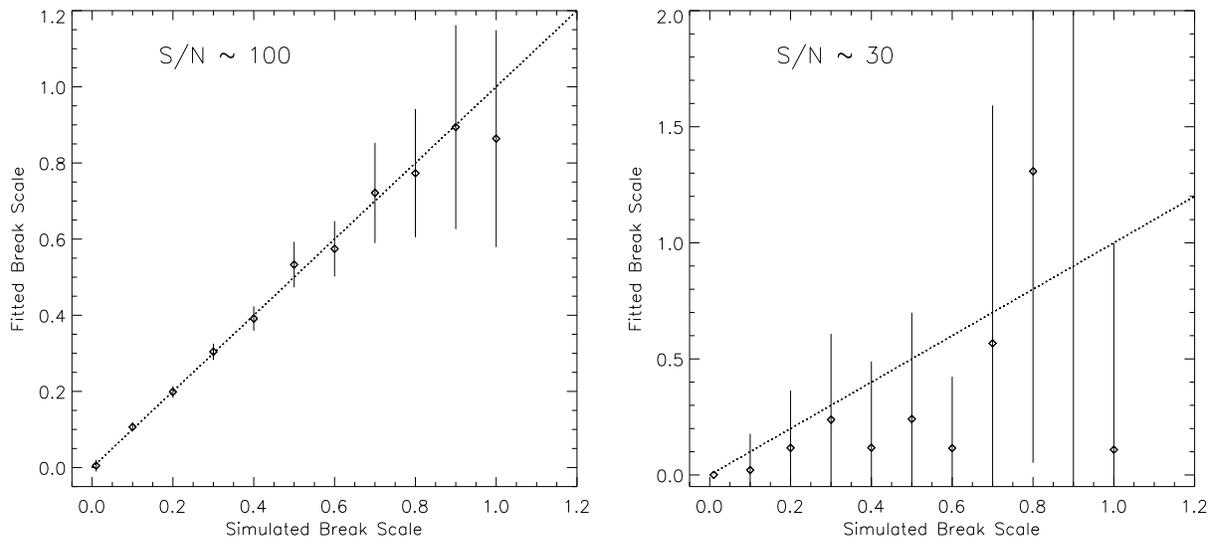}}
\caption{The break scale $\Lambda$ found by 5-parameter SBPL fits to
11 simulated spectra for bright ({\it left}) and faint ({\it right}) cases.  
Horizontal axis shows the simulated $\Lambda$ values
and vertical axis shows the fitted $\Lambda$ values.
The dotted line corresponds to the correct $\Lambda$.}
\label{fig:sbpl5par}
\end{figure}

In Figure \ref{fig:sbpl5par}, we show the $\Lambda$ values found by
the 5-parameter SBPL model fits, with $\Lambda$ as a free parameter. 
\placefigure{fig:sbpl5par}
For the high S/N case, the correct $\Lambda$ values were found up to 
$\Lambda \sim 0.6$, with relatively small dispersions. 
For the low S/N case, only the very sharp break ($\Lambda = 0.01$) was 
constrained by the 5-parameter fits.
As for the other parameters associated with the fits, we found that even for the 
bright cases, the 5-parameter fits resulted in relatively large uncertainties in 
all parameters, which worsened as $\Lambda$ became larger. 
This was also indicated by the large cross-correlation coefficients among all 
parameters, resulting from the fits.
This confirms that fitting five free parameters at once does not determine the
parameter with a good confidence, regardless of the S/N of spectrum, and 
therefore, the full 5-parameter fit is not favored.
It is, however, worth noting that despite the large errors, $\Lambda$ found by 
the 5-parameter fits may still provide a rough estimate of the break scale 
even for faint spectra, as a last resort.

Alternatively, we could employ the grid-search method \citep{bev03} using
a subset of 4-parameter SBPL fits with various fixed values of $\Lambda$ to 
determine the real $\Lambda$.
Having such a set of 4-parameter model fits to each spectrum enables us to 
construct a one-dimensional $\chi^2$ map for $\Lambda$, showing
$\chi^2$ as a function of $\Lambda$.
From the $\chi^2$ map, we can determine the most likely value of $\Lambda$
(where the $\chi^2$ is minimum) as well as the confidence interval, 
while having the other parameters still constrained.

\begin{figure}
\epsscale{0.6}
\centerline{
\plotone{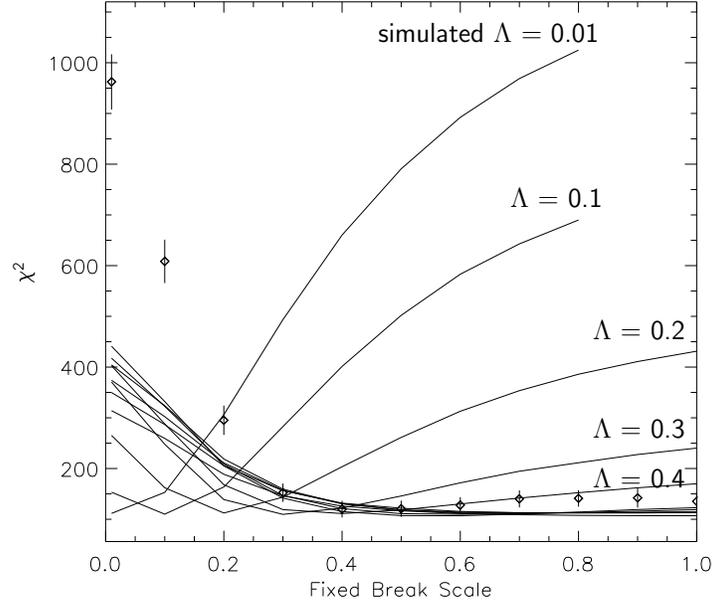}}
\caption{Median $\chi^2$ map of 4-parameter SBPL fits to 11 simulated spectra
(with $\Lambda$ = 0.01 to 1.0, from top to bottom; dof = 113) for the bright 
case (S/N $\sim$ 100).
Horizontal axis is the fixed $\Lambda$ values of the 4-parameter model.
The diamonds indicate the median $\chi^2$ from the BAND fits to 11
simulated spectra (dof = 112).}
\label{fig:sbpl_c2map}
\end{figure}

The $\langle \chi^2 \rangle$ map obtained from a 4-parameter model fitting 
of the bright (S/N $\sim$ 100) simulated spectra is shown in Figure 
\ref{fig:sbpl_c2map}.
\placefigure{fig:sbpl_c2map}
We find from the $\chi^2$ map that for $\Lambda \leq 0.4$ the set of 4-parameter 
fits yields a minimum for $\chi^2$ at the correct $\Lambda$ values with 1$\sigma$
uncertainties less than $0.01$.
However, for $\Lambda \geq 0.5$, the $\Lambda$ value could not be sufficiently 
constrained, especially at the upper ends, and the BAND model starts to give 
satisfactory fits that are statistically comparable to the SBPL model fits.
This suggests that in the case of $\Lambda \geq 0.5$ we can only determine 
the lower limit of $\Lambda = 0.5$ with confidence.
Furthermore, for $\Lambda > 0.6$, the uncertainties associated with other 
spectral parameters become large although they are still in agreement within 
the uncertainties with the simulated values.
As for the faint spectra (S/N $\sim$ 30), we found that the total change in 
$\chi^2$ for the entire set of $\Lambda$ values was only about 4, which is 
within the $2\sigma$ confidence interval for $\Delta {\rm dof} = 1$; therefore, 
the correct value of $\Lambda$ cannot be determined even with the use 
of 4-parameter fits, due to the low S/N.
In such cases, however, we also found that the $\Lambda$ determined from the
5-parameter SBPL fit can be used as an estimate.
In other words, the fit using a 4-parameter model with $\Lambda$ closest to the
$\Lambda$ found from the 5-parameter fit could yield parameters that are 
adequately constrained and still consistent with the actual simulated parameters.
The simulation was done both with 128-energy channel data and 16-energy channel
data, in order to investigate the possible effects that might arise from the
energy resolution issues.  We found no differences between the 128-channel data 
and the 16-channel data in determining the break scales.

Based on these simulation results, we concluded that the 4-parameter SBPL models 
with $\Lambda > 0.5$ do not contribute much additionally to our analysis;
therefore, we decided to use a set of 4-parameter SBPL models with $\Lambda =$ 
0.01, 0.1, 0.2, 0.3, 0.4, and 0.5, as well as 5-parameter SBPL model with 
$\Lambda$ varied, and $\Lambda$ fixed to the time-integrated fit value (for 
comparison with SP1).
With regard to cross-correlations among the spectral parameters, $E_{\rm b}$ and 
$\lambda_{2}$ are found to be always strongly anti-correlated in both 4-parameter 
and 5-parameter fits.
Moreover, the 4-parameter fits where $\Lambda$ is fixed tend to produce higher 
anti-correlation between $E_{\rm b}$ and $\lambda_{1}$ than the 5-parameter cases, 
which is expected for the fixed break scale cases.
Not surprisingly, in the 5-parameter fits, the $\lambda_{1}$ ($\lambda_{2}$) 
is more strongly correlated (anti-correlated) with $\Lambda$, as $\Lambda$ 
becomes larger.  There was no explicit difference found in these cross-correlations 
according to S/N, although in the faint case, the parameters were more 
difficult to be constrained.

\subsubsection{Comparison with BAND \& COMP}
\begin{figure}
\epsscale{1.0}
\centerline{
\plotone{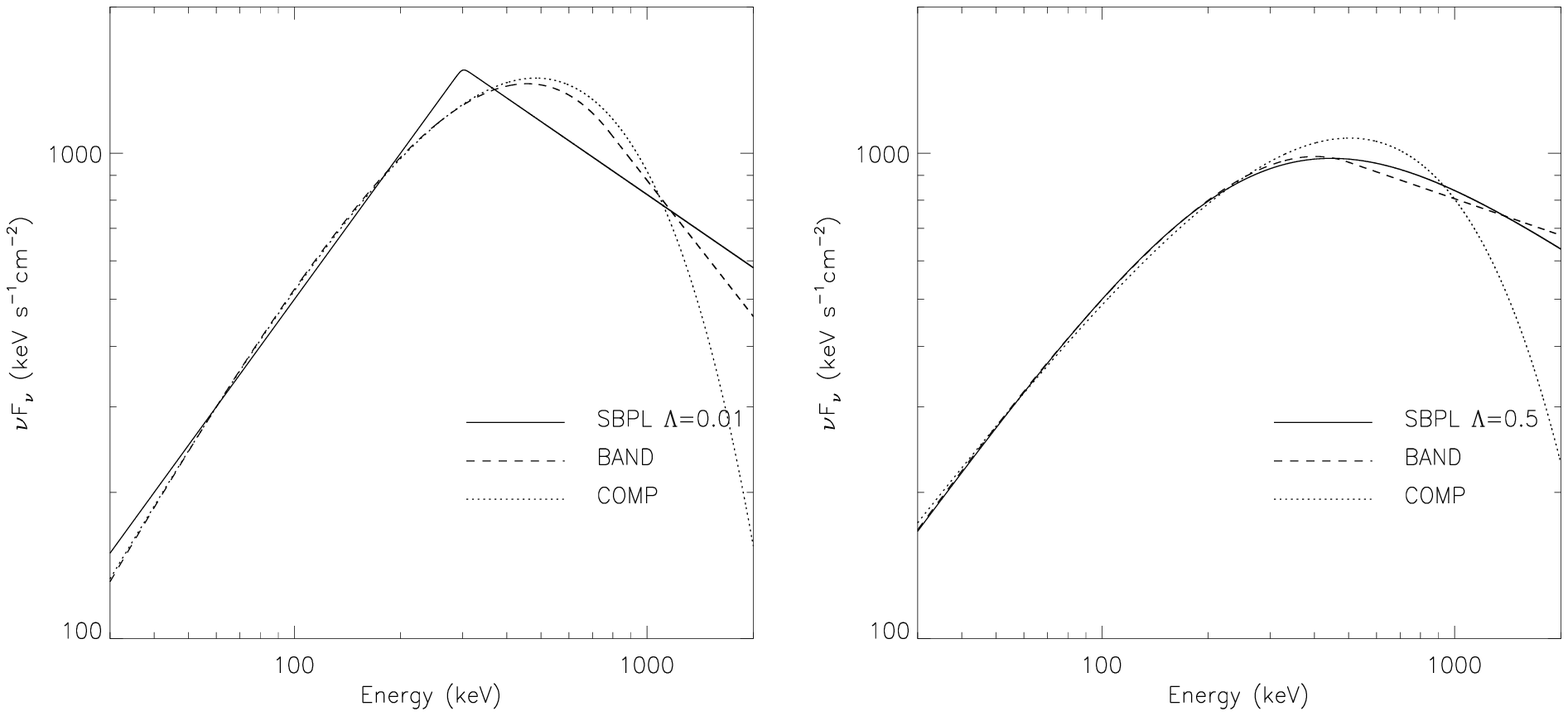}}
\caption{Example BAND (dashed lines) and COMP (dotted lines) fits to the 
simulated SBPL spectra (solid lines) with $\Lambda = 0.01$ ({\it left}) 
and with $\Lambda = 0.5$ ({\it right}).}
\label{fig:sbpl_band}
\end{figure}

The simulated SBPL spectra were also fitted with the BAND and COMP models.
The results of the BAND and COMP fits to the bright simulated SBPL spectra 
(i.e., S/N $\sim$ 100) are summarized in Table \ref{tab:sbpl_band_res}. 
\placetable{tab:sbpl_band_res}
As seen in the table (also in Figure \ref{fig:sbpl_c2map}), the BAND model is 
not able to adequately fit the SBPL spectra with relatively sharp break scale 
($\Lambda \lesssim 0.3$) because of its rather inflexible, smooth curvature.
The COMP model did not provide statistically acceptable fits, regardless
of the $\Lambda$ values, for this particular set of simulated spectra.
Generally, for sharply broken spectra with small $\Lambda$, the $E_{\rm peak}$ 
and $\alpha$ of BAND (also COMP) are larger than the SBPL ``$E_{\rm peak}$" 
and  $\lambda_1$, while the BAND $\beta$ is smaller than $\lambda_2$.
The opposite is true for smooth break spectra with large $\Lambda$.
The tendencies are clearly seen in the example $\nu F_{\nu}$ spectra in Figure 
\ref{fig:sbpl_band}, in which the BAND and COMP fits to the simulated SBPL 
spectra with $\Lambda = 0.01$ and 0.5 are shown.
\placefigure{fig:sbpl_band}
The BAND model seems to fit the SBPL spectra with $\Lambda \sim 0.4$ the best, 
at least for these given values of $E_{\rm b}$ and $\lambda_{1,2}$.  
The BAND fit to this spectrum, in fact, resulted in $E_{\rm peak}$ consistent 
with the SBPL ``$E_{\rm peak}$" and $\alpha$ larger than $\lambda_1$, which 
agrees with what was found from the SBPL fits to the simulated BAND spectra 
in an earlier section (\S \ref{sec:band_sim}).
We also observe in Table \ref{tab:sbpl_band_res} that the BAND fits yielded 
much smaller $E_{\rm peak}$ range ($\sim 390 - 460$~keV) than the simulated 
``$E_{\rm peak}$" ($\sim 300 - 670$~keV).
Although the COMP fits were significantly worse, the COMP $E_{\rm peak}$ was 
always higher and $\alpha$ was always softer than the BAND case, consistent 
with what we found in \S\ref{sec:band_sim}.

In terms of $\chi^2$ statistics, for the bright spectra with S/N $\sim$ 100, 
we find that the SBPL model can fit substantially better than the BAND model 
(at confidence level $>$ 99.9\%) to the simulated SBPL spectra with $\Lambda 
\leq 0.3$; however, for the higher values of $\Lambda$, BAND begins to fit 
statistically as well as SBPL fits.
Also in the bright spectrum case, the COMP model gave worse fits for all values 
of $\Lambda$, due to the lack of a high-energy component, and thus the SBPL 
fits were always better for this given set of the simulated spectral parameters.
In the case of dim spectra with S/N $\sim$ 30, on the other hand, the SBPL fitted 
better than the BAND or the COMP model for $\Lambda = 0.01$ and $0.1$, but only 
at confidence level of $\sim$ 90\%.  
Also for the dim spectra with $\Lambda > 0.1$, we found that the BAND and the 
COMP fits are statistically as good as the SBPL fit. \\

\section{Spectral Catalog and Analysis Results}\label{sec:results}
We fitted the five photon models (PWRL, BAND, BETA, COMP, SBPL; see
\S\ref{sec:models}) as well as SBPL with fixed break scales ($\Lambda =$ 0.01, 
0.2, 0.3, 0.4, 0.5, \& time-integrated value) to each of the 350 
time-integrated spectra and 8459 time-resolved spectra.
The spectral catalog containing all fit results is available electronically as 
a part of the public data archive at the High-Energy Astrophysics Science 
Archive Research Center (HEASARC)\footnotemark[3].
\footnotetext[3]{\url http://heasarc.gsfc.nasa.gov/docs/archive.html}
All the fit results are archived in the standard Flexible Image Transport 
System (FITS) format\footnotemark[4].\footnotetext[4]{\url http://fits.gsfc.nasa.gov}
The results of the comprehensive spectral analysis performed herein constitute 
the richest resource of GRB prompt emission spectral properties.
Therefore, careful examination of these results enables us to better constrain 
physical mechanisms for GRB prompt emission process.
These results also allows us to explore systematics that are internal to 
the spectral models employed.

The overall performance of each model in fitting all spectra is summarized in 
Table \ref{tab:goodfit}, in which the percentages of acceptable fits yielded by 
each model are shown.
The BETA model is excluded here because it is a special case of the BAND model
and was used only to investigate the constant-$\beta$ hypothesis in each burst.
The BETA model fits are explored in \S \ref{sec:beta_res}.
Also, the SBPL used here is the set of 4-parameter model fits, with the break 
scale $\Lambda$ determined according to minimum $\chi^2$ as described in \S 
\ref{sec:sbpl_sim}.
Therefore, we used the dof for 5-parameter fits to determine goodness of fits,
since we are indeed allowing the break scale (the fifth parameter) to vary.
From Table \ref{tab:goodfit}, it is clear that many spectra are adequately fitted 
with various photon models.
As determined solely by the $\chi^2$ of the fits, the SBPL model seems to be 
able to fit the data better than the other models, as seen in the 
Table \ref{tab:goodfit}, although the values are only slightly better than 
those obtained by the BAND model.
The time-resolved spectra provide better $\chi^2$ values, partially due to the 
lower S/N compared with the time-integrated spectra.
This is especially evident in the results for the COMP cases, possibly indicating
the existence of NHE spectra within high-energy bursts \citep{pen97}. 
As expected, the PWRL model resulted in poor fits for most of the spectra.
In the following sections, we look at the results of our spectral analysis
in terms of the parameter distributions, the model statistics, the comparison
between time-integrated and time-resolved spectra, correlations among the
spectral parameters, evolution of high-energy power law index ($\beta$), and
the comparison between short and long bursts.

\subsection{Spectral Parameter Distributions} \label{sec:par_dist}

The spectral parameters can be compared by two different aspects; namely, 
a comparison among parameters yielded by different models, and a comparison 
between time-integrated and time-resolved parameters. 
The comparison among the models reveals the internal characteristics of 
each model, whereas the comparison between the time-integrated and time-resolved 
parameters uncovers the differences internal to the spectra. 

Before comparing the fitted parameters of various models, there are some issues
to be discussed.  As mentioned in \S \ref{sec:models}, the parameterizations are 
different in each model.  
For clarity, the free parameters in each model are summarized in Table~\ref{tab:par}.
The main concern here is the difference in the low-energy spectral indices:
$\alpha$ of BAND and COMP and $\lambda_1$ of SBPL, where $\alpha$ is the 
asymptotic power-law index, while $\lambda_1$ is the index of the actual power 
law fit to the data.
The natural consequence of this is that $\alpha$ tends to be harder than 
$\lambda_1$ (i.e., $\alpha > \lambda_1$), when fitted to the same data, 
which was confirmed in the simulation study presented in \S\ref{sec:simulation}
(we note, however, $\alpha \sim \lambda_1$ if $\Lambda$ is large and/or 
$E_{\rm b}$ is low).
They are, therefore, not directly comparable.
In order to minimize the discrepancies, an ``effective" $\alpha$ 
($\alpha_{\rm eff}$) was introduced by \citet{pre98b}.
This is the tangential slope at 25 keV in a logarithmic scale, and is found to 
describe the data more accurately than the fitted asymptotic $\alpha$ value.
The 25 keV is the lower energy bound of LAD and therefore, $\alpha_{\rm eff}$ 
is the index of the low-energy power law within the data energy range.
Therefore, we employ the $\alpha_{\rm eff}$ instead of the fitted $\alpha$ 
values for BAND and COMP fits in the following parameter distribution 
comparisons. 
A detailed discussion of $\alpha_{\rm eff}$ can be found in Appendix 
\ref{ap:aleff}.

\begin{figure}
\centerline{
\plotone{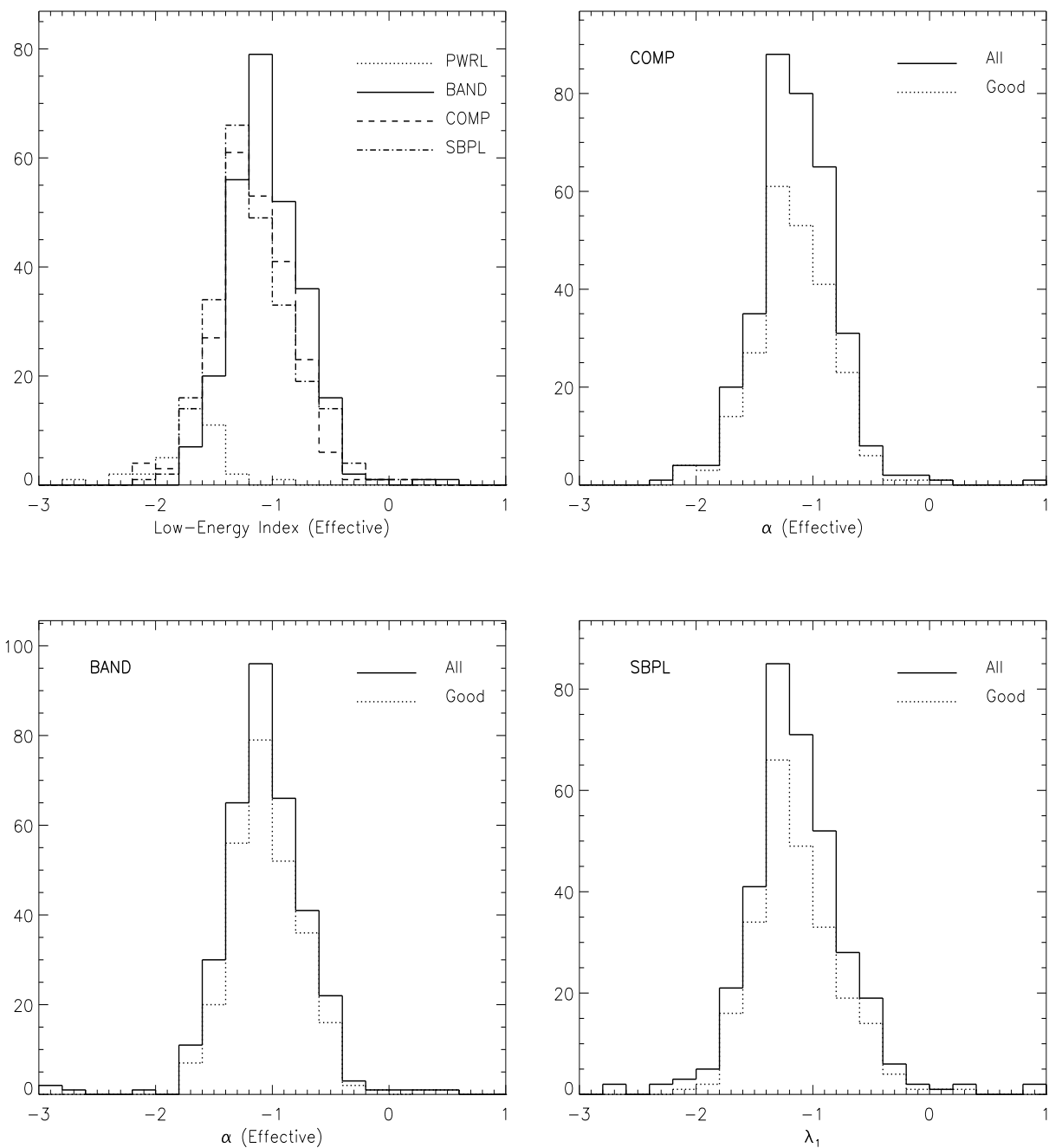}}
\caption{
Low-energy index distribution of 350 time-integrated spectra. 
[{\it Top Left}] {\it Good} parameters of all models.
Numbers of parameters included are 38 PWRL, 235 COMP, 271 BAND, and 241 SBPL.
The other plots show all (solid line) and {\it good} (dotted line) parameters 
of COMP, BAND, and SBPL.
The lowest (highest) bin includes values lower (higher) than the 
edge values.}
\label{fig:fpar_aleff}
\end{figure}

Another issue in comparing and presenting the spectral parameter distributions
is the uncertainty associated with each parameter.
The parameter distribution in a large sample can be effectively shown in a 
binned, histogram plot (e.g., Figure~\ref{fig:fpar_aleff}); however, such a 
plot does not include the uncertainties associated with each value.
Consequently, the reliability of the distribution is not evident.
One way of treating this problem is to only include well-constrained parameters
in the distribution.  Here, we will refer to these parameters as {\it good}
parameters.
This certainly results in providing a reliable probability distribution of a
given parameter, although it could also introduce some biases depending on how 
the {\it good} class of parameters are determined.
To interpret such distributions correctly, the criteria used to determine 
the {\it good} parameters are reviewed here.
We first applied a $\chi^2$ goodness-of-fit criteria of $> 3\sigma$ (99.7\%) 
for a given dof for each fit, to assure that the resulting parameters
are meaningful, before they can be considered {\it good}. \\

{\bf Low-Energy Indices.} 
Since each model handles spectral curvature differently, the condition under 
which the spectral indices can be determined with good confidence differs for 
each model.
The PWRL indices were found to be almost always constrained extremely well 
due to the simplicity of the model, so they can all be considered {\it good}.
On the other hand, $\alpha$ (BAND and COMP) is constrained only if the 
$e$-folding energy $E_0$ is sufficiently above the lower energy threshold of 
the data.  In our case, $E_0$ above 30 keV was considered acceptable.
For SBPL $\lambda_{1}$ to be constrained, the lowest energy that determines the
break scale must be above 30~keV.
Determined from the error distributions of all low-energy indices, we allow a 
maximum 1$\sigma$ uncertainty value of 0.4 in order for the parameters to be 
{\it good}.  
The value was selected so as to adequately constrain the parameters and still 
preserve more than 90\% of all low-energy index values regardless of models.

{\bf High-Energy Indices.}
In order for the BAND $\beta$ to be well determined, $E_{\rm 0}$ needs to be 
less than the upper energy limit of 1.5 MeV.  
For SBPL $\lambda_{2}$, similar to the low-energy index case above, the maximum 
break scale energy has to be less than 1.5 MeV.
In addition, we allow a maximum uncertainty of 1.0 for the high-energy indices,
again determined from the error distributions.  
More than 80\% of all $\beta$ and $\lambda_{2}$ provided uncertainties smaller 
than this value.

{\boldmath$E_{\rm peak}.$} For BAND and COMP, $E_{\rm peak}$ is the fitted value, 
whereas for SBPL, it is the calculated value (see Appendix \ref{ap:sbpl_ep} for 
calculation).  By definition, the parameter needs to represent the actual peak 
energy of the $\nu F_{\nu}$ spectrum.
In the cases where the high-energy power-law index $\geq -2$ or the low-energy
power-law index $\leq -2$, the fitted $E_{\rm peak}$ value is just a 
break and not the peak of $\nu F_{\nu} = E^2 f(E)$. 
Another case in which the fitted $E_{\rm peak}$ value is not the $\nu F_{\nu}$ 
peak is when the spectrum has a shape that is concave up.  
This could occur in two cases; (i) the COMP fits when $\alpha < -2$ (see 
\S\ref{sec:comp}) and (ii) the SBPL fits when $\lambda_{1} < \lambda_{2}$. 
The BAND model presumes $\alpha > \beta$.
Therefore, to obtain final $E_{\rm peak}$ distributions, BAND model fits with 
$\beta \geq -2$ and COMP fits with $\alpha \leq -2$ are excluded (for SBPL with 
$\lambda_{1} \leq -2$ and $\lambda_{1} < \lambda_{2}$, the $E_{\rm peak}$ was 
not calculated).  The maximum uncertainty allowed was 40\% of the parameter 
value for BAND and COMP, and 60\% for SBPL.

{\bf Break Energy.} Break energy includes the calculated $E_{\rm b}$ of BAND 
(Appendix \ref{ap:band_eb}) and the fitted $E_{\rm b}$ value of SBPL.
Unlike the case for the $E_{\rm peak}$, the break energy does not need to be 
the peak of the spectrum.
Therefore, the only requirement is that the parameter is within the data energy
range, namely, between 30 keV and 2 MeV.
This is justified by the fact that all BAND and SBPL fits with $E_{\rm b}$ 
values below 30 keV and above 2 MeV are associated with very large 
uncertainties in low-energy and high-energy indices, respectively, as well as 
in $E_{\rm b}$. 
The maximum allowed uncertainty of 70\% of the parameter value was set in order 
to include 85 -- 90\% of all break energies.  \\

\begin{figure}
\centerline{
\plotone{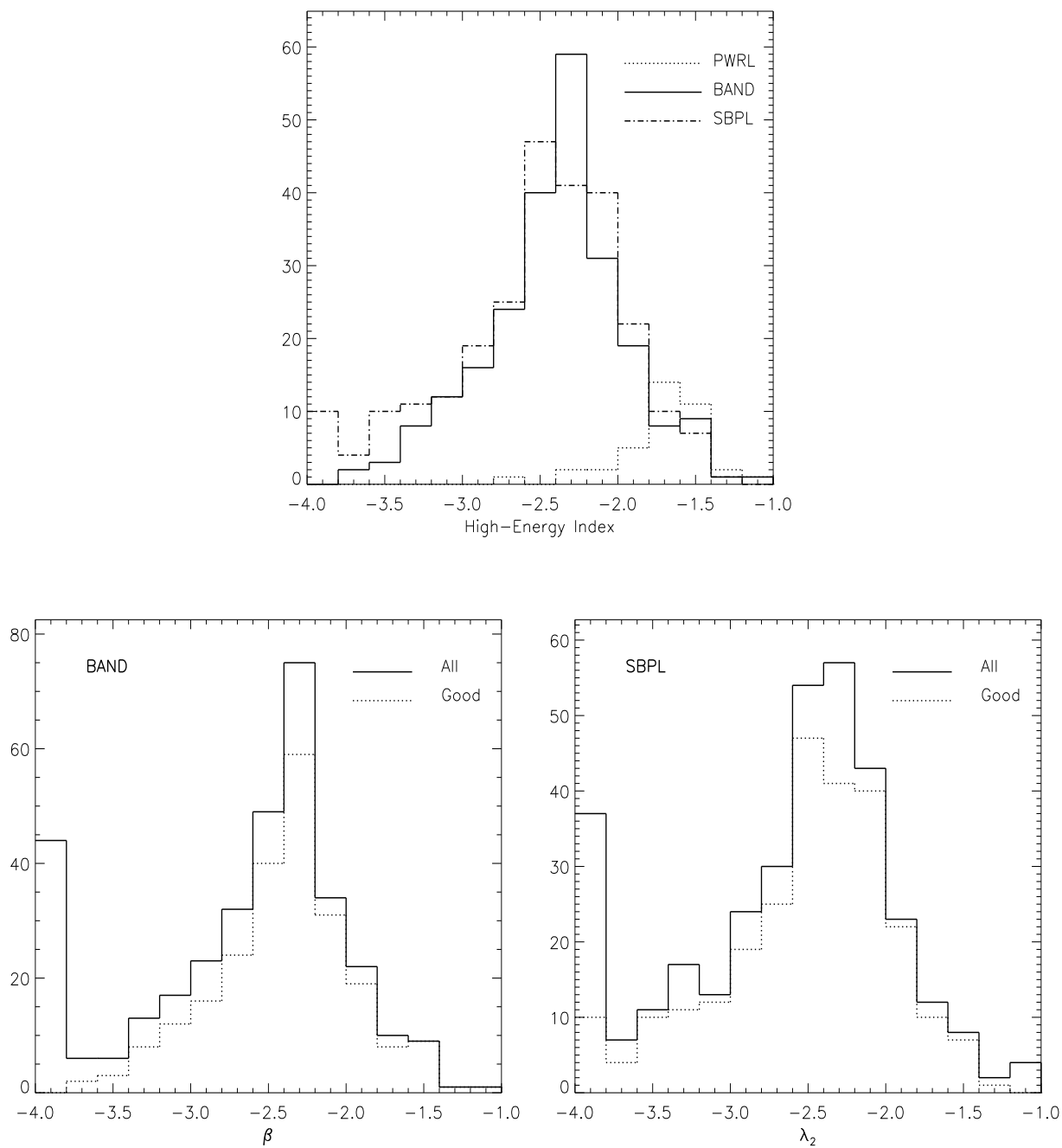}}
\caption{
High-energy spectral index distribution of 350 time-integrated spectra. 
[{\it Top}] {\it Good} parameters of all models.
Numbers of parameters included are 38 PWRL,  233 BAND, and 259 SBPL.
The other plots show all (solid line) and {\it good} (dotted line) parameters 
of BAND and SBPL.
The lowest (highest) bin includes values lower (higher) than the 
edge values.}
\label{fig:fpar_beta}
\end{figure}

\begin{figure}
\centerline{
\plotone{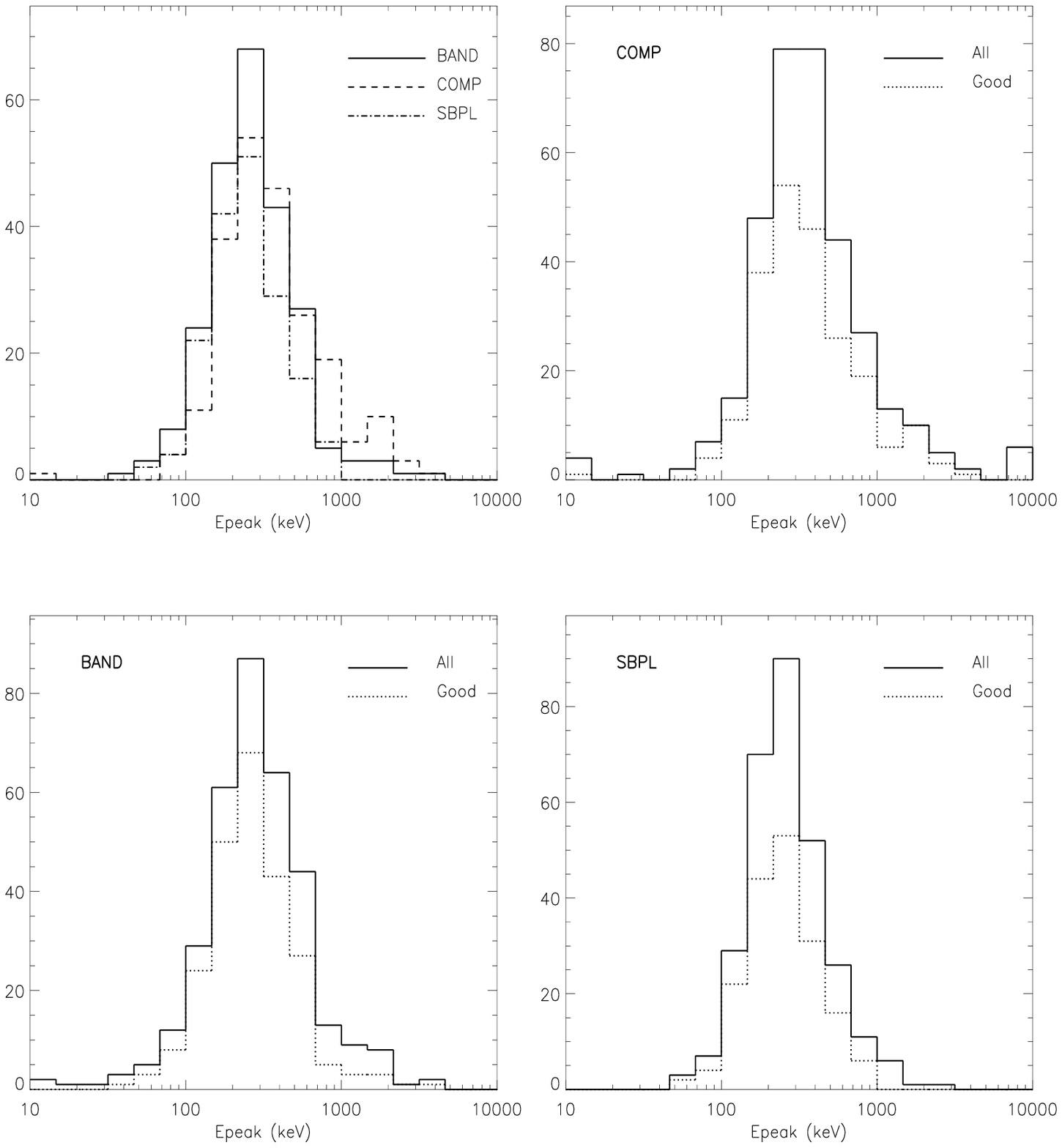}}
\caption{
$E_{\rm peak}$ distribution of 350 time-integrated spectra. 
[{\it Top Left}] {\it Good} parameters of all models.
Numbers of parameters included are 219 COMP, 237 BAND, and 172 SBPL.
The other plots show all (solid line) and {\it good} (dotted line) parameters 
of COMP, BAND, and SBPL.
The lowest (highest) bin includes values lower (higher) than the 
edge values.}
\label{fig:fpar_ep}
\end{figure}

\begin{figure}
\centerline{
\plotone{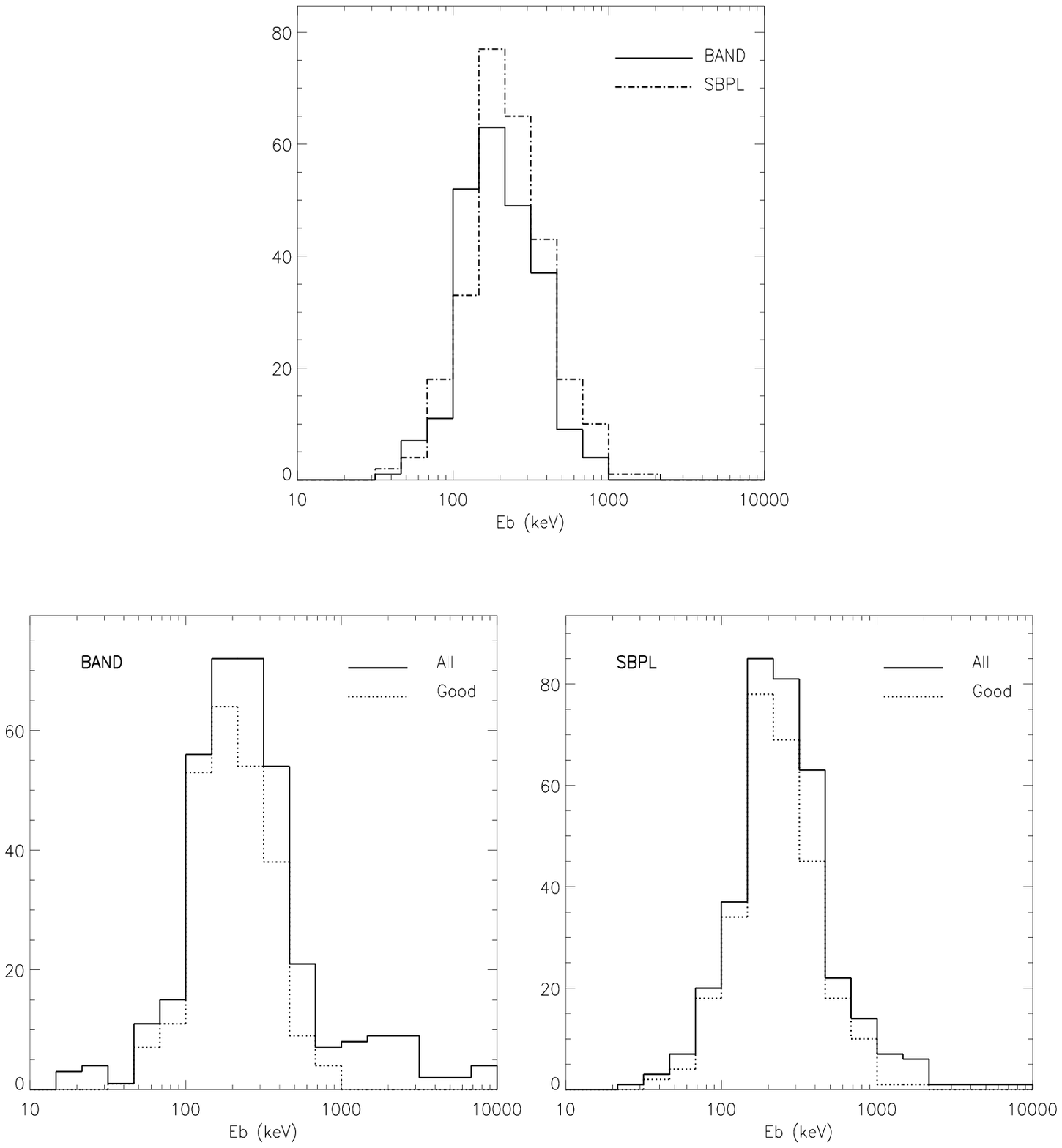}}
\caption{
Break Energy distribution of 350 time-integrated spectra. 
[{\it Top}] {\it Good} parameters of all models.
Numbers of parameters included are 233 BAND and 272 SBPL.
The other plots show all (solid line) and {\it good} (dotted line) parameters 
of BAND and SBPL.
The lowest (highest) bin includes values lower (higher) than the 
edge values.}
\label{fig:fpar_eb}
\end{figure}

\begin{figure}
\centerline{
\plotone{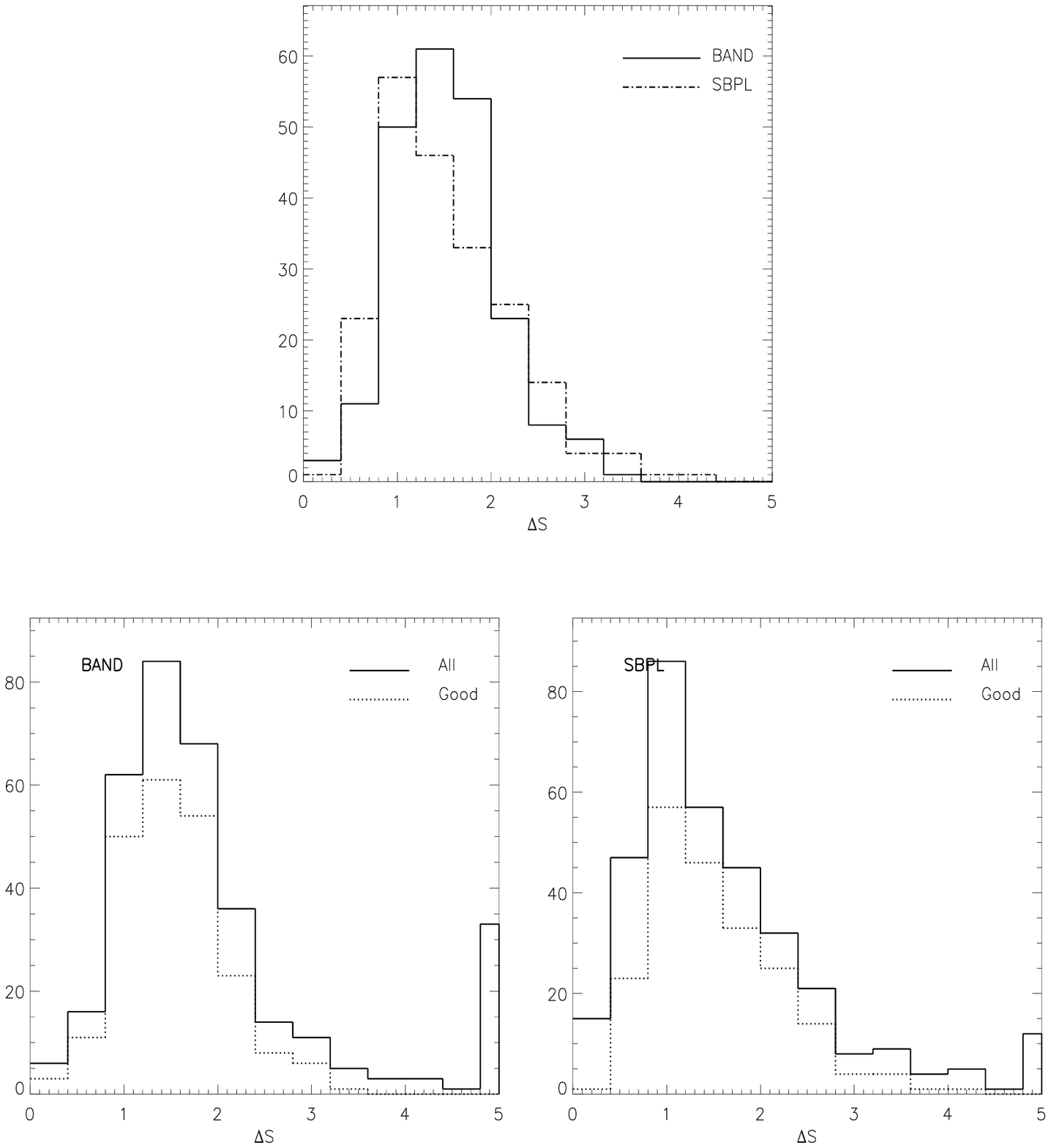}}
\caption{
$\Delta$S distribution of 350 time-integrated spectra. 
$\Delta$S is the difference between low-energy and high-energy indices.
[{\it Top}] {\it Good} parameters of all models.
Numbers of parameters included are 217 BAND and 209 SBPL.
The other plots show all (solid line) and {\it good} (dotted line) parameters 
of BAND and SBPL.
The lowest (highest) bin includes values lower (higher) than the 
edge values.}
\label{fig:fpar_ds}
\end{figure}

\begin{figure}
\centerline{
\plotone{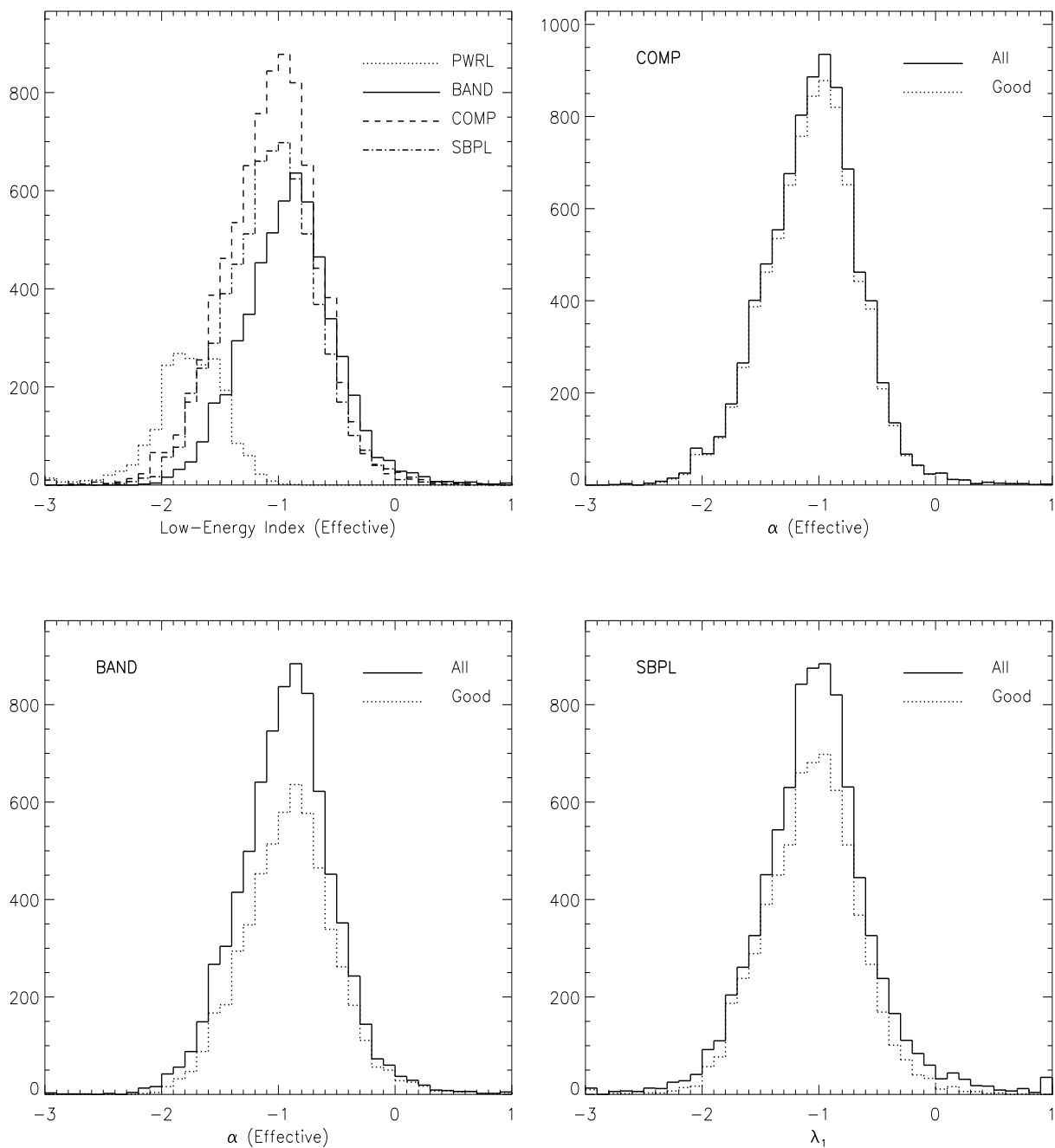}}
\caption{
Low-energy spectral index distribution of 8459 time-resolved spectra. 
[{\it Top Left}] {\it Good} parameters of all models.
Numbers of parameters included are 1971 PWRL, 8050 COMP, 5510 BAND, and 6533 SBPL.
The other plots show all (solid line) and {\it good} (dotted line) parameters 
of COMP, BAND, and SBPL.
The lowest (highest) bin includes values lower (higher) than the 
edge values.}
\label{fig:bpar_aleff}
\end{figure}

\begin{figure} 
\centerline{
\plotone{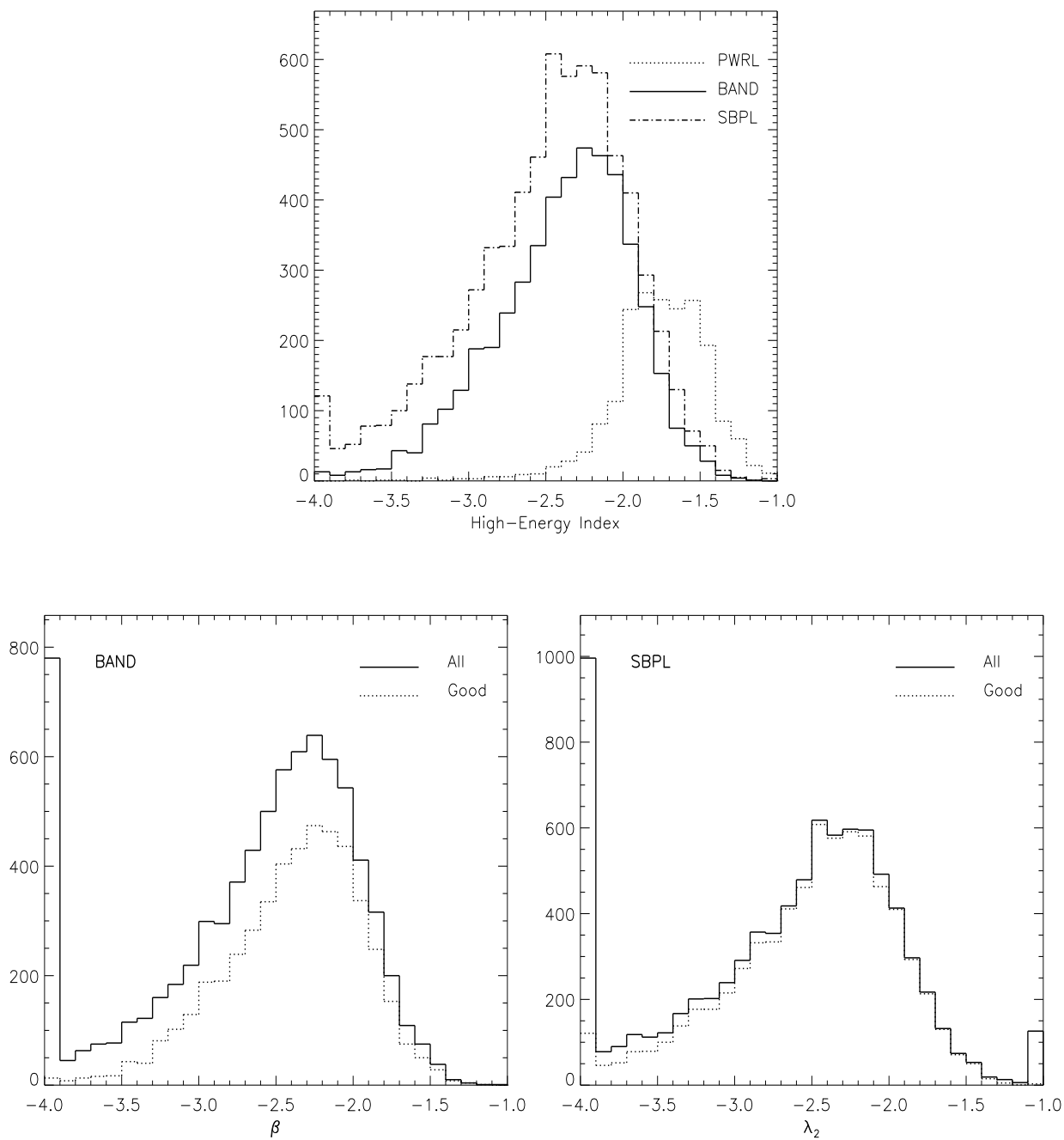}}
\caption{
High-energy spectral index distribution of 8459 time-resolved spectra. 
[{\it Top}] {\it Good} parameters of all models.
Numbers of parameters included are 1971 PWRL, 4810 BAND, and 7003 SBPL.
The other plots show all (solid line) and {\it good} (dotted line) parameters 
of BAND and SBPL.
The lowest (highest) bin includes values lower (higher) than the 
edge values.}
\label{fig:bpar_beta}
\end{figure}

\begin{figure} 
\centerline{
\plotone{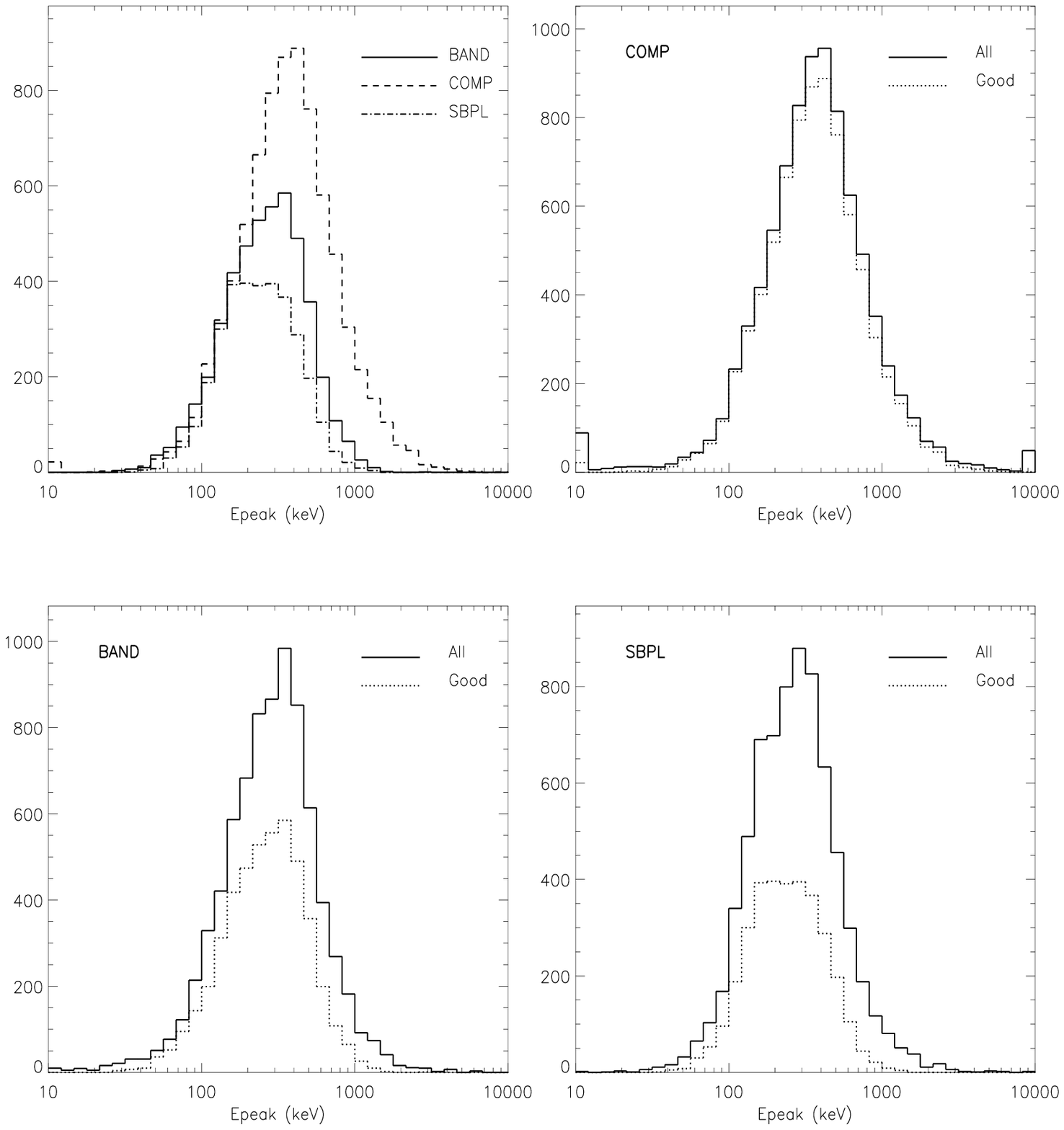}}
\caption{
$E_{\rm peak}$ distribution of 8459 time-resolved spectra. 
[{\it Top Left}] {\it Good} parameters of all models.
Numbers of parameters included are 7702 COMP, 4677 BAND, and 3291 SBPL.
The other plots show all (solid line) and {\it good} (dotted line) parameters 
of COMP, BAND, and SBPL.
The lowest (highest) bin includes values lower (higher) than the 
edge values.}
\label{fig:bpar_ep}
\end{figure}

\begin{figure} 
\centerline{
\plotone{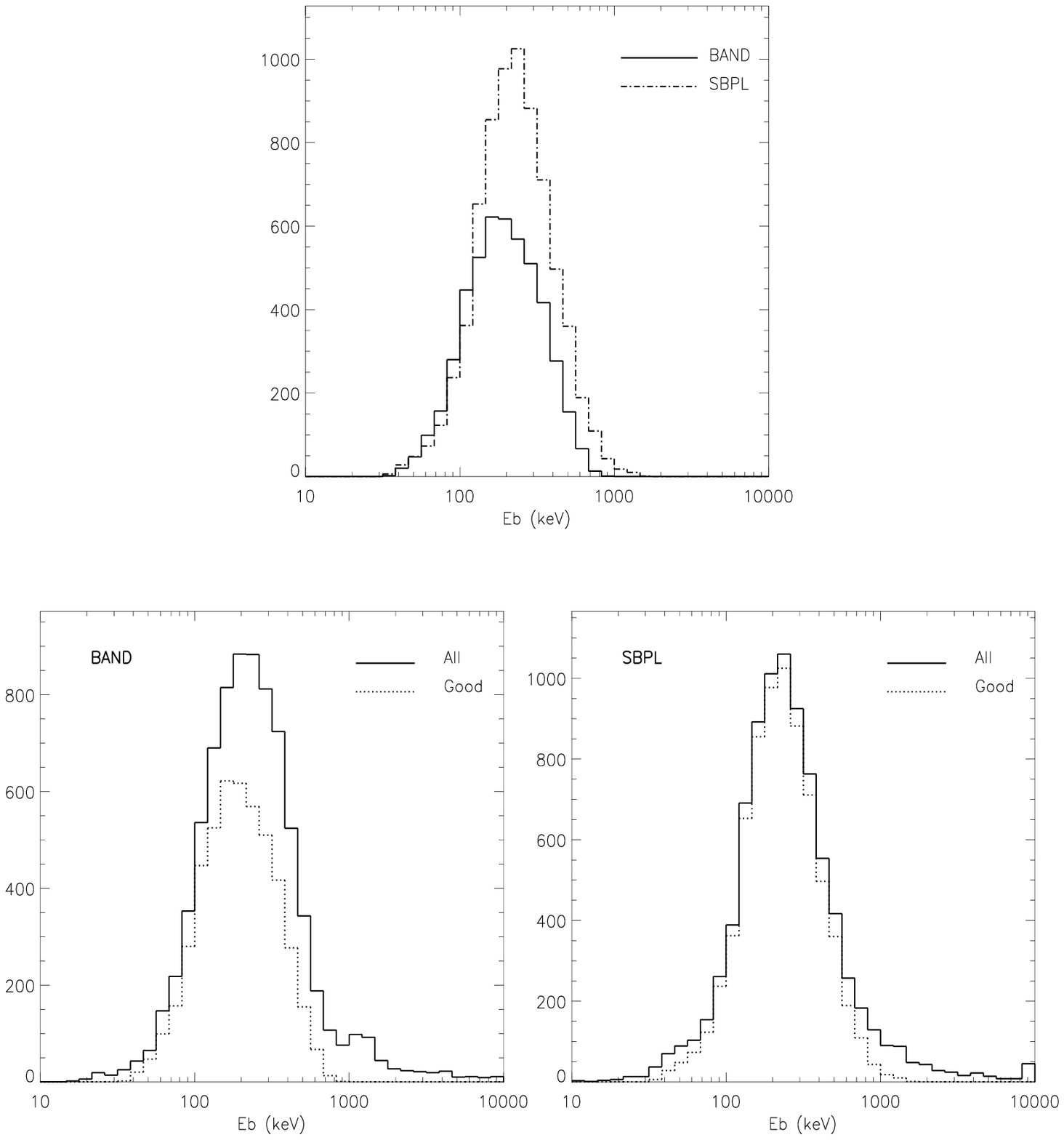}}
\caption{
Break energy distribution of 8459 time-resolved spectra. 
[{\it Top}] {\it Good} parameters of all models.
Numbers of parameters included are 4825 BAND and 7207 SBPL.
The other plots show all (solid line) and {\it good} (dotted line) parameters 
of BAND and SBPL.
The lowest (highest) bin includes values lower (higher) than the 
edge values.}
\label{fig:bpar_eb}
\end{figure}

\begin{figure} 
\centerline{
\plotone{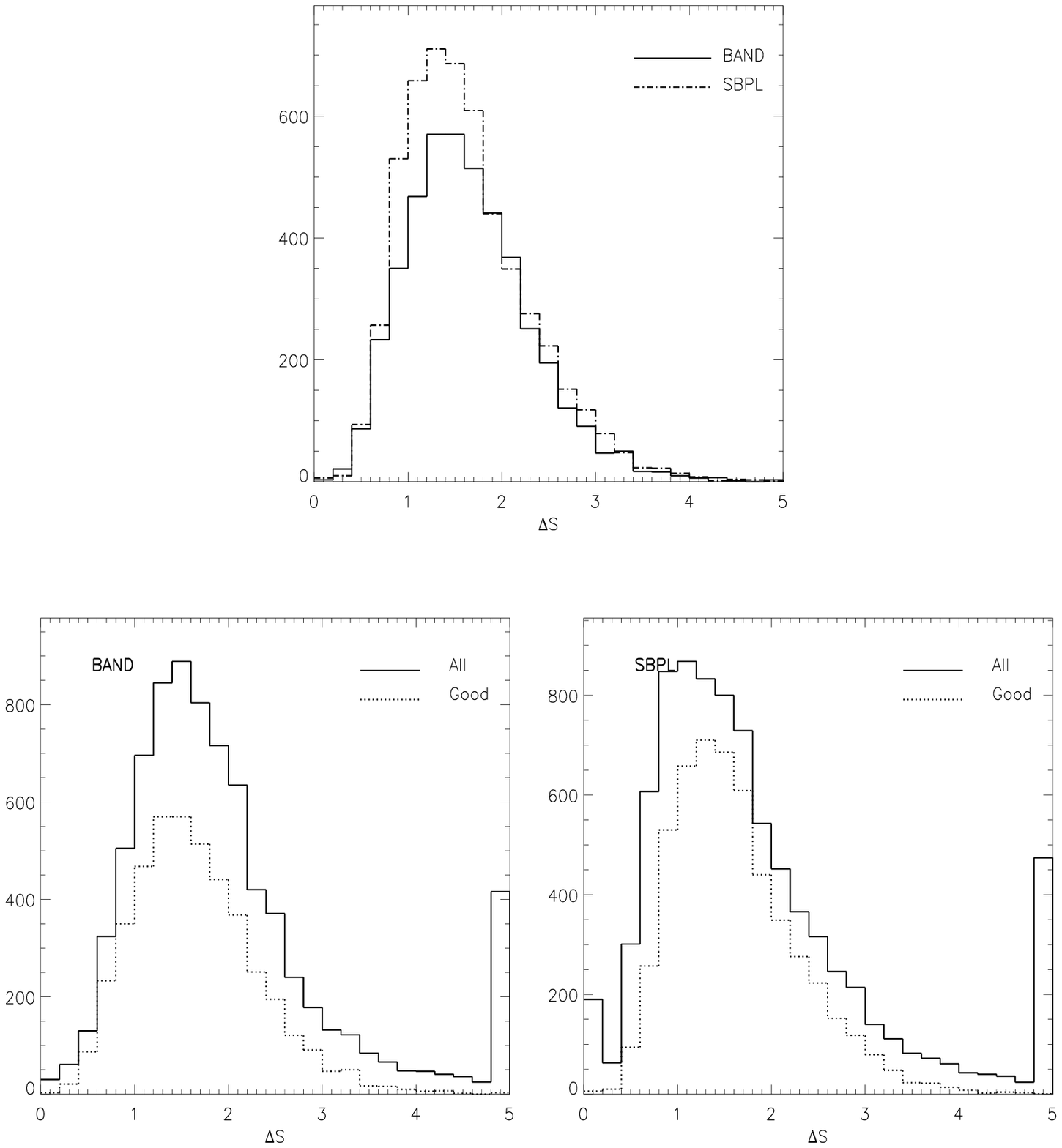}}
\caption{
$\Delta$S distribution of 8459 time-resolved spectra. 
$\Delta$S is the difference between low-energy and high-energy indices.
[{\it Top}] {\it Good} parameters of all models.
Numbers of parameters included are 4441 BAND and 5322 SBPL.
The other plots show all (solid line) and {\it good} (dotted line) parameters 
of BAND and SBPL.
The lowest (highest) bin includes values lower (higher) than the 
edge values.}
\label{fig:bpar_ds}
\end{figure}

With these in mind, we can now examine the spectral parameter distributions
for each model.
We first compare the spectral parameters obtained by fitting all spectra with 
all models, in order to explore the internal characteristics of the models.
In Figures \ref{fig:fpar_aleff} -- \ref{fig:fpar_ds}, we present distributions 
of the time-integrated spectral parameters, followed by the distributions of 
the time-resolved spectral parameters in Figures \ref{fig:bpar_aleff} -- 
\ref{fig:bpar_ds}, comparing all parameters and the {\it good} parameters 
obtained with each model.
In addition, the median values and the dispersion (quartile values)
of each distribution are listed in Table \ref{tab:mode_all}.
The PWRL indices are included in both low-energy and high-energy index
distributions because we cannot judge which power law the model represents.
They are found to cluster around $\sim -1.7$ by themselves, which is in between 
the typical low-energy and high-energy indices.
We note the index value agrees with the median power-law index ($\sim -1.6 
\pm 0.2$; index values taken from GCN circulars\footnotemark[5] and the Swift
GRB table\footnotemark[6]) of time-integrated GRB spectra 
observed with the Burst Alert Telescope on board {\it Swift}, in the
energy range of 15 -- 150 keV.
\footnotetext[5]{Gamma-ray burst Coordinate Network 
(http://gcn.gsfc.nasa.gov/gcn\_main.html)}
\footnotetext[6]{http://heasarc.gsfc.nasa.gov/docs/swift/archive/grb\_table.html}
\placefigure{fig:fpar_aleff}
\placefigure{fig:fpar_beta}
\placefigure{fig:fpar_ep}
\placefigure{fig:fpar_eb}
\placefigure{fig:fpar_ds}
\placefigure{fig:bpar_aleff}
\placefigure{fig:bpar_beta}
\placefigure{fig:bpar_ep}
\placefigure{fig:bpar_eb}
\placefigure{fig:bpar_ds}
\placetable{tab:mode_all}

As seen in these plots and the table, the corresponding parameters of different 
models are mostly consistent with each other within the dispersion.
With the use of $\alpha_{\rm eff}$, the low-energy index distributions (Figures 
\ref{fig:fpar_aleff} and \ref{fig:bpar_aleff}) determined by the BAND, COMP and 
SBPL models all seem to agree.
As for the high-energy indices (Figures \ref{fig:fpar_beta} and 
\ref{fig:bpar_beta}), the SBPL fits and the BAND fits also yielded consistent 
distributions.  
In the $E_{\rm peak}$ distributions (Figures \ref{fig:fpar_ep} and 
\ref{fig:bpar_ep}), we find that the COMP fits result in higher $E_{\rm peak}$ 
values and broader distribution than the BAND and SBPL ones.
Finally, the break energy distributions seen in Figures \ref{fig:fpar_eb} and
\ref{fig:bpar_eb} indicate that the typical break energies found by the BAND
fits are softer than the $E_{\rm peak}$ of the same model, due to the curvature
in the spectra.
The high-energy tail populations in the break energy distributions are more 
noticeable than the $E_{\rm peak}$ cases, although they are not constrained.
We also present in Figures~\ref{fig:fpar_ds} and \ref{fig:bpar_ds}, the change
in the spectral indices; $\Delta$S $\equiv \alpha_{\rm eff} - \beta$ (BAND),
$\lambda_1 - \lambda_2$ (SBPL).
This parameter has been previously examined by \citet{pre02} using the results 
of SP1, in order to probe the GRB emission process in the context of the
synchrotron shock model.
We show this here for the comparison with the SP1 results.
The distributions of $\Delta$S of the BAND and SBPL agree well, and are also
consistent with the distribution obtained by \citet{pre02}.

In terms of comparisons between the time-integrated and time-resolved spectral
parameters, the most obvious difference is seen in the low-energy index
distributions.  The indices of time-integrated spectra are found to be softer
than the time-resolved ones, except the PWRL case.
Peak energies, $E_{\rm peak}$, also tend to be slightly softer in the 
time-integrated spectra, especially when fitted with the COMP model.
The differences are due to the fact that $\alpha$ and $E_{\rm peak}$ are very 
often observed to strongly evolve during a burst \citep[e.g.,][]{for95, lia96, 
ban97, cri97, ryd99, cri99}. 
No significant difference in high-energy index distributions is seen.

\subsection{Model Comparison within Each Spectrum}\label{sec:bestmodel}
Since we have fitted the four models to all spectra, it is possible to 
statistically compare the models and determine the best-fit model to each 
individual spectrum.
The comparison is, however, not straightforward because each model provides 
a different dof.
A model that consists of a lower number of parameters is always preferred 
statistically over more complex models for the same $\chi^2$.
However, the data may require an extra parameter for a better fit, in which case
it should result in a significant improvement in $\chi^2$ when the extra parameter
is fitted.
Therefore, we look at the $\chi^2$ improvements in adding $N$ parameters ($\Delta
{\rm dof} = N$), starting from the simplest model (i.e., PWRL) to determine the
best-fit models.

The four models employed, namely PWRL, COMP, BAND, and SBPL, consist of 2, 3, 4, 
and 5 parameters, respectively.
For each spectrum, we take the $\chi^2$ of the PWRL fit as a reference.
We first compare the PWRL fit $\chi^2$ to the COMP fit $\chi^2$ for 
$\Delta {\rm dof} = 1$, and find the confidence level in $\chi^2$ improvement.
If the confidence level is greater than 99.9\%, the COMP fit is significantly 
better than the PWRL fit; whereas, for confidence levels between 
$80.0 - 99.9$\%, we cannot statistically determine the better model between 
the two. The best-fit model in such a case is classified as PWRL/COMP.
A confidence level lower than 80.0\% suggests that the PWRL is preferred.
Likewise, comparisons were made for all possible combinations of four models, 
with corresponding $\Delta$dof.
The confidence intervals of 99.9\% and 80.0\% were used for all comparisons 
involving PWRL, while 99.9\% and 68.3\% were used for the comparisons among 
the other models.
These threshold values for the confidence intervals were decided based upon
the simulation results described in full detail in \S \ref{sec:simulation}.
In the simulation, we fitted a simulated spectrum of a particular model by 
the other models and examined the degree of improvement in each fitting.

\begin{figure}
\centerline{
\plotone{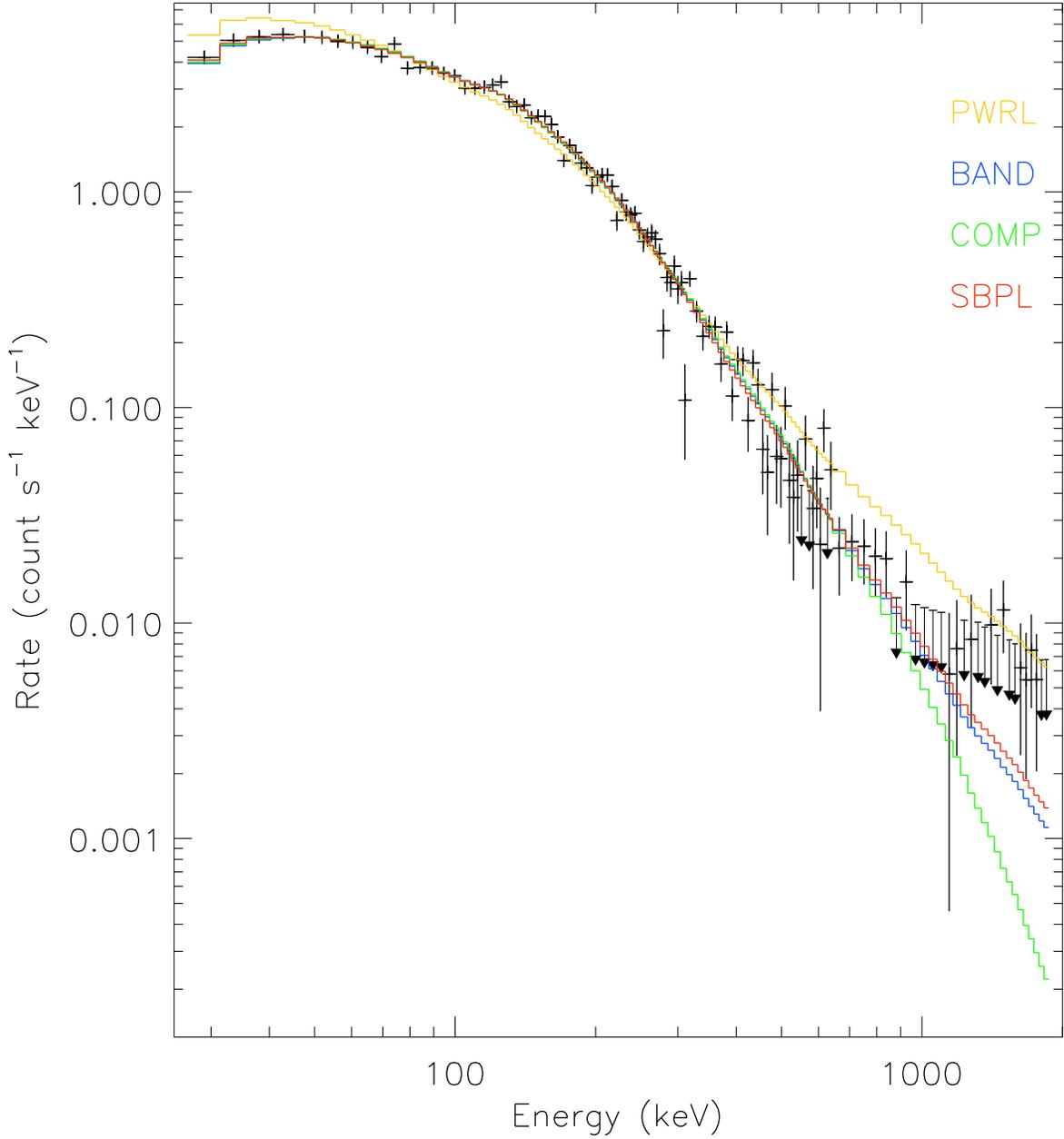}}
\caption{Count spectra of GRB 000429 (trigger number 8087).
Data points are shown as crosses and the color lines are convolved model counts.
COMP, BAND, and SBPL all fit statistically as good as each other.  
The BEST model determined by parameter constraints in this case is SBPL.}
\label{fig:8087_csp}
\end{figure}

In most cases, the $\chi^2$ probability was an adequate measure of the best fit
determination; however, we found some cases where the best-fit model
found by $\chi^2$ probability was actually not better than the other models
in terms of parameter constraints.
Therefore, in addition to the $\chi^2$ probabilities, the spectral parameters
were also checked, requiring reasonable constraints on the additional 
parameter(s).
If the best-fit model parameters are not constrained, the next statistically 
best-fit model with more constrained parameters was preferred.  
Finally, in case a preferred model could not be determined solely by 
changes in $\chi^2$, such as the PWRL/COMP case above, again their parameters 
were compared and a model with more constrained parameter set was designated 
as the best-fit model.
The results of the best-fit model determination are shown in Table
\ref{tab:models}.
\placetable{tab:models}
It is seen that for many spectra, the COMP, BAND, and SBPL models all resulted 
in comparably good fits (C/B/S case in the Table~\ref{tab:models}).
In most of these cases, the additional parameters were still well constrained, 
and the more complex models were meaningful.
Consequently, the SBPL fits were selected as the best-fit models,
despite the complexity of the model, in many of the C/B/S case.
To illustrate the goodness of fits of each model, a count spectrum of one 
C/B/S burst with all four model counts is plotted in Figure \ref{fig:8087_csp}.
\placefigure{fig:8087_csp}
Moreover, for the time-resolved spectra, in almost 30\% of the cases,
the COMP and BAND fits were both found to be acceptable (i.e., C/B).
For these spectra, the more complex BAND model was able to provide adequate 
parameter constraints, similar to the SBPL model in the C/B/S case above.
In these cases, low-energy indices from all models were usually found in 
agreement within 1$\sigma$ uncertainties (if $\alpha_{\rm eff}$ is used).
We also find that COMP tends to be preferable in fitting time-resolved spectra,
because of the existence of more spectra without high-energy component, as well as
the lower S/N in each spectrum compared with the time-integrated spectra.
In addition, those best-fit by PWRL ($\sim$ 5\% of all spectra) were indeed 
among the dimmer, low S/N spectra.
In the case of SBPL fit spectra, many spectra provided smaller 
break scales (i.e., sharper breaks).

\begin{figure}
\centerline{
\plotone{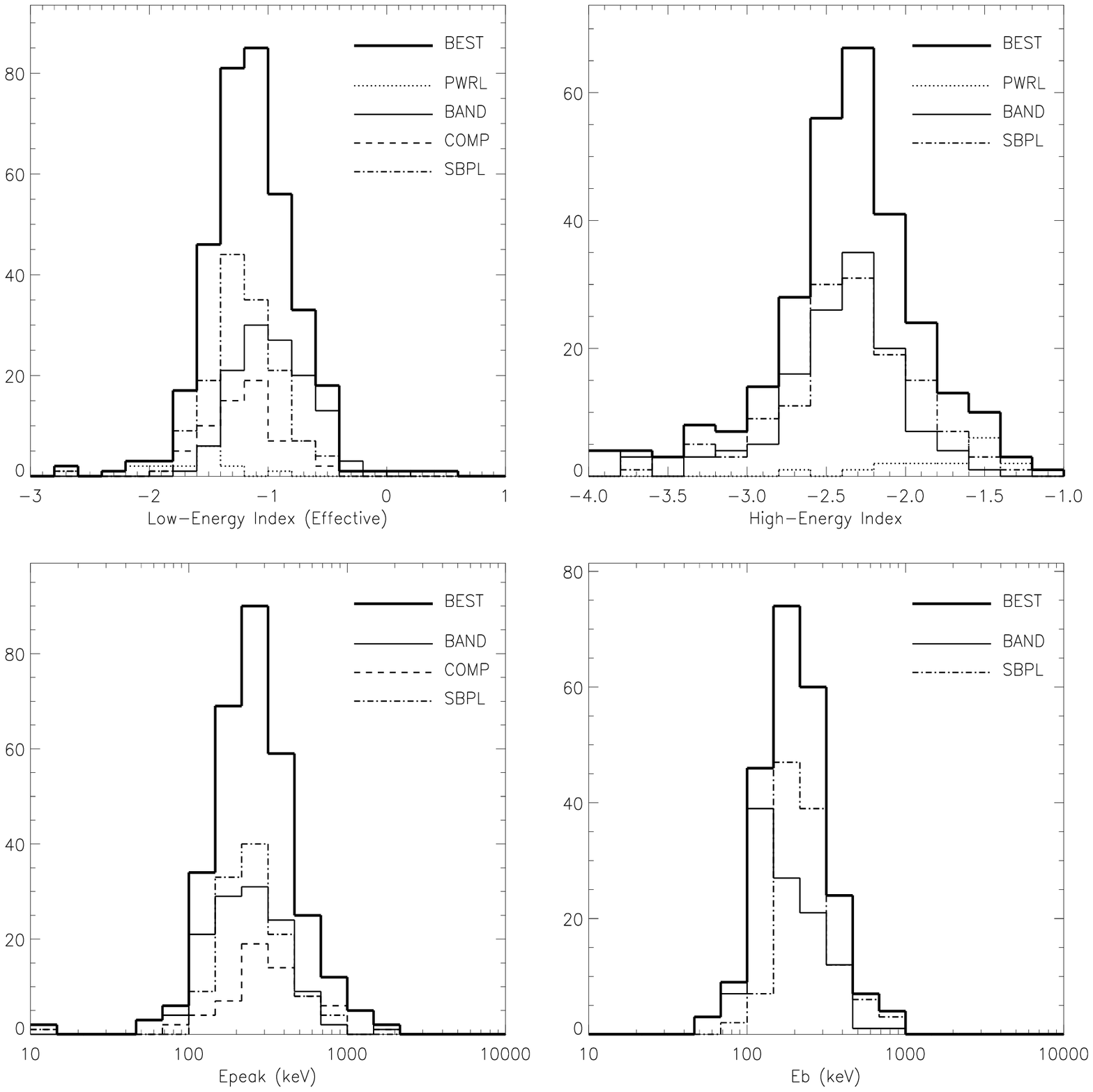}}
\caption{
BEST model parameter distributions of the time-integrated spectra. 
The bold black lines show the total distributions, and the constituents are shown
in various lines.
The lowest (highest) bin includes values lower (higher) than the edge values.}
\label{fig:fpar_best1}
\end{figure}

\begin{figure} 
\centerline{
\plotone{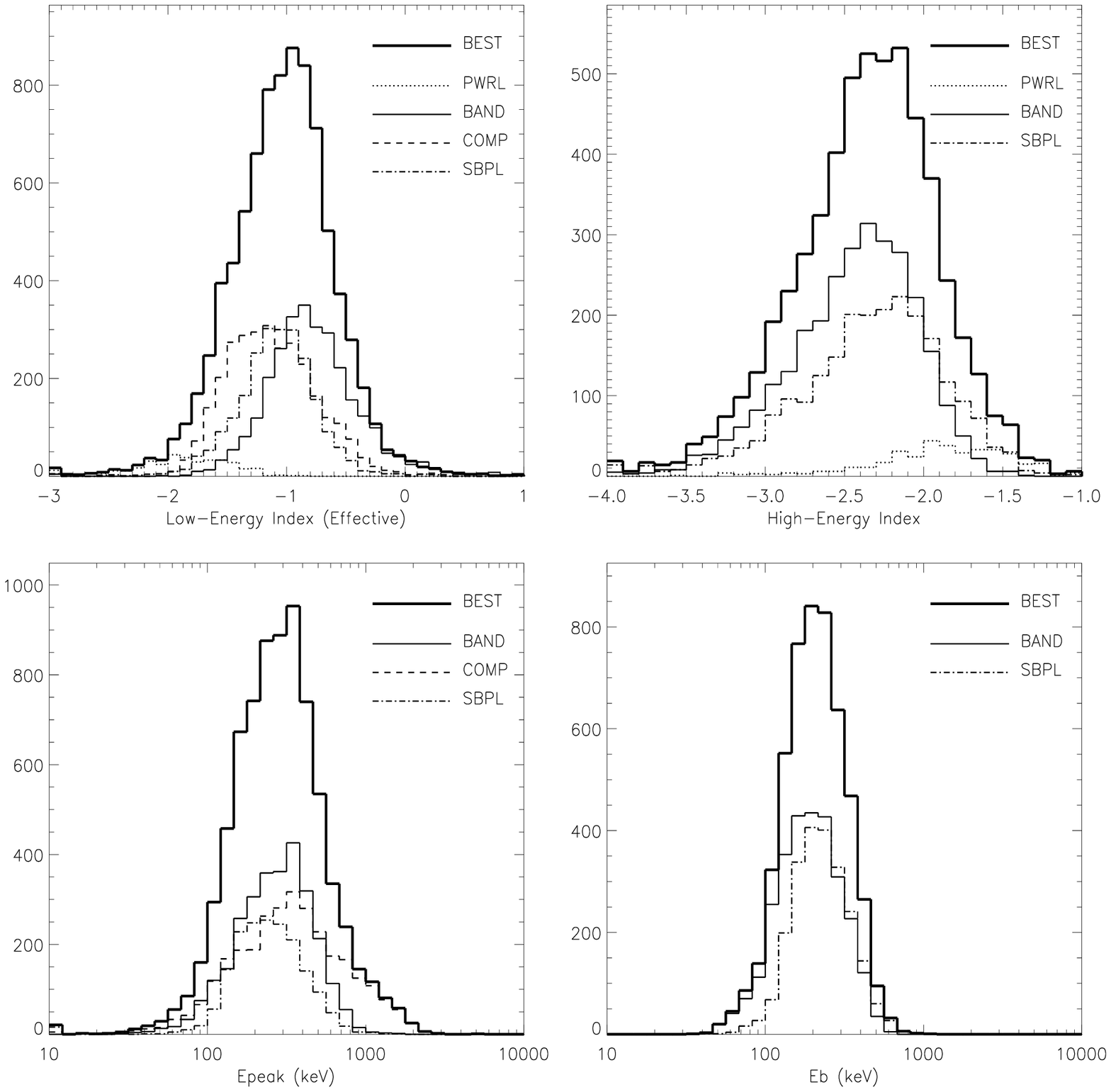}}
\caption{
BEST model parameter distributions of the time-resolved spectra. 
The bold solid lines show the total distributions, and the constituents are shown
in various lines.
The lowest (highest) bin includes values lower (higher) than the edge values.}
\label{fig:bpar_best1}
\end{figure}

\begin{figure}
\centerline{
\plotone{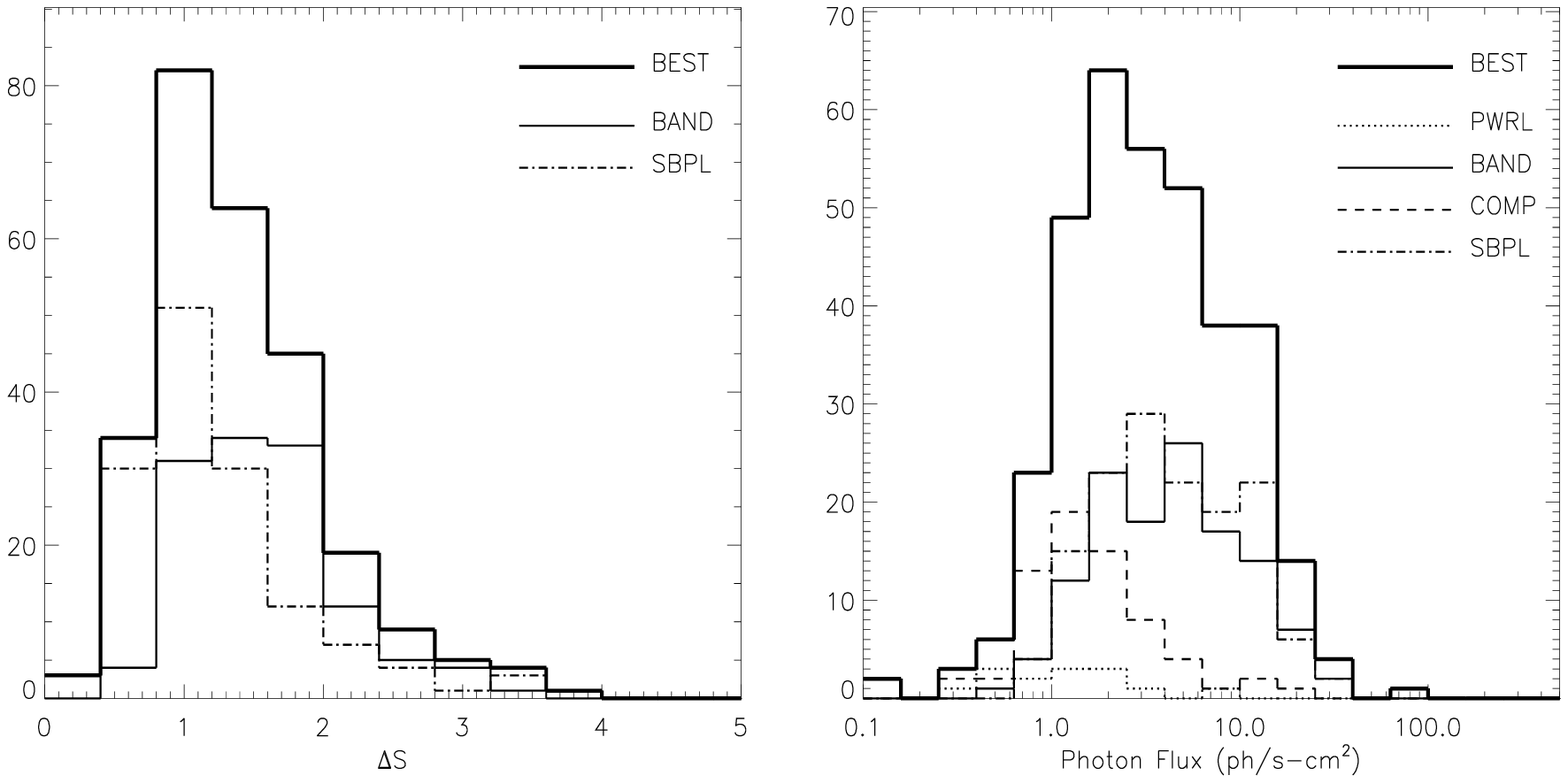}}
\caption{
BEST model spectral index difference ($\Delta$S) and average photon flux 
distributions of the time-integrated spectra. 
The bold black lines show the total distributions, and the constituents are shown
in various lines.
The lowest (highest) bin includes values lower (higher) than the edge values.}
\label{fig:fpar_best2}
\end{figure}

\begin{figure}
\centerline{
\plotone{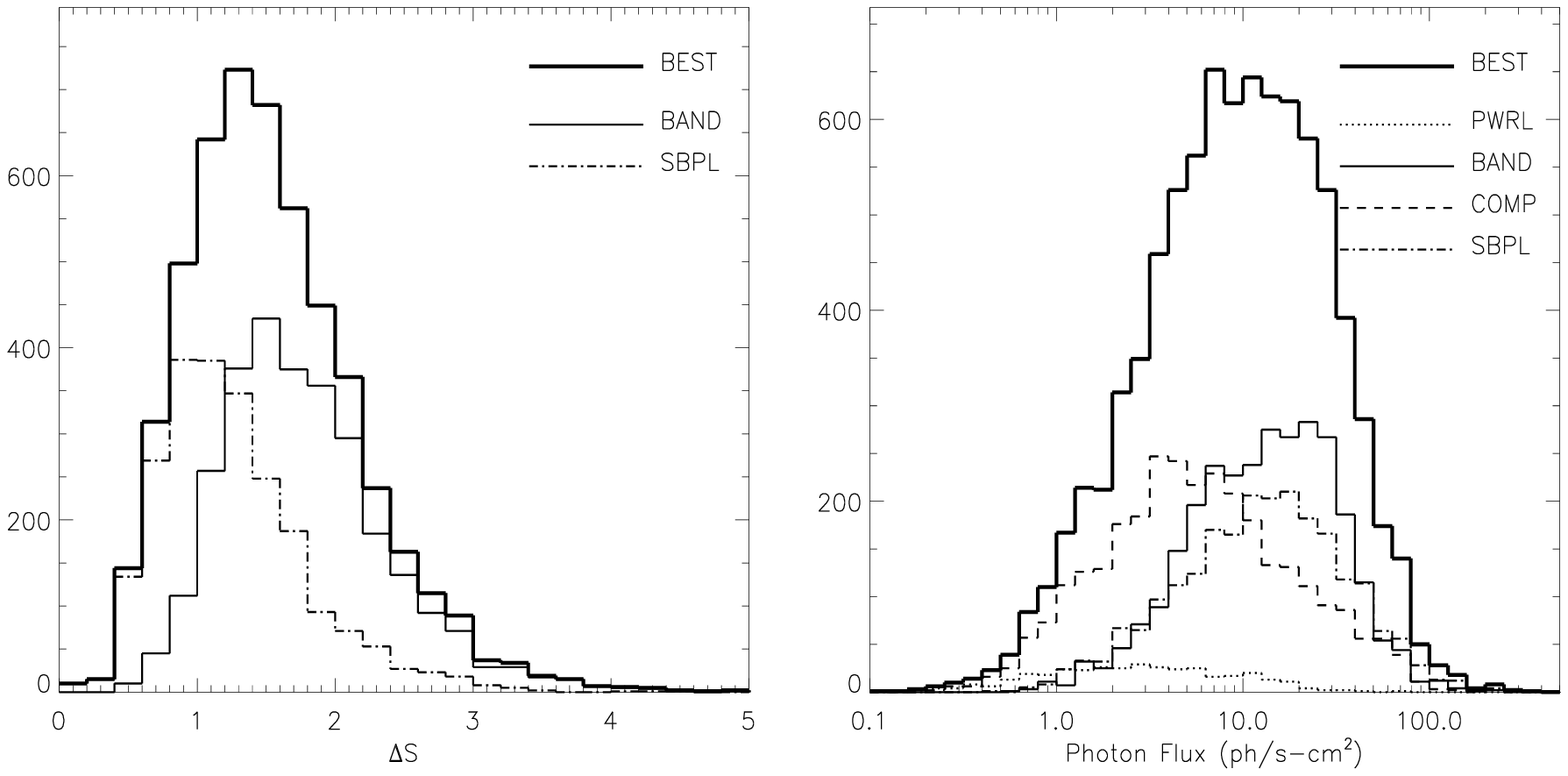}}
\caption{
BEST model spectral index difference ($\Delta$S) and average photon flux 
distributions of the time-resolved spectra. 
The bold solid lines show the total distributions, and the constituents are shown
in various lines.
The lowest (highest) bin includes values lower (higher) than the 
edge values.}
\label{fig:bpar_best2}
\end{figure}

Henceforth, we refer to this set of models as the ``BEST" models, consisting of 
PWRL, COMP, BAND, and SBPL, each with numbers presented in the 
{\small{\bf TOTAL}} row of Table \ref{tab:models}. 
A collection of parameters obtained with the BEST models,
with the use of $\alpha_{\rm eff}$ and calculated $E_{\rm peak}$ and $E_{\rm b}$,
can be thought as a well-constrained, model-unbiased set of parameters best 
describing each spectrum in our sample.
The time-integrated spectral fit results obtained with the BEST models are 
presented in Table \ref{tab:results}, for each burst.
\placetable{tab:results}
Additionally, we show the spectral parameter distributions of the BEST models 
in Figures \ref{fig:fpar_best1} -- \ref{fig:bpar_best2}, for both the 
time-integrated and time-resolved cases.
In these figures, the distributions of each parameter of the BEST model and 
the contribution of each model that consists the BEST model are shown.
Also, the median values and the dispersion of these distributions are summarized 
in Table \ref{tab:mode_best}.
\placefigure{fig:fpar_best1}
\placefigure{fig:bpar_best1}
\placefigure{fig:fpar_best2}
\placefigure{fig:bpar_best2}
\placetable{tab:mode_best}

We find the BAND model tends to fit spectra with harder low-energy indices, 
and especially in the time-resolved case, the COMP model seems to fit those 
with much softer indices. 
The spectra with positive low-energy indices are dominated by the BAND model 
fits.  Most of the time-resolved COMP spectra with low-energy indices less than 
$-1.5$ were found to have low $E_{\rm peak} \lesssim 300$ keV. 
The value of low-energy index $-1.5$ corresponds to the fast-cooling case of 
the synchrotron emission spectrum \citep{ghi99}.
Out of the 8459 time-resolved spectra, 407 spectra resulted in the low-energy
indices above the synchrotron limit of $- 2/3$ \citep{kat94, pre98b} by more than
$3\sigma$.
Meanwhile, for the high-energy spectral index distribution of time-resolved 
spectra, it seems those with harder high-energy indices are better fit by 
the SBPL model, while those with softer indices are better fit by the BAND 
model.  We also find that the majority of SBPL with high-energy indices larger 
than $-2.0$ have small differences in low-energy and high-energy spectral 
indices ($\Delta$S $<$ 1).  

In the distributions of $E_{\rm peak}$ and break energy, we observe no secondary 
populations at energy below 100 keV, which was previously indicated by the 
break energy distribution of time-resolved spectra in SP1.
The distributions of $E_{\rm peak}$ and break energy are different mainly 
due to the contribution of $E_{\rm peak} \gtrsim 1$ MeV that is dominated 
by COMP spectra.
These represent very hard spectra with the actual $E_{\rm peak}$ values lying
very close to or higher than the upper energy bound of our datasets 
($\gtrsim$ 1 MeV).
For such spectra, BAND or SBPL do not provide well-constrained high-energy
spectral indices, and PWRL does not accommodate well the curvature of the 
spectra; therefore, the COMP fits were preferred. 
The $\Delta$S distributions in Figures \ref{fig:fpar_best2} and 
\ref{fig:bpar_best2} (left panels) show that the SBPL spectra generally have 
much smaller changes in the indices.  
Most of those spectra with small differences in indices $\Delta$S, 
have sharp breaks ($\Lambda \leq 0.1$), and more than a half of them have
high-energy indices $> -2.0$, indicating the existence of another break energy 
at higher energy.
Finally, from the photon flux plots (right panels in Figures \ref{fig:fpar_best2} 
and  \ref{fig:bpar_best2}), it is also found that the simpler models (PWRL and
COMP) are best fitted to weaker spectra, compared with  more complex models 
(BAND and SBPL), as expected.

\begin{figure} 
\centerline{
\plotone{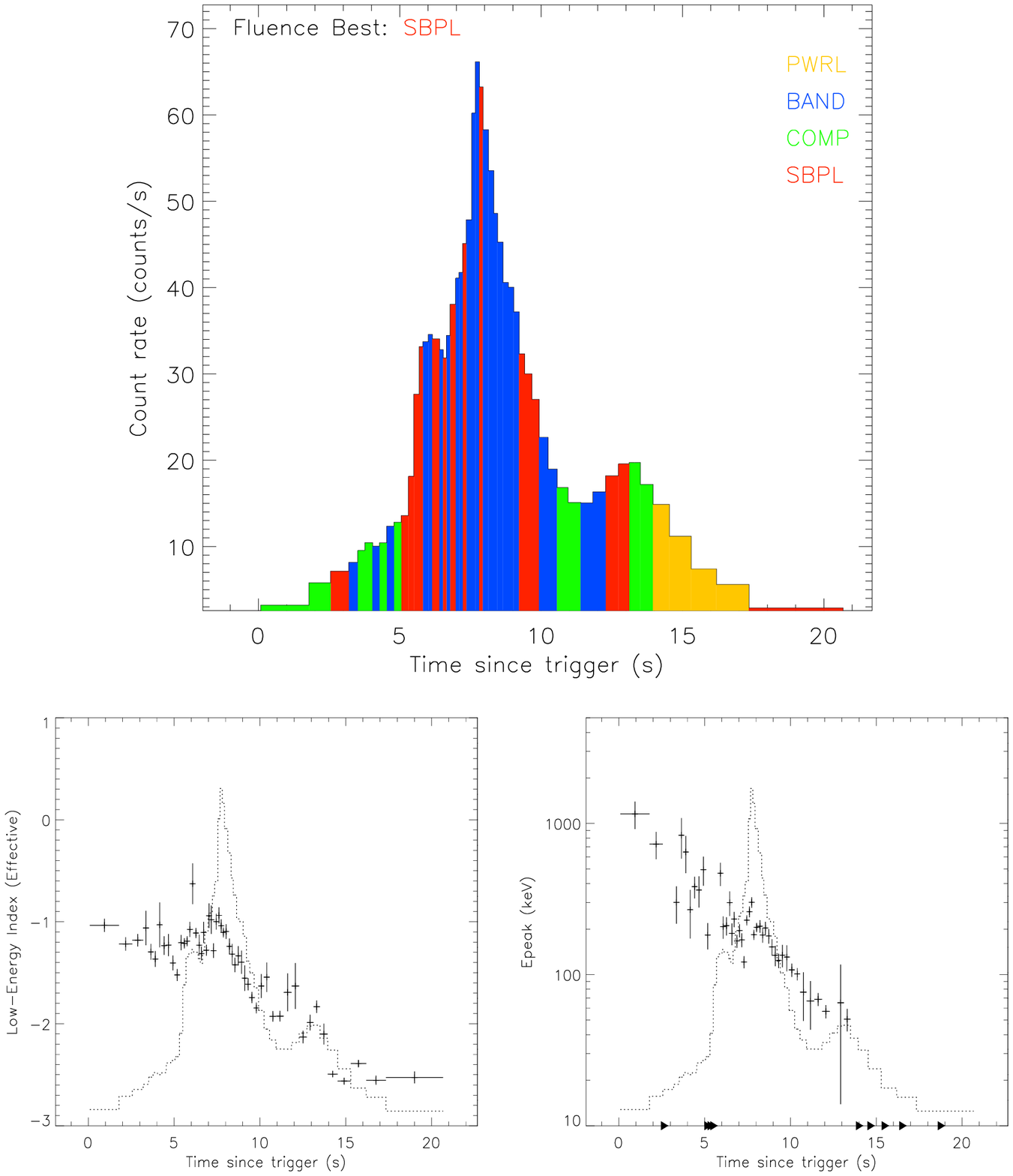}}
\caption{[{\it Top}] BEST model history of GRB 950403 (BATSE trigger number 3491).
The BEST model of the time-integrated spectrum for this burst is SBPL.
[{\it Bottom}] Evolutions of the low-energy index (effective) and $E_{\rm peak}$
for the same event.  The arrowheads in $E_{\rm peak}$ plot indicate where the
$E_{\rm peak}$ values cannot be determined.}
\label{fig:3491_bestmdl}
\end{figure}

\begin{figure} 
\centerline{
\plotone{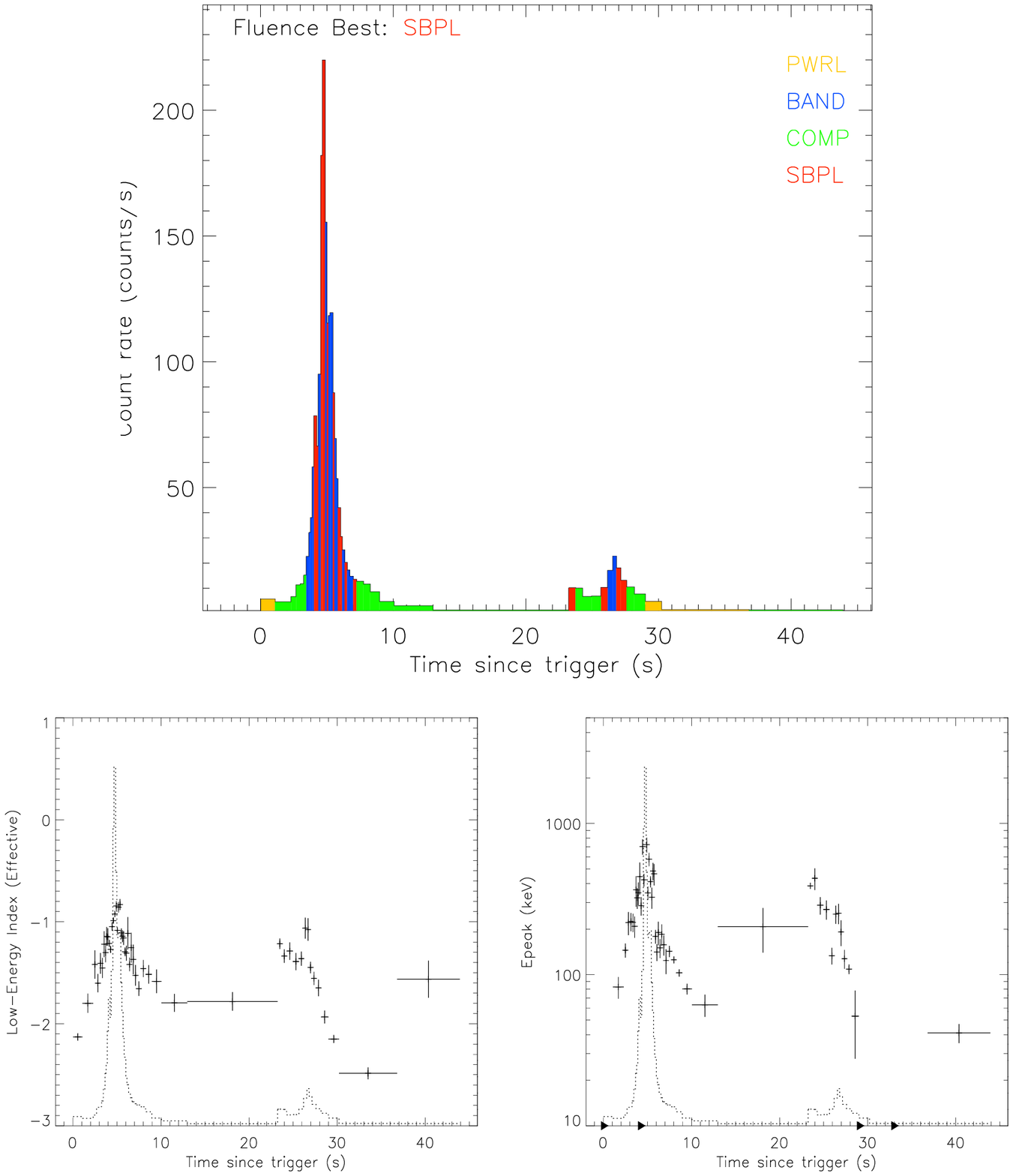}}
\caption{[{\it Top}] BEST model history of GRB 950403 (BATSE trigger number 3492).
The BEST model of the time-integrated spectrum for this burst is SBPL.
[{\it Bottom}] Evolutions of the low-energy index (effective) and $E_{\rm peak}$
for the same event.  The arrowheads in $E_{\rm peak}$ plot indicate where the
$E_{\rm peak}$ values cannot be determined.}
\label{fig:3492_bestmdl}
\end{figure}

Using the set of BEST models determined for each time-resolved spectrum, 
we can also examine which parts of the burst are well fit by which models.
This in turn reveals the evolution of spectral shapes within bursts.
In Figures \ref{fig:3491_bestmdl} and \ref{fig:3492_bestmdl}, we present the 
BEST models as a function of time for two example bursts that exhibit the 
hard-to-soft and the ``tracking" behaviors.  
Both behaviors are defined by $E_{\rm peak}$ and/or low-energy index evolution, 
which are commonly observed in GRBs \citep{for95, cri97}.
The evolution of the low-energy index and $E_{\rm peak}$ are also shown
along with the BEST model histories.
In the hard-to-soft burst (Figure \ref{fig:3491_bestmdl}), the first two-thirds 
is best fit by BAND, SBPL, or the COMP with high $E_{\rm peak}$. 
The softer tail spectra of the burst are well fit by either PWRL with index 
less than $-2$ (indicating $E_{\rm peak} \lesssim 30$ keV) or COMP with low 
$E_{\rm peak}$.
The last time interval of this burst was best fit by SBPL with $E_{\rm b}
= 177 \pm 55$ keV, but with $\lambda_1 > \lambda_2$ (i.e., concave-up shape),
due to a few sigma excess counts in one energy channel at $\sim$ 440 keV.
On the other hand, in the tracking burst (Figure \ref{fig:3492_bestmdl}),
the rise and fall of each peak are well fit by the low-$E_{\rm peak}$ COMP 
or PWRL.
\placefigure{fig:3491_bestmdl}
\placefigure{fig:3492_bestmdl}

\subsection{Time-Integrated and Time-Resolved Spectra}\label{sec:flnc_vs_res}
\begin{figure} 
\centerline{
\plotone{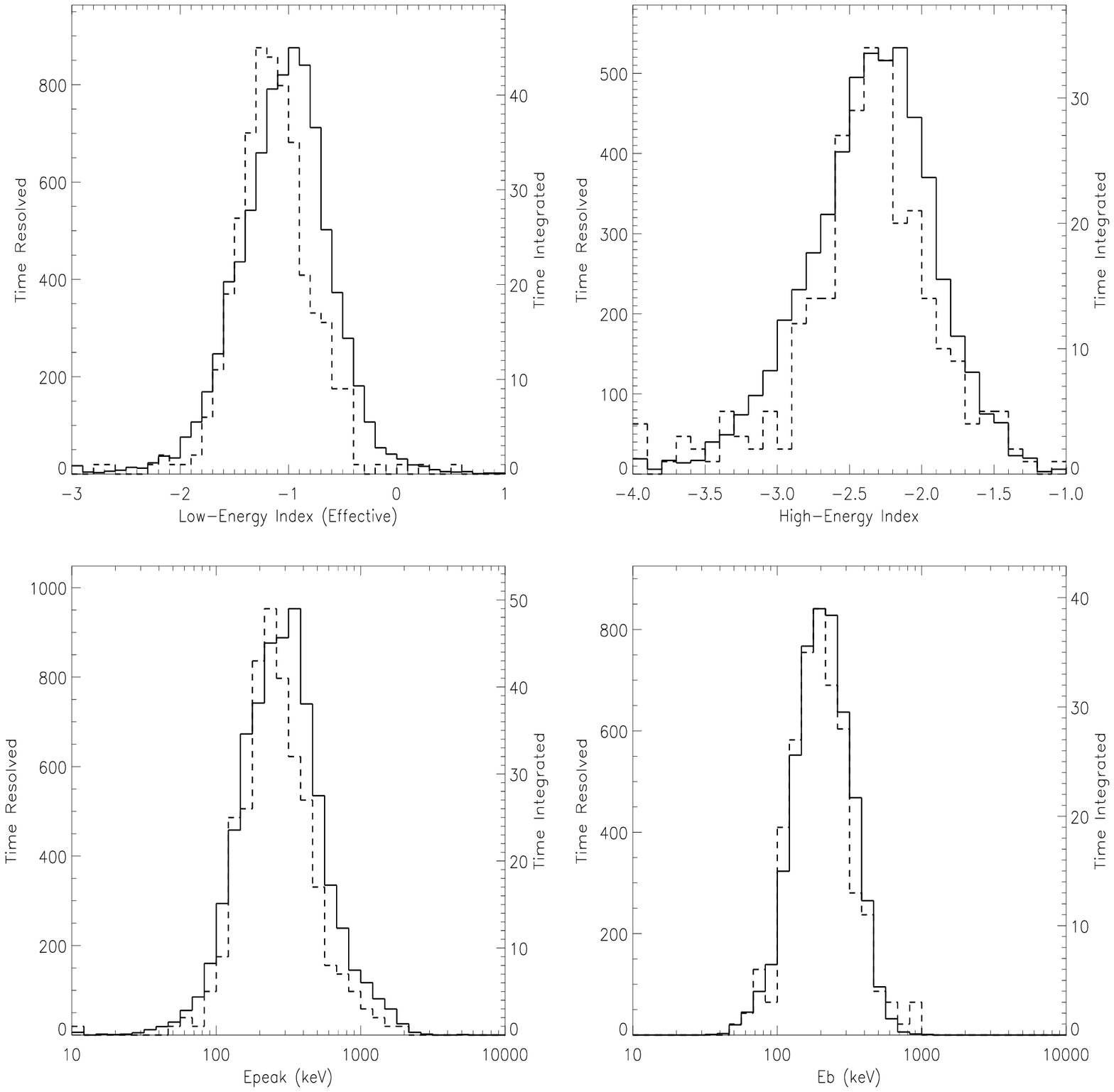}}
\caption{
Comparisons of the BEST model parameter distributions of time-integrated 
(dashed; right axis) and time-resolved spectra (solid; left axis). 
The lowest (highest) bin includes values lower (higher) than the 
edge values.}
\label{fig:fb_best1}
\end{figure}

For the comparison between the time-integrated and the time-resolved spectral
parameters of the BEST models, we plot the total distributions of time-integrated 
and resolved parameters in Figure \ref{fig:fb_best1}.
\placefigure{fig:fb_best1}
These are the total distributions presented in Figures \ref{fig:fpar_best1} and 
\ref{fig:bpar_best1}, but are now plotted on top of each other.
Also the median values of each parameter are summarized in Table
\ref{tab:spec_param}, along with the dispersion of the distributions.
\placetable{tab:spec_param}
In order to determine if the distributions of the time-integrated and 
time-resolved spectral parameters are consistent, we employed the 
Kolmogorov-Smirnov (K-S) test \citep{pre92}.
The K-S test determines the parameter $D_{\rm{\scriptscriptstyle KS}}$, 
which measures the maximum difference in the cumulative probability 
distributions over parameter space, and the significance probability 
$P_{\rm{\scriptscriptstyle KS}}$ for the value of 
$D_{\rm{\scriptscriptstyle KS}}$.
The null hypothesis is that the two datasets are drawn from the same parent 
distributions; therefore, a small $P_{\rm{\scriptscriptstyle KS}}$ rejects the 
hypothesis and indicates that the datasets are likely to be different.
Determined by the K-S statistics, we find a significant difference in the 
low-energy spectral index distributions of time-integrated and time-resolved 
spectra, with $P_{\rm{\scriptscriptstyle KS}}$ of 10$^{-16}$ 
($D_{\rm{\scriptscriptstyle KS}}$ = 0.23).
Also, a less significant difference was found between the $E_{\rm peak}$
distributions with $P_{\rm{\scriptscriptstyle KS}}$ of 8 $\times$ 10$^{-3}$ 
($D_{\rm{\scriptscriptstyle KS}}$ = 0.10).
The distributions of the high-energy index and the break energy were consistent.
This is in agreement with what we observed in the parameter distributions of 
individual model fits earlier (\S\ref{sec:par_dist}; Figures 
\ref{fig:fpar_aleff} -- \ref{fig:bpar_eb} and Table \ref{tab:mode_all}),
and is, again, due to the spectral evolution within each burst.

\begin{figure} 
\centerline{
\plotone{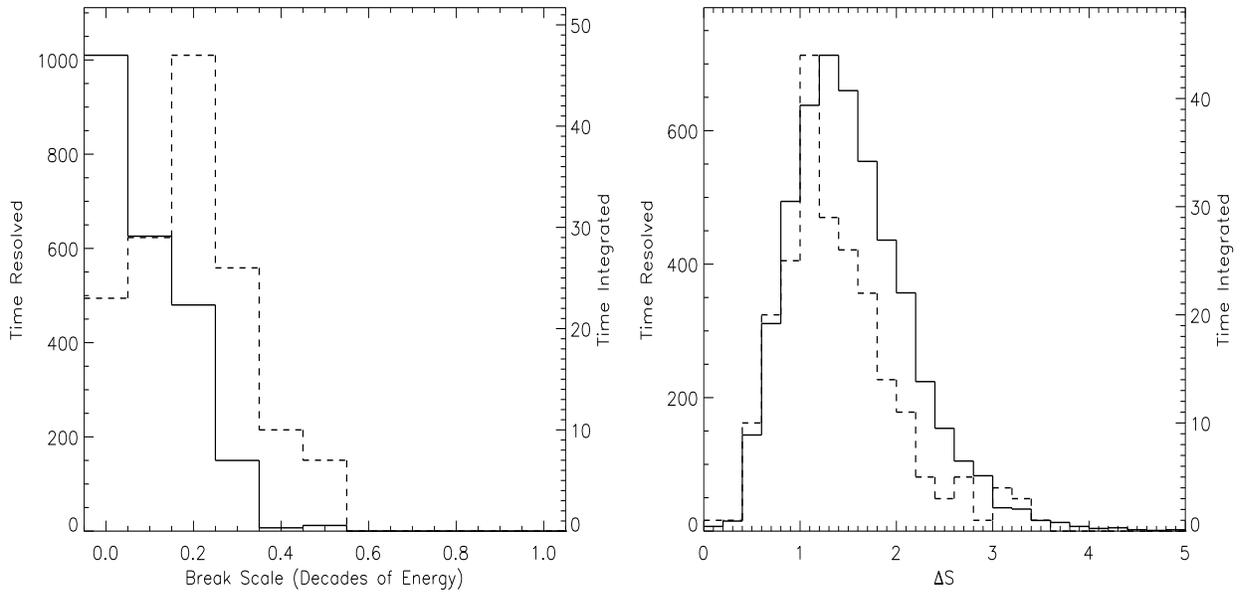}}
\caption{
Comparisons of the BEST model break scale ($\Lambda$) and $\Delta$S 
distributions of time-integrated (dashed; right axis) and time-resolved spectra 
(solid; left axis). 
The break scale values are for spectra fitted with SBPL only, while $\Delta$S
values are for those fitted with BAND and SBPL.}
\label{fig:fb_best2}
\end{figure}

We also compare the time-integrated and time-resolved distributions of SBPL 
break scales $\Lambda$ and $\Delta$S in Figure \ref{fig:fb_best2}.
\placefigure{fig:fb_best2}
It can be seen from the break scale comparison that the time-integrated SBPL 
spectra are smoother (larger break scales) than the time-resolved ones.
This is naturally expected because the integrated spectra are summations of 
resolved spectra with possibly various $E_{\rm b}$ values, and therefore, any
sharp break features may be smeared over.
For the same reason, the tendency of smaller $\Delta$S observed in integrated
spectra is expected as well.

It has been a common practice to fit the same broken-power law model (BAND in 
particular) to both time-integrated and time-resolved spectra.  
The model usually yields sufficiently good fits to both types of spectra; however, 
if the time-integrated spectrum consists of a set of broken-power law spectra with
evolving break energy and indices, it is possible that the time-integrated 
spectra deviates from a broken-power law shape.
Often, we do observe some indication of such deviations in the residual 
patterns obtained from time-integrated spectral fits to broken-power law models.
To probe this, we explored how each time-resolved spectrum contributes to the
time-integrated spectrum.  The time-integrated spectra are usually considered
as ``average" spectra; however, averaging (whether error/intensity-weighted 
or not) the best-fit spectral parameters from each time-resolved spectrum 
does not yield the best-fit time-integrated spectral parameters.
This is because the time-integrated spectra are averaged over count space
rather than parameter space.
Consequently, if each of the fitted time-resolved spectral models are 
indeed accurately representing the observed spectral data, summing over the 
model photon counts should reveal more accurate time-integrated photon spectrum.

\begin{figure}
\epsscale{0.6}
\centerline{
\plotone{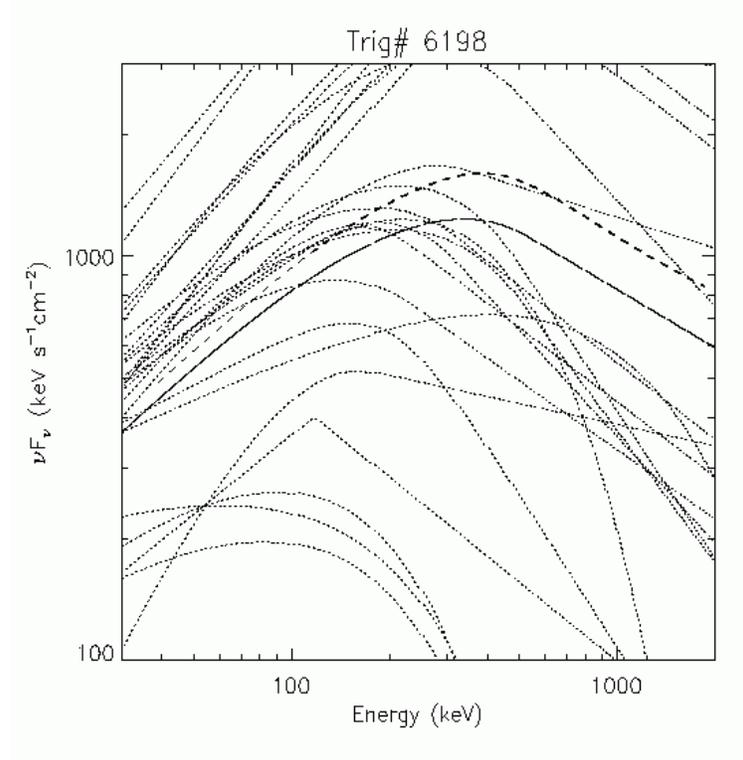}}
\caption{
Spectra of GRB 970420 (BATSE trigger number 6198).
Time-integrated BEST model is plotted as a solid curve.
The dashed curve shows the $\bar{{\mathcal F}}_{fluence}$ spectrum, which is
different from the BEST model.
The time-resolved BEST models are shown in dotted curves.
}
\label{fig:6198_spec}
\end{figure}

\begin{figure}
\epsscale{1.0}
\centerline{
\plotone{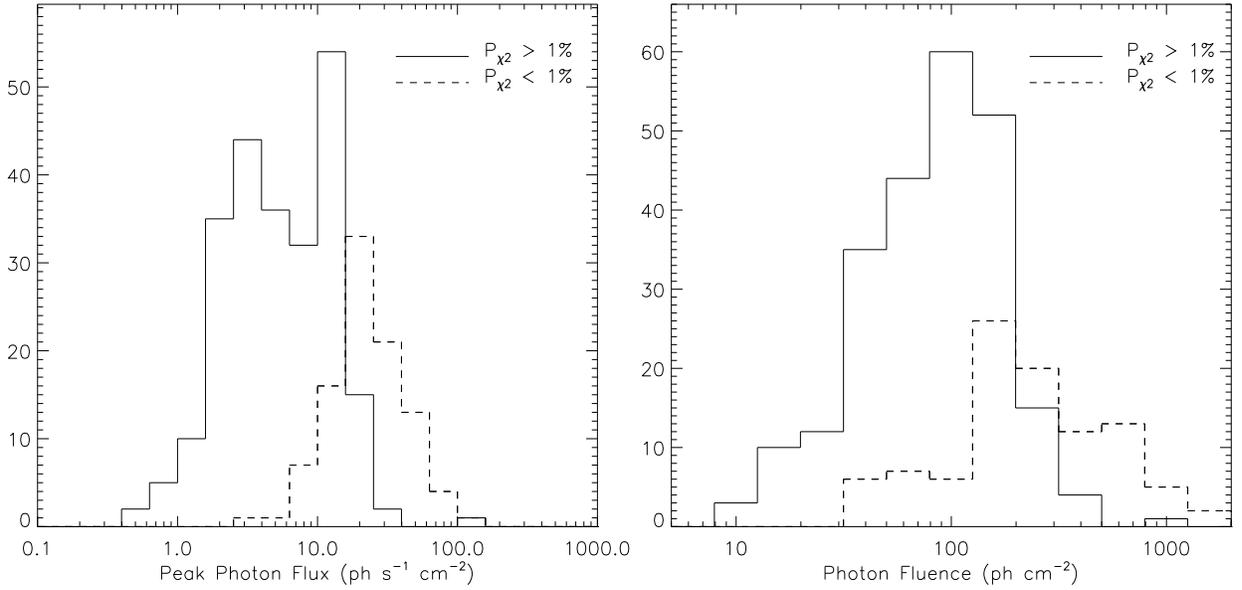}}
\caption{Distribution of peak photon flux (256 ms, 30 -- 500 keV; {\it left}) 
and total photon fluence values (25 -- 2000 keV; {\it right}) of 333 bursts.
The bursts with $P_{\chi^2_f} > 1$\% (solid line) have lower flux and fluence 
than those with $P_{\chi^2_f} < 1$\% (dashed line).}
\label{fig:maxp}
\end{figure}

Therefore, for each of 333 bursts (excluding 17 GRBs with single spectra), 
we obtained the time-integrated photon flux by summing the BEST model 
photon counts in the time-resolved spectra and dividing by their total 
durations;
\begin{equation}\label{eqn:sumspec}
\bar{{\mathcal F}}_{fluence}(E) = 
\frac{\sum_i{f_i(E) \Delta t_i}}{\sum_i\Delta t_i} .
\end{equation}
Here, $f_i(E)$ is the BEST model photon flux of each time-resolved spectrum 
as a function of photon energy, and $\Delta t_i$ is the accumulation time of 
each spectrum.
The initial visual comparisons of $\bar{{\mathcal F}}_{fluence}$ plotted over 
the best-fit time-integrated spectral models suggested that they agree 
remarkably well for most of the GRBs in our sample.
For quantitative comparisons of $\bar{{\mathcal F}}_{fluence}$ to the actual 
BEST model photon flux, we employed the $\chi^2$ statistic \citep{pre92}: 
\begin{equation}
\chi^2_f = \sum_{j = 1}^{N}  
   \frac{(\bar{{\mathcal F}}_{j} - {\mathcal F}_{j})^2}{{\sigma_j}^2},
\end{equation}
where $N$ is a number of energy bins, $\bar{{\mathcal F}}_{j}$ denotes
$\bar{{\mathcal F}}_{fluence}$ per energy bin, ${\mathcal F}_{j}$ is 
the actual time-integrated BEST model photon flux per energy bin, and 
${\sigma_j}^2$ is the variance associated with ${\mathcal F}_{j}$ obtained from 
the fits.
For each burst, the $\chi^2_f$ statistics and the corresponding significance
probability, $P_{\chi^2_f}$, were determined (for $N$ dof).  
Based on the statistic, 29\% of 333 bursts were rejected at 99\% confidence 
level (i.e., $P_{\chi^2_f} < 1$\%) due to large $\chi^2$, and 
therefore, $\bar{{\mathcal F}}$ was different from ${\mathcal F}$.
An example of such bursts is shown in Figure \ref{fig:6198_spec}.
\placefigure{fig:6198_spec}
We found that these bursts with very small $P_{\chi^2_f}$ values were of the 
brightest of the burst sample in terms of peak photon flux 
(Figure \ref{fig:maxp}).
\placefigure{fig:maxp}
Many of these bursts resulted in large $\chi^2_f$ because the brightest 
spectrum within the burst tends to dominate in the summation process.  
This caused $\bar{{\mathcal F}}$ to consistently have much larger flux than the 
BEST model photon flux, as seen in Figure \ref{fig:6198_spec}.
The time-integrated spectral fits for most of these bursts also resulted in
statistically poor spectral fits, with $\chi^2$ goodness of fit worse than 
$3\sigma$, even with the BEST models.
In very bright bursts, the contribution of systematic uncertainties in the 
data is generally expected to be more evident (see Figure \ref{fig:system}).
As a result, the best-fit photon models may not represent the actual
spectral shapes for these bursts.
We also found that the majority of the bursts for which MER data were used fall
into the group of small $P_{\chi^2_f}$.
This is likely because there are more systematic uncertainties present in the
multiple-detector MER data than the single-detector HERB or CONT data.
In addition, the MER bursts in this group
tend to consist of a larger number of spectra.
This is because the MER data type was used when the corresponding HERB data were 
incomplete, which occurs frequently in the case of very bright bursts.

\begin{figure}
\centerline{
\plotone{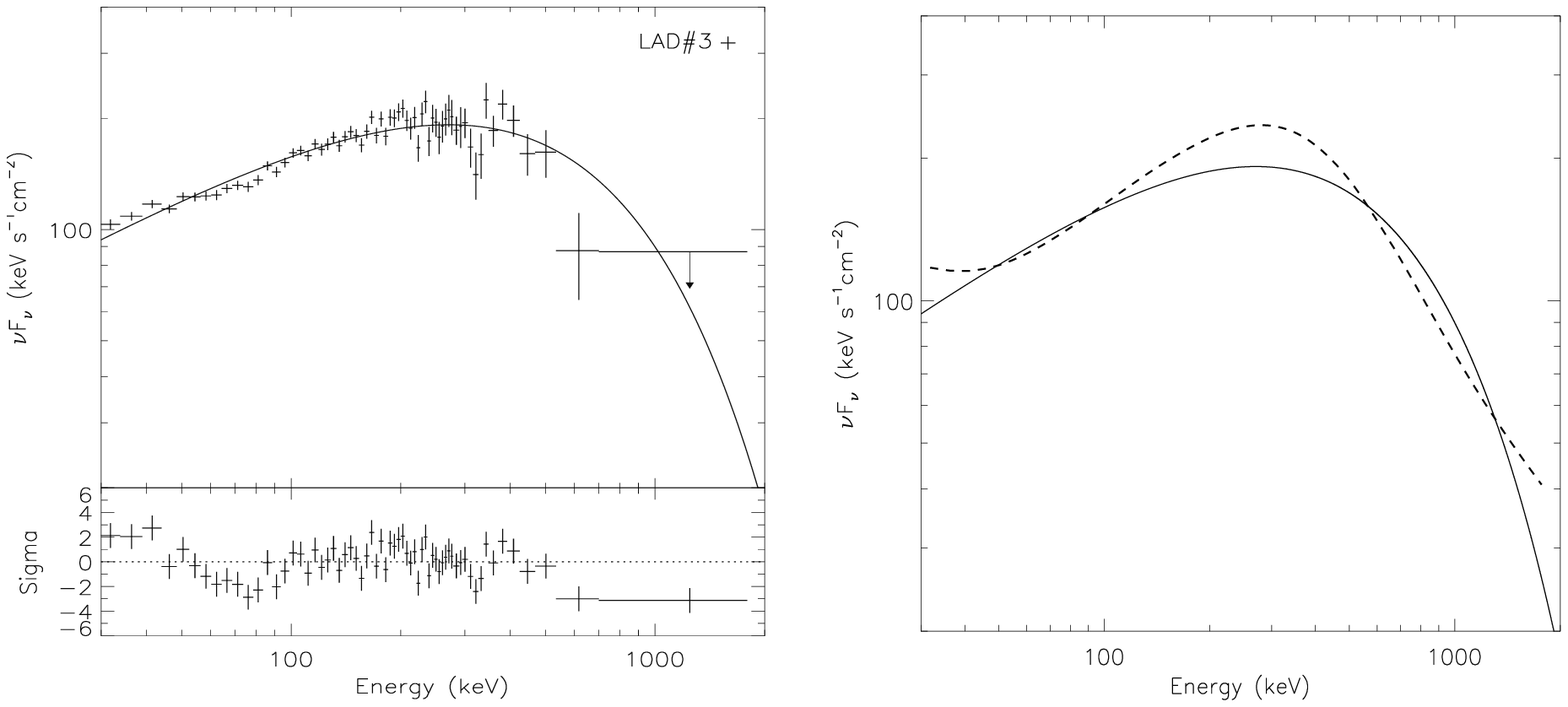}}
\caption{[$Left$] The actual deconvolved time-integrated photon data (crosses) 
and the best-fit model of GRB 980306.  
[$Right$] The constructed photon flux $\bar{\mathcal{F}}$ of GRB 980306 (dashed 
curve).  In both plots, the time-integrated BEST model (COMP) is plotted as 
solid curves.}
\label{fig:6630flnc_sp}
\end{figure}

Furthermore, among the small-$P_{\chi^2_f}$ group, there are some bursts where
the constructed photon flux $\bar{\mathcal{F}}$ seems to follow the actual
time-integrated photon data points (deconvolved with BEST models) very well.
One such example is shown in the Figure \ref{fig:6630flnc_sp}, for GRB 980306
(BATSE Trigger number 6630).
Although there seems be a slight offset in amplitude, it is remarkable that
the shape of constructed photon flux ($\bar{{\mathcal F}}$; dashed line) 
traces the actual deconvolved data points of the time-integrated spectrum.
The time-integrated spectral fit for this burst is still acceptable
($\chi^2/{\rm dof} = 159.6/110$; see Table \ref{tab:results}); however,
the deviations of the real spectral shape from a smoothly-broken power-law
model are evident.

From these, we infer that the time-integrated spectrum can be constructed 
by summing over the photon model counts of each time-resolved spectra within 
the burst (Equation \ref{eqn:sumspec}), provided that the BEST model in each 
time-resolved spectrum is an accurate representation of the observed spectral 
shape.  
The actual time-integrated spectrum may not be as simple as the BEST models 
fit to the spectrum, due to spectral evolution.
The bursts with peak photon flux larger than $\sim 30$ photons s$^{-1}$ 
cm$^{-2}$ will likely suffer from large systematic uncertainties, with the 
results that the constructed spectrum does not represent the actual spectral 
shape, or has a shape that differs greatly from the best-fit model.
Therefore, the fitted spectral parameters of the time-integrated spectra
may only represent the best parameter possibly fit by our simple models that 
are limited in their spectral shapes. The spectral parameter distributions
should be interpreted with this in mind.
This may also be applicable to the time-resolved spectra given that the cooling
timescales (both radiative and dynamical) can be much shorter than the 
integration times of each spectrum.

\subsection{Correlations Among Spectral Parameters}\label{sec:corr}
Some empirical correlations among GRB spectral parameters, such as between 
$E_{\rm peak}$ and low-energy index, have previously been reported with smaller 
samples \citep[e.g.,][]{cri97,llo02}.
The correlations were found either within individual bursts or for a collection 
of time-resolved parameters.
From inspection of scatter plots of the BEST parameters, we found no indication 
of strong global correlations among the time-integrated spectral parameters.
On the other hand, among the time-resolved spectral parameters, broad positive 
trends between $E_{\rm peak}-$low-energy index, $E_{\rm b}-$low-energy index, 
and $E_{\rm peak}-$high-energy index, were visible in the scatter plots.
To probe the existence of correlations among the spectral parameters derived 
from our sample, we calculated the Spearman rank-order correlation coefficients, 
$r_s$, and the associated significance probabilities, $P_{rs}$ \citep{pre92},
for several combinations of spectral parameters of the BEST models.
The null hypothesis is that no correlation exists; therefore, a small $P_{rs}$
indicates a significant correlation.

To eliminate the possible effects of a dispersion of $E_{\rm peak}$ and 
$E_{\rm b}$ due to the cosmological redshift ($z$) that varies from burst to burst, 
it is best to look for parameter correlations within individual bursts.
For each burst with a sufficient number of data points ($N \geq 10$) in our 
sample, we examined the correlations between combinations of low-energy and 
high-energy indices, $E_{\rm peak}$, and $E_{\rm b}$, by determining $r_s$ and 
$P_{rs}$. 
We use the BEST model parameters of each time-resolved spectrum within bursts,
and also, for simplicity, we denote the low-energy and high-energy indices of 
the BEST model as $\alpha$ and $\beta$ respectively, in this section.
The number of parameter pairs ($N$) in each burst was less than or equal to 
the number of time-resolved spectra; however, this varied according to which 
parameter pair was considered since we used the BEST model parameters and 
certain parameters are not obtainable for certain resolved spectra.
For example, within the same burst, it is possible that for an 
$E_{\rm peak} - \alpha$ correlation, $N = 20$ pairs were available, but for 
$E_{\rm peak} - \beta$ only $N = 15$ were available.
For most of the cases, $N$ was $\gtrsim$ 80\% of the number of spectra and was 
always more than 10.  
We note that the $P_{rs}$ values take into account the actual $N$ in the 
calculation that determines each correlation significance.
The resulting numbers of bursts considered for each correlation were summarized 
in Table \ref{tab:corr_sum} along with percentages of bursts with 
significant correlations between each pair of parameters.
\placetable{tab:corr_sum}
We considered the correlation significant if $P_{rs} < 10^{-3}$ ($ > 3\sigma$).
The coefficient $r_s$ and the associated $P_{rs}$ of the bursts that showed 
significant correlations in at least one of the parameter pairs are listed in 
Table \ref{tab:corr}.
\placetable{tab:corr}

\begin{figure} 
\epsscale{0.9}
\centerline{
\plotone{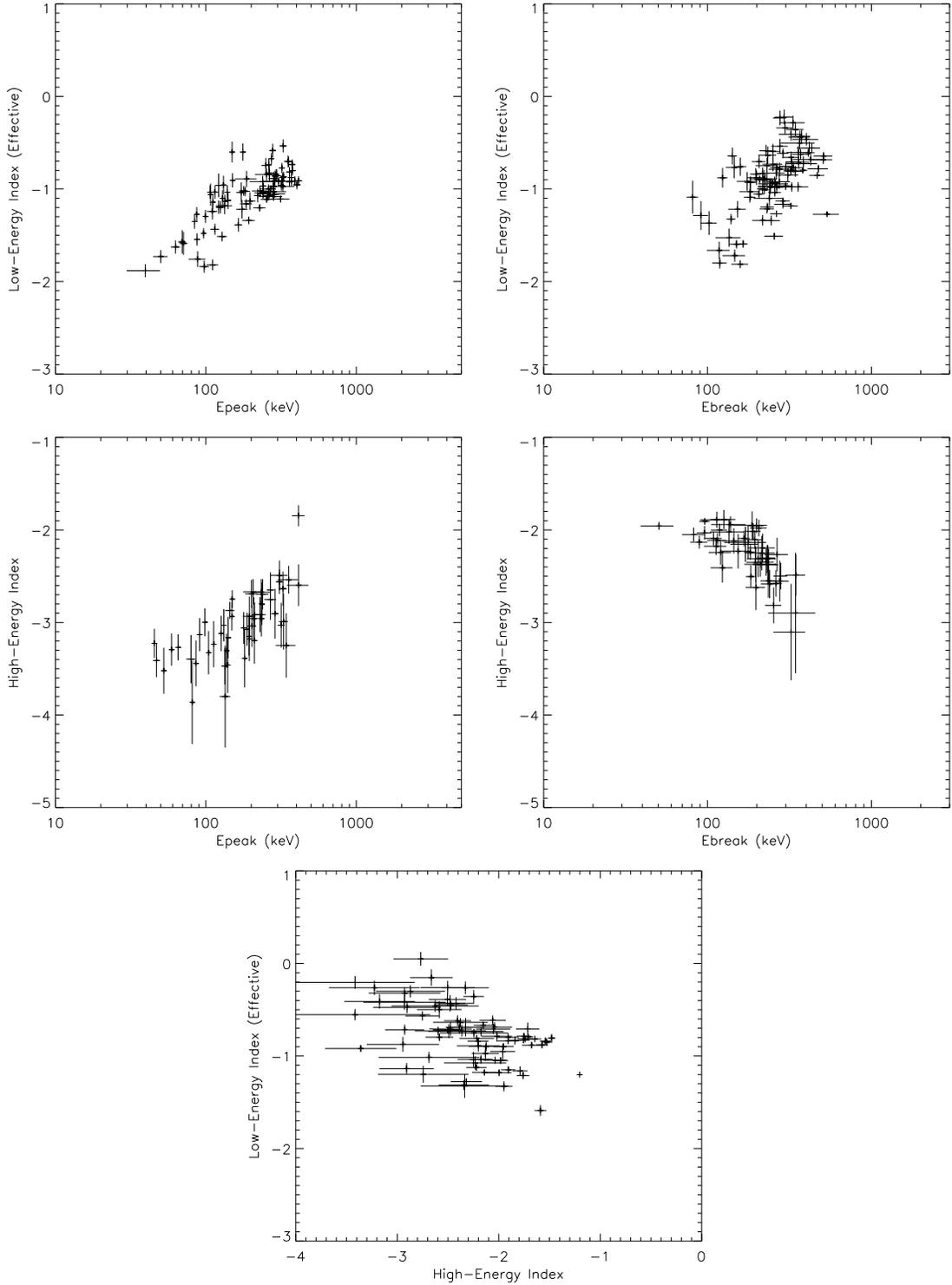}}
\caption{Example scatter plots of spectral parameter pairs showing correlations.
[{\it Top}] $E_{\rm peak}-\alpha$ of GRB 941020 (trigger number 3253; 
{\it left}) and $E_{\rm b}-\alpha$ of GRB 931204 (2676; {\it right}). 
Both show strong positive correlations.
[{\it Middle}] $E_{\rm peak}-\beta$ of GRB 911118 (trigger number 1085; 
{\it left}) and $E_{\rm b}-\beta$ of GRB 980203 (6587; {\it right}).
[{\it Bottom}] $\alpha-\beta$ of GRB 920824 (trigger number 1872). 
Negative correlations are evident in $E_{\rm b}-\beta$ and $\alpha-\beta$.}
\label{fig:corr}
\end{figure}

To illustrate some of the strong correlations, we show example scatter plots 
of each type of parameter pairs in Figure \ref{fig:corr}.
\placefigure{fig:corr}
The strongest positive correlation is found between $E_{\rm peak}$ and 
$\alpha$, in 26\% of the bursts.  In many cases, the chance probabilities of 
these correlations are extremely low. 
Positive correlations between $E_{\rm b}-\alpha$ are also evident, although
only in half of the fraction of the $E_{\rm peak}$ cases.
We note that the results were not significantly altered when unadjusted
low-energy index (regular $\alpha$ for BAND and COMP) were used, instead of
$\alpha_{\rm eff}$ that we use here for the BEST model parameters.
The $E_{\rm b}$ value is equal to the $E_{\rm peak}$ value only when a 
spectrum has a sharp break.  
Otherwise, a break scale or curvature determines the relation between them.  
Therefore, the same degree of correlation is not expected for $E_{\rm b}$
and $E_{\rm peak}$.
On the contrary, the $E_{\rm peak}-\beta$, $E_{\rm b}-\beta$ and $\alpha-\beta$
correlations are found in a much lower fraction of the bursts than the 
$E_{\rm b}$/$E_{\rm peak}$ and $\alpha$ cases, and are mostly negative.
It must be noted, however, that the uncertainty associated with each parameter 
is not taken into account in determining the rank-order correlation, and 
$\beta$ is usually associated with relatively larger uncertainties than the 
other parameters.  This can be seen in the scatter plots in Figure
\ref{fig:corr}.
The measurement uncertainties may have masked an actual correlation.
We have also found significant positive correlations between $E_{\rm peak}$ and
the time-resolved photon flux in 28\% of GRBs, and between $\alpha$ and the
photon flux (42\%), indicating the tracking behavior.  
About 18\% of GRBs with the $E_{\rm peak}-$flux correlation also belong to the 
group with the $\alpha-$flux correlation, and therefore show strong tracking 
behavior in both $E_{\rm peak}$ and $\alpha$. 
Some of these also display overall hard-to-soft behavior, indicated by strong 
correlations of the parameters with time.

These correlations among the spectral parameters should be accounted for within
an emission model.  The observed correlations between fitted model parameters
indicate possible correlations between physical parameters at the GRB emitting 
region.
For example, according to the synchrotron shock model, $E_{\rm peak}$ should
be related to the magnetic field strength and the minimum energy of accelerated
electrons \citep{tav96b}, whereas low-energy index may depend on the
electron pitch-angle distribution or density of the absorbing medium \citep{llo02}.
On the other hand, in the jitter radiation model, $E_{\rm peak}$ varies with the
ratio between deflection and beaming angles of an emitting electron, and 
low-energy index depends on the ratio of strengths of large-scale and small-scale
magnetic fields existing at the shock \citep{med00}.

\begin{figure}
\epsscale{0.6}
\centerline{
\plotone{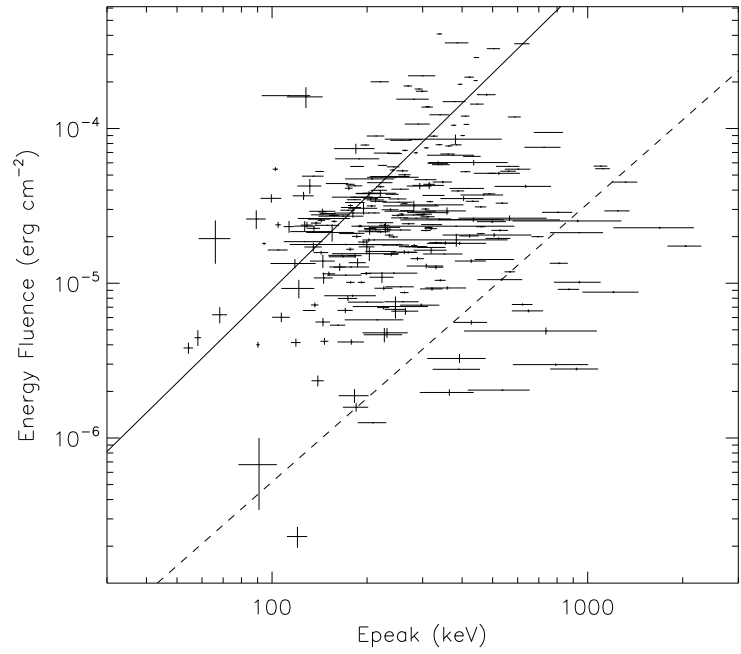}}
\caption{Energy fluence in 25 -- 2000 keV vs.~$E_{\rm peak}$ plot.
The Amati relation limit is shown as a solid line and Ghirlanda 3$\sigma$ limit
is shown as a dashed line.  Bursts below these lines are 
inconsistent with the relation.  The uncertainties are 1$\sigma$.}
\label{fig:eflnc_ep}
\end{figure}

Finally, global correlations between the time-integrated $E_{\rm peak}$, $\alpha$, 
$\beta$, and energy or photon flux/fluence values (including peak photon flux)
were also investigated.
We found only one relatively significant correlation, between $E_{\rm peak}$ and 
total energy fluence, with $r_s = 0.20$ and $P_{rs} = 2.5 \times 10^{-4}$.
One of the most highly cited empirical correlations is the $E_{\rm peak} - 
E_{\rm iso}$ correlation found by \citet{ama02}. 
The Amati relation is $E_{\rm p,0} \propto E_{\rm iso}^{0.5}$, where
$E_{\rm p,0}$ is the peak energy in the source rest frame, namely, 
$(1 + z)E_{\rm peak}$, and $E_{\rm iso}$ is an isotropic equivalent total
emitted energy for a given distance.
The relation was found by using a dozen GRBs observed with {\it BeppoSAX}
with known redshift values $z$.
If true, such a correlation can provide strong constraints on the GRB emission 
mechanism and the fundamental nature of GRBs.
It has recently been shown, however, that the Amati relation suffers from
a strong selection effect and is greatly inconsistent with a larger set of
GRB data obtained with BATSE \citep{nak05b, ban05}.
This was indicated by a limit in observed $E_{\rm peak}$ and energy fluence values
(and therefore independent of redshift) that is implied by the Amati relation.
Responding to the results, \citet{ghi05} argued that taking into account
the intrinsic scatter of the relation, the BATSE bursts may still be
consistent.  This claim, however, has also been challenged \citep{nak05c}.
To show how $E_{\rm peak}$ and energy fluence values in our sample are correlated,
we plot $E_{\rm peak}$ vs.~total energy fluence in Figure \ref{fig:eflnc_ep}. 
\placefigure{fig:eflnc_ep}
On the scatter plot, we also show the limit on the Amati relation and the maximum 
3$\sigma$ limit derived by \citet{ghi05}.
The bursts below these lines are inconsistent with the relation.
Note that the GRB sample used to test the consistency in \citet{ban05} includes
many more dimmer bursts than bursts in our sample.
Despite the fact that our sample only consists of bright GRBs, we confirm that 
most of our bursts are significantly inconsistent with the Amati limit.
Even for the 3$\sigma$ Ghirlanda limit, we observe more outliers than they found 
in their work.
Our well-constrained parameters strongly indicate that the Amati relation is 
only valid for a small sample of selected bursts.

\subsection{BETA Model Fit Results}\label{sec:beta_res}
Since the BETA model is a variation of the BAND model, we did not include the BETA
model in obtaining the overall model statistics above.  
The purpose of this model is to test whether the high-energy power-law index, 
$\beta$, stays constant for the entire duration of a burst, which is expected in 
the simplest shock acceleration and the GRB emission scenario.
Earlier work by \citet{pre98a} found that 34\% of 122 GRBs investigated were 
inconsistent with the constant-$\beta$ hypothesis.
Note that most GRBs in their sample overlaps with the sample presented herein.
In \citet{pre98a}, the $\beta$ values that were used as the constant values were 
the error-weighted average of $\beta$ obtained from fits to each of time-resolved 
spectra within bursts.
These values are different from those obtained by the time-integrated fits, due 
to the following reasons. 
Spectra with higher S/N usually have smaller error associated with their fits.
Therefore, they have more weight in determining the error-weighted average of 
$\beta$.
These spectra, however, may have much shorter integration times, because of the
binning by S/N as well as the characteristics of HERB data (also used in 
\citet{pre98a}).
Despite their large photon flux values, they may not provide large photon fluence
counts due to the short time intervals.
As discussed in the earlier section (\S\ref{sec:flnc_vs_res}), the time-integrated 
spectrum can be obtained by summing over the photon counts (not count rate) 
rather than error-weighted averaging of parameters.
This resulted in the values used to test the constancy of the index in 
\citet{pre98a} to be higher than those obtained from fits to the time-integrated 
spectra, which are used in this work.  

In order to have sufficient time samples to study the time evolution of $\beta$, 
as well as to be consistent with the \citet{pre98a} analysis, our sample was 
limited to bursts that consist of at least eight spectra.
Additionally, there were some cases in which the time-integrated $\beta$ values 
were less than $-5$ by more than 1$\sigma$, and thus lacking the high-energy 
power-law component (i.e., COMP-like spectra).
In these cases, fitting the time-resolved spectra with these small values
is not meaningful for our purpose here, and therefore, those bursts were also
excluded.  The remaining total number of bursts in the sample was 210.

To determine whether $\beta$ is consistent with being constant throughout a 
burst, we checked the resulting $\chi^2$ of the BETA model fit to each 
time-resolved spectrum and calculated the corresponding goodness of fit.
The maximum goodness of fit allowed was set to an equivalent chance probability 
of one per number-of-spectra in each burst in order to take into account the
number of trials in each case: in other words, one spectrum in 
each burst was statistically expected not to give an acceptable BETA fit. 
Consequently, the goodness-of-fit allowed was always at least $87.5$\% 
(corresponding to a minimum number of spectra of eight).
Then, if all but one spectrum within a burst resulted in $\chi^2$ probabilities 
less than the given maximum goodness-of-fit level, the burst was considered to 
be consistent with having a constant $\beta$.
We call these bursts ``constant-$\beta$ GRBs" and the others ``varying-$\beta$
GRBs" hereafter.

\begin{figure}
\epsscale{0.6}
\centerline{
\plotone{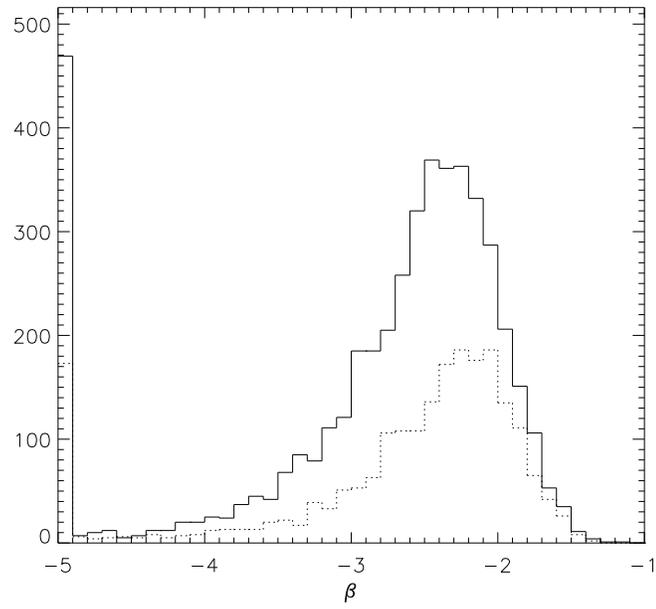}}
\caption{
The distributions of the BAND fit $\beta$ values of time-resolved spectra within
varying-$\beta$ (solid line) and constant-$\beta$ (dotted line) GRBs.
The lowest bin includes values lower than --5.}
\label{fig:const_beta}
\end{figure}

\begin{figure}
\epsscale{0.6}
\centerline{
\plotone{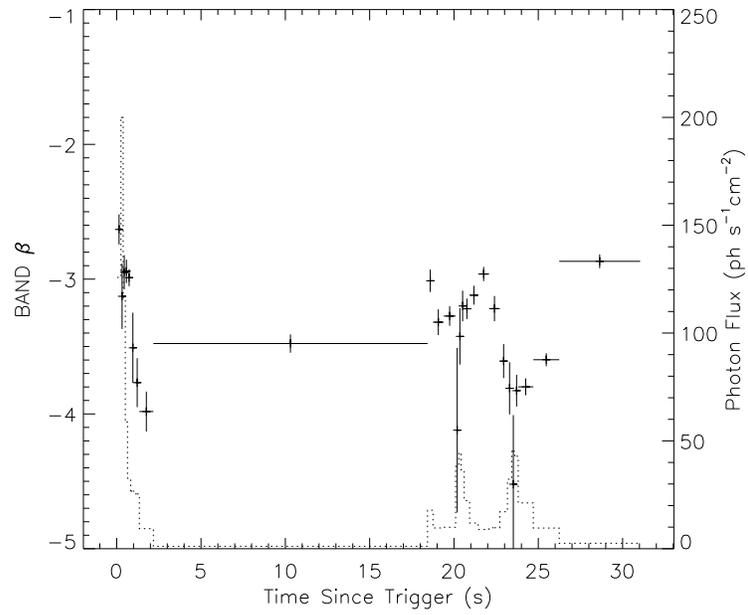}}
\caption{
Evolution of the BAND fit $\beta$ values of an example varying-$\beta$ GRB
(GRB 970201, BATSE trigger number 5989).
The photon flux history is overplotted with dotted lines (right axis).}
\label{fig:5989_beta}
\end{figure}

We found that nearly a half of the sample (108 out of 210 GRBs) are 
varying-$\beta$ GRBs. 
The result differs from the result of 34\% that was obtained by \citet{pre98a},
who searched for a correlation between $\beta$ and time.
The time-integrated spectra of the varying-$\beta$ GRBs are more 
likely to be best fit with the COMP model, which has a high-energy cutoff.
In Figure \ref{fig:const_beta}, the distributions of the BAND fit $\beta$ values
(i.e., fit with $\beta$ as a free parameter) 
for the time-resolved spectra within the varying-$\beta$ and constant-$\beta$ GRBs 
are compared.
\placefigure{fig:const_beta}
We also show the BAND $\beta$ evolution of an example varying-$\beta$ GRB in
Figure \ref{fig:5989_beta}.
\placefigure{fig:5989_beta}
A larger fraction of the varying-$\beta$ spectra are fitted by smaller $\beta$ 
values than the constant-$\beta$ spectra.
It may be the case that the varying-$\beta$ GRBs contain high-energy cutoff 
spectra where $\beta$ is essentially less than $-5$, which in turn fails the 
BETA fits with $\beta > -5$.
The median value of the time-resolved BAND $\beta$ for the varying-$\beta$ and 
constant-$\beta$ spectra are $-2.51^{+0.32}_{-0.52}$ and $-2.37^{+0.30}_{-0.49}$, 
respectively.
No differences in the $E_{\rm peak}$ and the $\alpha$ distributions are 
evident between the constant and varying $\beta$ bursts.
For the constant-$\beta$ GRBs, the median value for the BETA high-energy index 
(the value that were used as the constant value) was $-2.39^{+0.19}_{-0.34}$.

\subsection{Short GRBs}\label{sec:17short}
\begin{figure}
\epsscale{1.0}
\centerline{
\plotone{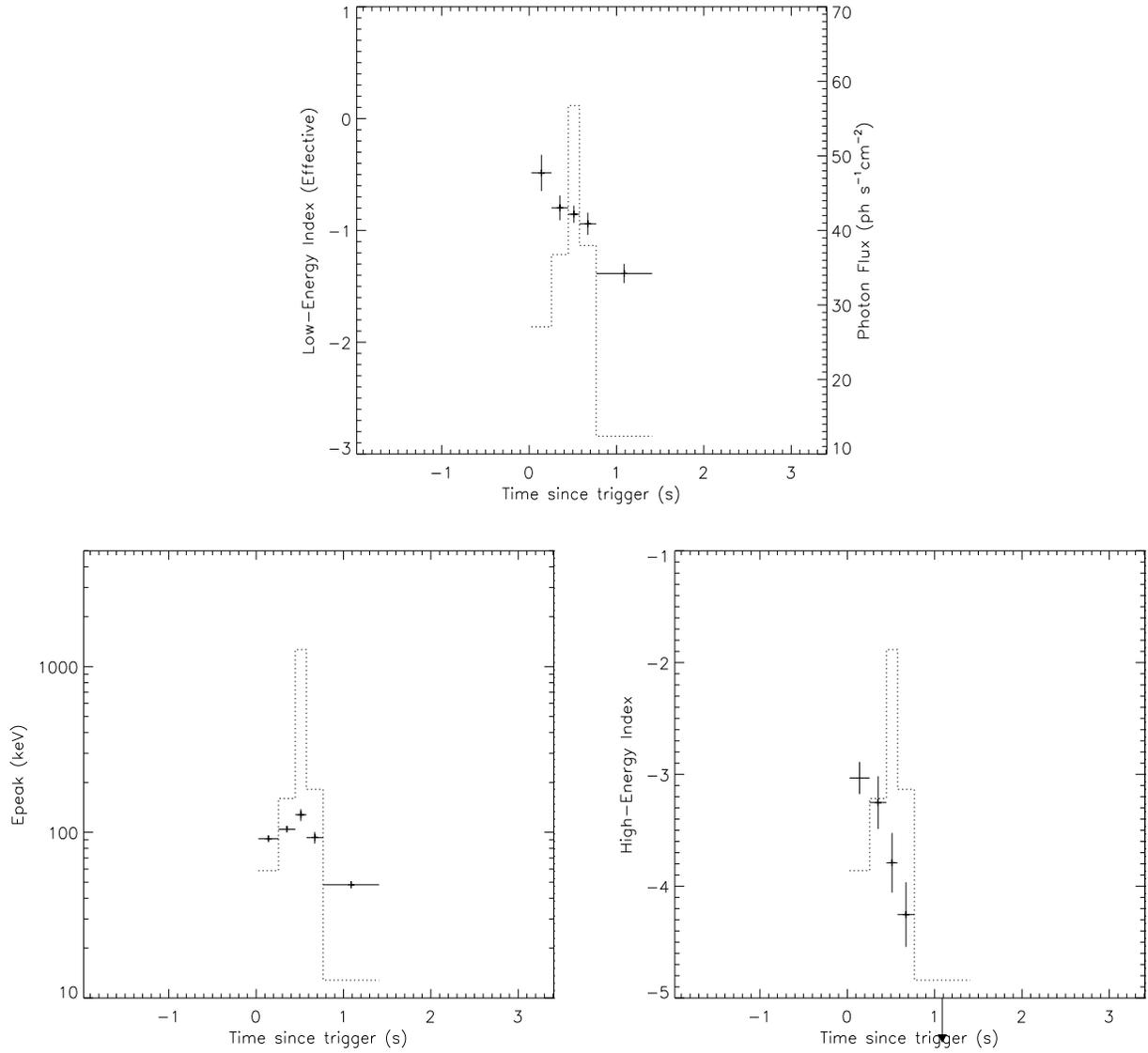}}
\caption{Parameter evolution of a short event, GRB 000326 (trigger number 8053).
The photon flux histories are over-plotted with dotted lines (top panel, right
axis).  $E_{\rm peak}$ tracks the photon flux while the indices evolve from 
hard to soft.}
\label{fig:8053}
\end{figure}

Our sample includes 17 short bursts ($T_{90} \lesssim 2$ s), which are listed
in Table \ref{tab:short}. 
\placetable{tab:short}
Short GRBs were previously shown to be spectrally harder than long bursts in 
terms of the spectral hardness ratio \citep{dez92, kou93} and spectral parameters 
\citep{pac03}.
The parameter comparison of short GRBs and long GRBs in our sample is summarized 
in Table \ref{tab:short_long}.
\placetable{tab:short_long}
There is no significant difference between the spectral parameters of the short 
bursts and the rest (long bursts), in both time-integrated and time-resolved 
spectra.
This may be expected since our sample is limited to the brightest bursts,
whereas the sample used in \citet{pac03} did not have such selection effects.
\citet{mal95} found that brighter bursts tend to be harder, and our sample
here belongs to the brightest of the five groups defined by \citet{mal95}.
Besides, we used mostly HERB data for the short bursts, which may be
missing a substantial portion of the emission of the short bursts due to the 
accumulation start time of the HERB data ($\ge$ 64 ms after burst onset time). 
Three short bursts, GRBs 920414, 980228, and 000326 (BATSE trigger numbers 1553, 
6617, and 8053), were bright enough to provide three to five time-resolved 
spectra, and they all clearly exhibited spectral evolution.
As an example, the spectral evolution of GRB 000326 is shown in Figure 
\ref{fig:8053}.
\placefigure{fig:8053}
In two cases (GRBs 920414 and 980228), the spectra evolved hard to soft in all
low-energy \& high-energy indices and $E_{\rm peak}$, whereas in the other case 
(GRB 000326; Figure \ref{fig:8053}) we observed tracking behavior in 
$E_{\rm peak}$ while both indices monotonically decayed. 
If short GRBs generally evolve hard to soft, the HERB data of short bursts will
lack a coverage of the hardest portions of the bursts, biasing the spectral 
parameters of the short GRBs in our sample to softer values. \\

\section{Summary and Discussion} \label{sec:summary}
We have analyzed the large sample of bright BATSE GRBs with high-energy resolution 
and high-time resolution, using five different photon models. 
For both time-integrated and time-resolved spectra, we have presented the 
distributions of each spectral parameter; low-energy and high-energy power law
indices, $E_{\rm peak}$, and break energy, determined with the best-fit model to 
each spectrum, with good statistics.

We confirmed, using a much larger sample, that the most common value for the 
low-energy index is $\sim -1$ \citep[SP1;][]{ghi02}.
The overall distribution of this parameter shows no clustering or distinct 
features at the values expected from various emission models, such as --2/3 
for synchrotron \citep{kat94, tav96b}, 0 for jitter radiation \citep{med00}, 
or --3/2 for cooling synchrotron \citep{ghi99}.
About 5\% of the time-resolved spectra are found to have the low-energy indices
significantly above the synchrotron limit of --2/3.
The median value for the high-energy indices is found to be $\sim -2.3$ and
the parameter can be constrained only if the value is larger than $\sim -5$ with
the BATSE LAD data.
The dispersions in the low-energy and high-energy index distributions were found to
be comparable, $\sigma \sim 0.25$.
For the first time, $E_{\rm peak}$ and break energy, at which the two power laws 
are joined, have been made distinct.
We presented the $E_{\rm peak}$ and break energy distributions separately, and 
found that $E_{\rm peak}$ tends to be harder than the break energy.
This is due to the existence of curvature in spectra.
$E_{\rm peak}$ becomes break energy only in a sharply-broken power-law spectrum. 
The $E_{\rm peak}$ distribution peaks at $\sim 300$ keV while that of the break 
energy peaks at $\sim 200$ keV, both with very narrow width of $\lesssim 100$
keV.  We now know that there exist fainter, softer GRBs with lower 
$E_{\rm peak}$ that BATSE would not have detected \citep[e.g.,][]{sak05}; 
therefore, the actual observed $E_{\rm peak}$ distribution of GRBs should extend 
to a lower energy.
The narrowness of the distributions, if real, implies an extremely
narrow intrinsic distribution of these parameters, and poses a challenge
for the internal shock model of GRB prompt emission \citep{zha02}.
There are small populations of GRBs (7\% of time-resolved spectra) that have 
high-energy indices above --2 by more than 3$\sigma$.
This indicates that there may be a small tail population of spectra with 
$E_{\rm peak} \gtrsim 2$ MeV.

It must be noted that the time-resolved spectral analysis results obtained here
may be biased by the brighter (higher photon flux) portions of each burst because
the HERB data type was designed to sample more frequently during the intense 
episodes.
Since the $E_{\rm peak}$ values tend to be correlated with the photon flux in 
general \citep{mal95}, the spectra with higher $E_{\rm peak}$ might be 
over-sampled whereas those with lower $E_{\rm peak}$ are likely to be 
under-sampled.
This may partially explain the differences observed in the $E_{\rm peak}$ 
distributions of time-integrated and resolved spectra.
Similarly, the low-energy indices also tend to correlate with $E_{\rm peak}$; 
therefore, the same type of bias may exist in the low-energy index distribution 
as well.

We have also explored time-integrated spectra by reconstructing them from
the time-resolved spectra, and identified cases where the actual 
integrated spectra deviate from the simple broken power-law model, due to 
spectral evolution within the integration time.
Therefore, fine time resolution data are crucial to more accurately represent the 
spectral shape using a relatively simple model.
Significant correlations among parameters are also observed in some GRBs.
The most evident positive correlation is found between $E_{\rm peak}$ and 
the low-energy index, in 26\% of GRBs considered.  
Since our parameter sets are well constrained and unbiased by photon models, the 
correlations found are more likely to reflect intrinsic properties of GRBs, 
rather than the instrumental or other systematic effects, as has been pointed out
\citep[e.g.,][]{llo02}.
In addition, a mild global correlation between $E_{\rm peak}$ and energy fluence in our 
sample was found.  The significance of the correlation is much lower than
what was found previously by \citet{llo+00}, with a larger sample including many
more dimmer bursts.

In its simplest picture, the synchrotron shock model assumes that the electrons 
are accelerated by the first-order Fermi mechanism to a power-law distribution 
($N(\gamma) \propto \gamma^{-p}$), which does not evolve in time.
The electron index $p$ is then related to $\beta$ either as $\beta = - (p + 2) / 2$
in case of the fast-cooling synchrotron spectrum, or as $\beta = - (p + 1) / 2$ 
in case of non-cooling synchrotron \citep{sar98}.
Our constant $\beta \sim -2.4$ then indicates $p \sim 2.8$ or $\sim 3.8$, 
respectively.
The ultra-relativistic Fermi-type acceleration results in $p = 2.2 - 2.3$
\citep{gal02}; therefore, if indeed the electron distribution remain unchanged
within a GRB, the fast radiative cooling of the electrons are more likely to be 
in effect at the source, as has been suggested \citep{pre02, llo02}.
However, we found, using the BETA model, that 51\% of 210 GRBs considered are 
inconsistent with the constant-$\beta$ hypothesis.
Finally, we found no significant difference between the spectral parameters of 17 
short bursts and long bursts in our sample, which is possibly due to the
instrumental effects.
In addition, spectral evolutions, hard-to-soft or tracking, were found in three 
of the short bursts with enough number of time-resolved spectra.
These evolution patterns are often observed in individual peaks of long GRBs.

It is important to note that the analysis presented here only includes bright
GRBs, which tend be spectrally harder in general than dim GRBs.
The global correlations we studied here between the spectral hardness and the 
burst intensity (flux and fluence), would probably change somewhat, if 
dimmer bursts were included.
Another sample selection bias exists against very short GRBs, due to the fact that
we used the peak photon flux in 256 ms to determine the brightness of the bursts.
Many bright short GRBs have duration much shorter than 256 ms, in which case the
photon flux with 256-ms integration time may smear out the brightness.
The short GRBs included in our analysis, therefore, have duration $\gtrsim$ 256 ms,
and only comprises less than 5\% of our sample of 350 bursts.
This is much less than the fraction of short GRBs in the entire BATSE GRB sample 
($\sim$ 19\%).

The GRB spectral database obtained in this work is derived from the most
sensitive and largest database to date.
Therefore, these results set a standard for spectral properties of GRB prompt
emission with exceptional statistics.
Our analysis results can provide reliable constraints for 
existing and future theoretical models of the particle acceleration and 
GRB emission mechanisms.
The spectral database also allows for the spectral evolutions within 
each burst to be taken into consideration when constraining such models. \\

\begin{appendix}
\section{Derivation of the Smoothly Broken Power Law}\label{ap:sbpl_model}

\begin{figure}
\epsscale{0.6}
\centerline{
\plotone{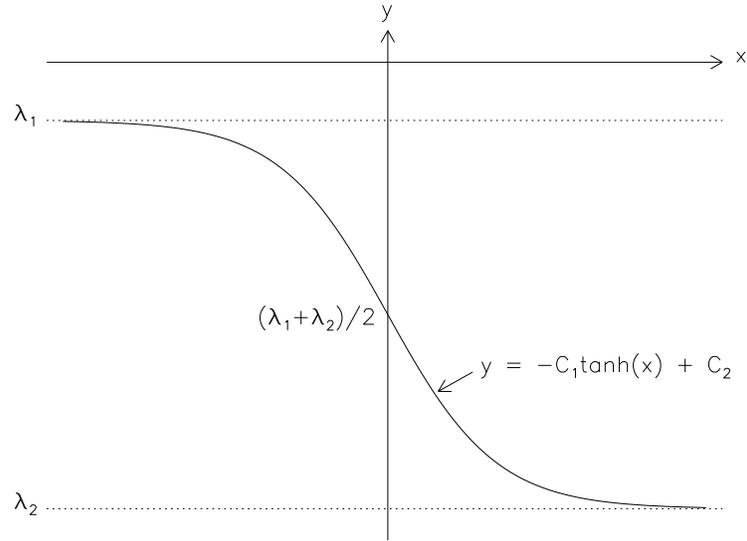}}
\caption{Plot of Equation \ref{eqn:tanh}.  
The index changes from $\lambda_1$ to $\lambda_2$ smoothly.}
\label{fig:tanh}
\end{figure}

The basic concept is that the power-law index changes smoothly from
the low-energy index ($\lambda_1$) to high-energy index ($\lambda_2$).
Since $\lambda_1 > \lambda_2$ in general, we assume the index change 
is described by a negative hyperbolic tangent function, namely
$y = -C_1 \tanh (x) + C_2$, as plotted in Figure \ref{fig:tanh}.
$C_1$ and $C_2$ are positive constant.
\placefigure{fig:tanh}

Here, $y = d(\log f(E))/d(\log E)$, and it is evident from Figure 
\ref{fig:tanh} that
$C_1 = \frac{\textstyle \lambda_1 - \lambda_2}{\textstyle 2}$ and 
$C_2 = \frac{\textstyle \lambda_1 + \lambda_2}{\textstyle 2}$.
So we have
\begin{equation}\label{eqn:tanh}
\frac{d(\log f)}{d(\log E)} = -\frac{\lambda_1 - \lambda_2}{2}
   \tanh (x) + \frac{\lambda_1 + \lambda_2}{2},
\end{equation}
where
\begin{displaymath}
x \equiv \frac{\textstyle \log (E/E_{\rm b})}{\textstyle \Lambda}.
\end{displaymath}
$x$ is defined so that $x = 0$ at a break energy $E_{\rm b}$, and also the 
break scale, $\Lambda$, determines the width of the transition region from 
$\lambda_1$ to $\lambda_2$.

To obtain the photon flux, $f(E)$, first integrate Equation \ref{eqn:tanh} with 
respect to $\log E$, and we get
\begin{eqnarray*}
\log f(E) &=& \int \left[ \frac{\lambda_2 - \lambda_1}{2} \tanh (x) 
      + \frac{\lambda_1 + \lambda_2}{2} \right] d(\log E) \\
&=& \frac{\lambda_2 - \lambda_1}{2} \ln (\cosh (x)) \frac{d(\log E)}{dx}
      + \frac{\lambda_1 + \lambda_2}{2} \log E + \mathbb{C}  \\
&=& \frac{\lambda_2 - \lambda_1}{2} 
      \ln \left[ \cosh \frac{\log (E/E_{\rm b})}{\Lambda} \right] \Lambda
      + \frac{\lambda_1 + \lambda_2}{2} \log E + \mathbb{C}, \\
\end{eqnarray*}
where $\mathbb{C}$ is the integration constant.

Therefore, the photon flux is
\begin{equation}\label{eqn:f0}
f(E) = 10^{\textstyle \frac{\lambda_2 - \lambda_1}{2} \Lambda 
      \ln \left[ \cosh \frac{\log (E/E_{\rm b})}{\Lambda} \right]}
      E^{\textstyle \frac{\lambda_1 + \lambda_2}{2}} \mathbb{C}'.
\end{equation}

A boundary condition is that $f(E) = A$ when $E = E_{\rm piv}$, where $A$ is an
amplitude of the photon flux determined at the pivot energy $E_{\rm piv}$, 
so we find
\begin{eqnarray*}
f(E = E_{\rm piv}) = 10^{\textstyle \frac{\lambda_2 - \lambda_1}{2} \Lambda 
      \ln \left[ \cosh \frac{\log (E_{\rm piv}/E_{\rm b})}{\Lambda} \right]}
      E_{\rm piv}^{\frac{\lambda_1 + \lambda_2}{2}} \mathbb{C}' = A  \\
\rightarrow \mathbb{C}' = A~10^{-\frac{\lambda_2 - \lambda_1}{2} \Lambda 
      \ln \left[ \cosh \frac{\log (E_{\rm piv}/E_{\rm b})}{\Lambda} \right]}
      E_{\rm piv}^{\textstyle -\frac{\lambda_1 + \lambda_2}{2}}.
\end{eqnarray*}

Then the Equation \ref{eqn:f0} becomes
\begin{equation}\label{eqn:f}
f(E) = A~10^{\textstyle \frac{\lambda_2 - \lambda_1}{2} \Lambda  \left[
      \ln \left( \cosh \frac{\log (E/E_{\rm b})}{\Lambda} \right)
      - \ln \left( \cosh \frac{\log (E_{\rm piv}/E_{\rm b})}{\Lambda} \right) \right]}
      \left( \frac{E}{E_{\rm piv}} \right)^{\textstyle \frac{\lambda_1 + \lambda_2}{2}} \\
\end{equation}

Rewriting this, we obtain the smoothly broken power law model:
\begin{equation}
f_{\rm SBPL}(E) = A \left(\frac{\textstyle E}{\textstyle E_{\rm piv}}\right)^b 
   10^{(a - a_{\rm piv})}, 
\end{equation}
where
\begin{eqnarray*}
a = m \Lambda \ln{\left(
   \frac{\textstyle e^{q} + e^{-q}}{\textstyle 2}\right)},
&a_{\rm piv} &= m \Lambda \ln{\left(
   \frac {\textstyle e^{q_{\rm piv}} + e^{-q_{\rm piv}}}{\textstyle 2}
   \right)}, \\
q = \frac {\textstyle \log {(E/E_{\rm b})}} {\textstyle \Lambda}, 
&q_{\rm piv} &= \frac {\textstyle \log {(E_{\rm piv}/E_{\rm b})}} 
   {\textstyle \Lambda}, \nonumber \\
m = \frac{\textstyle \lambda_2 - \lambda_1}{\textstyle 2}, 
\rm{ and }
&b &= \frac{\textstyle \lambda_1 + \lambda_2}{\textstyle 2}. 
\end{eqnarray*}

\section{SBPL Model $E_{\rm \lowercase{peak}}$}\label{ap:sbpl_ep}
The SBPL model is parameterized with $E_{\rm b}$, which is the energy at which 
the low-energy power law joins with the high-energy power law; this is clearly 
different from the $E_{\rm peak}$ parameters of the BAND and COMP models.
Therefore, $E_{\rm b}$ and $E_{\rm peak}$ are not directly comparable with each 
other.  
However, it is easy to derive ``$E_{\rm peak}$" (i.e., the peak in
$\nu F_{\nu}$ spectrum) for a spectrum fitted with SBPL, which,
in turn, can be compared to the $E_{\rm peak}$ of the other models directly.

\subsection{Finding $E_{\rm peak}$}
Starting from Equation \ref{sbpl_model}, the $\nu F_{\nu}$ flux can be 
written as
\begin{displaymath}
\nu F_{\nu} \equiv E^2 f(E) = \frac{A}{E_{\rm piv}^b} E^{(b + 2)} 
                                    10^{(a - a_{\rm piv})},
\end{displaymath}
where 
\begin{eqnarray*}
a = m \Lambda \ln{\left(
   \frac{\textstyle e^{q} + e^{-q}}{\textstyle 2}\right)},
&a_{\rm piv} &= m \Lambda \ln{\left(
   \frac {\textstyle e^{q_{\rm piv}} + e^{-q_{\rm piv}}}{\textstyle 2}
   \right)}, \\
q = \frac {\textstyle \log {(E/E_{\rm b})}} {\textstyle \Lambda}, 
&q_{\rm piv} &= \frac {\textstyle \log {(E_{\rm piv}/E_{\rm b})}} 
   {\textstyle \Lambda}, \nonumber \\
m = \frac{\textstyle \lambda_2 - \lambda_1}{\textstyle 2}, 
\textrm{ and }
&b &= \frac{\textstyle \lambda_1 + \lambda_2}{\textstyle 2}.
\end{eqnarray*}
To find the peak in the $\nu F_{\nu}$ spectrum, we need
\begin{displaymath}
\left.\frac{d}{dE} (\nu F_{\nu}) \right| _{E_{\rm peak}}= 0,
\end{displaymath}
and we find
\begin{equation}
\frac{d}{dE} (\nu F_{\nu}) = 
   \frac{A}{E_{\rm piv}^b} E^{(b + 1)}
   10^{(a -a_{\rm piv})} ( b + 2 + m~ \tanh(q)). \nonumber
\end{equation}

\noindent Set this equation to 0 at $E = E_{\rm peak}$, 
and we obtain
\begin{equation}
\tanh(q) = \frac{- (b + 2)}{m}. \nonumber
\end{equation}
But $q (E_{\rm peak}) 
   = \frac {\textstyle \log {(E_{\rm peak}/E_{\rm b})}} {\textstyle \Lambda}$
so the equation becomes
\begin{equation}
\tanh\left( \frac{\textstyle \log{(E_{\rm peak}/E_{\rm b})}}{\textstyle \Lambda}\right) 
      = \frac{- (b + 2)}{m}, \nonumber
\end{equation}
and solving this for $E_{\rm peak}$, we obtain
\begin{equation}\label{eqn:sbpl_epeak}
{\it E}_{\rm peak} = \
   E_{\rm b}~ 10 ^ {\left[{\textstyle \Lambda \tanh^{-1} }
   \left( \frac{\textstyle \lambda_{1} + \lambda_{2} + 4}
   {\textstyle \lambda_{1} - \lambda_{2}} \right)
   \right]}.
\end{equation}

We note here that since we must have $\left| \frac{\lambda_{1} + \lambda_{2} + 4}
{\lambda_{1} - \lambda_{2}} \right| < 1$,
this is only valid for $\lambda_{1} > -2$ and $\lambda_{2} < -2$ in order
for the $\nu F_{\nu}$ spectrum to have a peak within the spectral energy range.      

\subsection{Error Propagation ($\sigma_{E_{\rm peak}}^{}$)}
\label{ap:sbpl_err}
In order to calculate the uncertainties associated with the derived $E_{\rm peak}$ 
values, the errors associated with each parameter involved have to be propagated 
correctly.

Generally, the variance of a function, $y$, of $N$ parameters ($x_1$, $x_2$,...,
$x_N$) can be found by:
\begin{equation}\label{eqn:var}
{\rm Var}(y) = 
   \sum_{i=1}^N
   \left[\left(\frac{\partial y}{\partial x_i}\right)^2 {\rm Var}(x_i) \right]
   + 2 \frac{\partial y}{\partial x_1}
       \frac{\partial y}{\partial x_2} {\rm Cov}(x_1, x_2)
   + 2 \frac{\partial y}{\partial x_2}
       \frac{\partial y}{\partial x_3} {\rm Cov}(x_2, x_3)
   + ...,
\end{equation}
where Cov$(x_j, x_k) = R_{jk} \sigma_{x_j}\sigma_{x_k}$ and $R_{jk}$ is the 
correlation coefficient between $x_j$ and $x_k$.
Unfortunately, the covariance matrices for individual fits were not stored
in the parameter files in our database here, and therefore, we could
only estimate the uncertainties of the derived $E_{\rm peak}$ by neglecting
the cross terms in Equation \ref{eqn:var}.
The general effect of excluding the terms may be tested by examining 
cross-correlations between each parameter involved.

For the SBPL $E_{\rm peak}$, we have Equation \ref{eqn:sbpl_epeak} as a
function of three fitted parameters, $E_{\rm b}$, $\lambda_{1}$, and $\lambda_{2}$,
with the uncertainties associated with each parameter,
$\sigma_{E_{\rm b}}, \sigma_{\lambda_{1}}, \sigma_{\lambda_{2}}$.
First, to make the equation simpler, define
\begin{equation}\label{def}
s \equiv \Lambda \tanh^{-1} u \quad \rm{~and~} \quad
u \equiv \frac{\lambda_{1} + \lambda_{2} + 4}{\lambda_{1} - \lambda_{2}},
\end{equation}
and rewrite Equation \ref{eqn:sbpl_epeak} as
\begin{displaymath}
f(E_{\rm b}, \lambda_{1}, \lambda_{2}) = 
E_{\rm peak} = E_{\rm b}~ 10 ^ s.
\end{displaymath}

\noindent Then, the variance of $E_{\rm peak}$ is
\begin{equation}\label{eqn:err1}
\sigma_{E_{\rm peak}}^2 =
   \sigma_{E_{\rm b}}^2 \left(\frac{\partial f}{\partial E_{\rm b}}\right)^2
 + \sigma_{\lambda_{1}}^2 \left(\frac{\partial f}{\partial \lambda_{1}}\right)^2
 + \sigma_{\lambda_{2}}^2 \left(\frac{\partial f}{\partial \lambda_{2}}\right)^2,
\end{equation}
considering only the uncorrelated terms of Equation \ref{eqn:var}.

Now, the derivatives are
\begin{eqnarray*}
\frac{\partial f}{\partial E_{\rm b}} = 10^s \hspace{1in} \\
\frac{\partial f}{\partial \lambda_{1}} =
   E_{\rm b} 10^s ~\ln10~ \frac{\Lambda}{(1 + u)(\lambda_{1} - \lambda_{2})} \\
\frac{\partial f}{\partial \lambda_{2}} = 
   E_{\rm b} 10^s ~\ln10~ \frac{\Lambda}{(1 - u)(\lambda_{1} - \lambda_{2})}.
\end{eqnarray*}   

Therefore, Equation \ref{eqn:err1} becomes
\begin{displaymath}
\sigma_{E_{\rm peak}}^2 =
   \sigma_{E_{\rm b}}^2 (10^s)^2 +
    \left(E_{\rm b} 10^s ~\ln10~ \frac{\Lambda}{\lambda_{1} - \lambda_{2}}\right)^2
    \left(\frac{\sigma_{\lambda_{1}}^2}{(1+u)^2}
            + \frac{\sigma_{\lambda_{2}}^2}{(1-u)^2}\right),
\end{displaymath}
and finally, the $E_{\rm peak}$ uncertainty is
\begin{equation}\label{eqn:sbpl_err}
\sigma_{E_{\rm peak}}^{} =10^s \sqrt{\sigma_{E_{\rm b}}^2 + 
   \left(E_{\rm b} ~\ln10 ~\frac{\Lambda}{\lambda_{1} - \lambda_{2}}\right)^2
    \left(\frac{\sigma_{\lambda_{1}}^2}{(1+u)^2}
            + \frac{\sigma_{\lambda_{2}}^2}{(1-u)^2}\right)},
\end{equation}
where 
\begin{displaymath}
s = \Lambda \tanh^{-1} u \quad \rm{~and~} \quad
u = \frac{\lambda_{1} + \lambda_{2} + 4}{\lambda_{1} - \lambda_{2}}.
\end{displaymath}

The covariance terms that were excluded are
\begin{eqnarray}\label{eqn:sbpl_cov}
2 \left[ {\rm Cov}(E_{\rm b},\lambda_{1}) 
   \frac{\partial f}{\partial E_{\rm b}}\frac{\partial f}{\partial \lambda_{1}}
   + {\rm Cov}(E_{\rm b},\lambda_{2}) 
   \frac{\partial f}{\partial E_{\rm b}}\frac{\partial f}{\partial \lambda_{2}}
   + {\rm Cov}(\lambda_{1},\lambda_{2}) 
   \frac{\partial f}{\partial \lambda_{1}}\frac{\partial f}{\partial \lambda_{2}}   
\right] \\ \nonumber
= \frac{2 (10^{2x}) E_{\rm b} \ln10 \Lambda}{\lambda_{1}-\lambda_{2}}
   \left[
   \frac{R_{E_{\rm b},\lambda_{1}}\sigma_{E_{\rm b}}\sigma_{\lambda_{1}}}{1+u}
   + \frac{R_{E_{\rm b},\lambda_{2}} \sigma_{E_{\rm b}}\sigma_{\lambda_{2}}}
   {1-u}
   + R_{\lambda_{1}, \lambda_{2}} \sigma_{\lambda_{1}}\sigma_{\lambda_{2}}
   \frac{E_{\rm b} \ln10 \Lambda(2 + \lambda_{2})}
      {(1 - u^2)(\lambda_{1} - \lambda_{2})}
\right],
\end{eqnarray}
where $s$ and $u$ are defined above.
To understand the overall consequence of disregarding the covariance terms, 
we investigated the cross-correlations between the parameters (i.e., $E_{\rm b}$, 
$\lambda_{1}$, and $\lambda_{2}$, with $\Lambda =$ fixed).
This was done by using the results of 4-parameter SBPL fits to a set of simulated 
SBPL spectra with several different break scales described in \S \ref{sec:sbpl_sim}.
Although the correlation coefficients change slightly as a function of
fixed $\Lambda$ values, we found $E_{\rm b}-\lambda_{1}$ and 
$E_{\rm b}-\lambda_{2}$ are always strongly anti-correlated.
The corresponding average correlation coefficients were 
$\langle R_{E_{\rm b},\lambda_{1}}\rangle \sim -0.8$ and 
$\langle R_{E_{\rm b},\lambda_{2}}\rangle \sim -0.9$.
On the other hand, $\lambda_{1}-\lambda_{2}$ was always positively correlated with
$\langle R_{\lambda_{1},\lambda_{2}}\rangle \sim 0.5$.

Using the real set of parameters in our analysis, along with the
average correlation coefficients found from the simulation above, we found that
the correlated terms (Equation \ref{eqn:sbpl_cov}) are almost always negative,
as long as two of the correlation coefficients are negative.
Therefore, it is likely that the errors calculated using Equation 
\ref{eqn:err1} is overestimated for many spectra.

\section{BAND Model $E_{\rm \lowercase{b}}$}
\label{ap:band_eb}

Although a break energy is not parameterized in the BAND model, the model 
has a broken power-law shape, and therefore, we could find a spectral break energy 
equivalent to the $E_{\rm b}$ of SBPL for a direct comparison.
We stress again, as we did in \S\ref{sec:sbpl_model}, that this $E_{\rm b}$ 
is not the characteristic energy in Equation~\ref{band_model}; 
$E_{\rm c} = (\alpha - \beta) E_{\rm peak} / (2 + \alpha)$, which is often called 
the \textit{break energy} \citep[e.g., SP1;][]{sak04a} but is not the 
energy where the power law changes.  
Rather, the energy $E_{\rm c}$ corresponds to where the low-energy 
power law with an exponential cutoff ends and the pure high-energy power law 
starts; therefore, it is always greater than the $E_{\rm b}$ we are 
trying to find here.

\subsection{Finding $E_{\rm b}$}
Since $E_{\rm b} < E_{\rm c}$ always, we only use the $E < E_{\rm c}$ case in 
Equation \ref{band_model}.  
To find $E_{\rm b}$, we consider the change in slope of logarithmic tangential
lines from $\alpha$ to $\beta$, which is smooth and continuous.
We start by writing Equation \ref{band_model} in logarithmic scale;
\begin{equation}
   {\rm log}f(E) = {\rm log}A - 2 \alpha + \alpha {\rm log}E 
       - \frac{\textstyle (2 + \alpha) E}{\textstyle E_{\rm peak}} 
       {\rm log} e. \nonumber
\end{equation}
Then, the derivative with respect to log$E$ is
\begin{equation}\label{slope}
   \frac{d{\rm log}f(E)}{d{\rm log}E} = 
         \alpha - \frac{\textstyle (2 + \alpha) E}{\textstyle E_{\rm peak}}, 
\end{equation}
which is the equation for the slope as a function of $E$.

A break region can be defined as the energy range where the slope is between 
$\alpha$ and $\beta$.
By definition of the model, the slope is equal to $\alpha$ when $E = 0$ and 
$\beta$ when $E = E_{\rm c}$, then the break region is from 0 to 
$E_{\rm c}$ in keV.
However, since the fitted $\alpha$ value is the tangential slope at $E = 0$
and does not exactly represent the low-energy spectral shape, the actual slopes 
at energies $E > 0$ is always less than $\alpha$ due to the exponential term in 
the model.
In order to take this into account, we use ``effective" low-energy power law 
index, $\alpha_{\rm eff}$, described in Appendix \ref{ap:aleff}.
This is essentially the slope (Equation \ref{slope}) at a fixed energy of 25 keV.
Accordingly, we use the $\alpha_{\rm eff}$ in calculating the $E_{\rm b}$,
and the break region is from 25 keV to $E_{\rm c}$. 

The $E_{\rm b}$ then should be at the center of the break region,
namely, the energy at which the slope value differs from both 
$\alpha_{\rm eff}$ and $\beta$ by the same amount, 
$(\alpha_{\rm eff} - \beta)/2$.
In other words, the slope at $E_{\rm b}$ must equal 
$(\alpha_{\rm eff} + \beta)/2$.
Since the slope is a linear function, this is just the center energy between
25 keV and $E_{\rm c}$, and is
\begin{equation}\label{eqn:band_eb}
   E_{\rm b} = \frac{E_{\rm c} - 25}{2} + 25
       = \frac{\textstyle \alpha - \beta}{\textstyle 2} 
         \frac{\textstyle E_{\rm peak}}{\textstyle 2 + \alpha} + 12.5. 
\end{equation}

\subsection{Error Propagation ($\sigma_{E_{\rm b}}^{}$)}\label{sec:band_err}
The general error propagation methodology was discussed in \S \ref{ap:sbpl_err}.
For the same reason, we do not include the cross terms of Equation \ref{eqn:var} 
in the error calculation here.
The BAND $E_{\rm b}$ (Equation \ref{eqn:band_eb}) is a
function of three fitted parameters, $E_{\rm peak}$, $\alpha$, and $\beta$, with
uncertainties, $\sigma_{E_{\rm peak}}$, $\sigma_{\alpha}$, and $\sigma_{\beta}$,
respectively.  The $E_{\rm peak}$ is noted as $E_{\rm p}$ in the following 
equations for simplicity.
Let $f(E_{\rm p}, \alpha, \beta) = E_{\rm b}$, then the variance of $E_{\rm b}$ 
can be written as
\begin{equation}\label{eqn:band_var}
\sigma_{E_{\rm b}}^2 =
   \sigma_{E_{\rm p}}^2 \left(\frac{\partial f}{\partial E_{\rm p}}\right)^2
 + \sigma_{\alpha}^2 \left(\frac{\partial f}{\partial \alpha}\right)^2
 + \sigma_{\beta}^2 \left(\frac{\partial f}{\partial \beta}\right)^2,
\end{equation}
without the covariance terms. The derivatives are
\begin{eqnarray*}
\frac{\partial f}{\partial E_{\rm p}}&
   = \frac{\alpha - \beta}{2 (2 + \alpha)}  \\ \\
\frac{\partial f}{\partial \alpha}&
   = \frac{E_{\rm p} (2 + \beta)}{2 (2 + \alpha)^2}    \\  \\
\frac{\partial f}{\partial \beta}&
   =  - \frac{E_{\rm p}}{2 (2 + \alpha)}.
\end{eqnarray*}

\noindent Therefore, Equation \ref{eqn:band_var} becomes
\begin{equation}
\sigma_{E_{\rm b}}^2 =
   \sigma_{E_{\rm p}}^2 \left(
   \frac{\alpha - \beta}{2 (2 + \alpha)} \right)^2
 + \sigma_{\alpha}^2 \left(
   \frac{E_{\rm p} (2 + \beta)}{2 (2 + \alpha)^2} \right)^2
 + \sigma_{\beta}^2 \left(
    - \frac{E_{\rm p}}{2 (2 + \alpha)} \right)^2 , \nonumber
\end{equation}
and we find the uncertainty
\begin{equation}
\sigma_{E_{\rm b}} = \frac{1}{2 (2 + \alpha)} \sqrt{
   (\alpha - \beta)^2 \sigma_{E_{\rm p}}^2 
   + E_{\rm p}^2 \left( \left(\frac{2 + \beta}{2 + \alpha}\right)^2
   \sigma_{\alpha}^2 + \sigma_{\beta}^2 \right) }.
\end{equation}

Similar to the SBPL $E_{\rm peak}$ case, the correlations between the parameters 
involved in Equation \ref{eqn:band_eb} (i.e., $E_{\rm peak}$, $\alpha$, and 
$\beta$) were investigated using the fits to the same set of the simulated SBPL 
spectra that was used to explore the SBPL parameter correlations
in \S \ref{ap:sbpl_err} for consistency, as well as the simulated BAND
spectra.
We found that both $E_{\rm peak}-\alpha$ and $E_{\rm peak}-\beta$ are
always highly anti-correlated regardless of the $E_{\rm peak}$ values or
the SBPL break scale, with correlation coefficients of 
$\langle R_{E_{\rm p},\alpha}\rangle \sim -0.9$ and 
$\langle R_{E_{\rm p},\beta}\rangle \sim -0.7$.  
The correlation between $\alpha$ and $\beta$ tends to be weaker, though always 
positively correlated, with a correlation coefficient of 
$\langle R_{\alpha,\beta}\rangle \sim 0.5$.

The cross terms of $\sigma_{E_{\rm b}}^2$ are
\begin{eqnarray}\label{eqn:band_cov}
2 \left[ {\rm Cov}(E_{\rm p},\alpha) 
   \frac{\partial f}{\partial E_{\rm p}}\frac{\partial f}{\partial \alpha}
   + {\rm Cov}(E_{\rm p},\beta) 
   \frac{\partial f}{\partial E_{\rm p}}\frac{\partial f}{\partial \beta}
   + {\rm Cov}(\alpha,\beta) 
   \frac{\partial f}{\partial \alpha}\frac{\partial f}{\partial \beta}   
\right] \\ \nonumber
= \frac{2 E_{\rm p}}{4 (2 + \alpha)^2}
   \left[R_{E_{\rm p},\alpha} \sigma_{E_{\rm p}}\sigma_{\alpha}
   \frac{(2 + \beta) (\alpha - \beta)}{2 + \alpha}
   + R_{E_{\rm p},\beta} \sigma_{E_{\rm p}}\sigma_{\beta} (\alpha - \beta)
   + R_{\alpha, \beta} \sigma_{\alpha}\sigma_{\beta}
   \frac{E_{\rm p}(2 + \beta)}{2 + \alpha}
\right].
\end{eqnarray}
Using the actual spectral parameters in our sample, with the average
correlation coefficients found above, most of the time the cross term was found 
to be negative, as long as the parameters are more strongly anti-correlated.
Therefore, we believe the uncertainties found in Equation \ref{eqn:band_var}
without the covariance terms tend to be overestimated.

\section{Effective $\alpha$}
\label{ap:aleff}

\begin{figure}
\epsscale{0.6}
\plotone{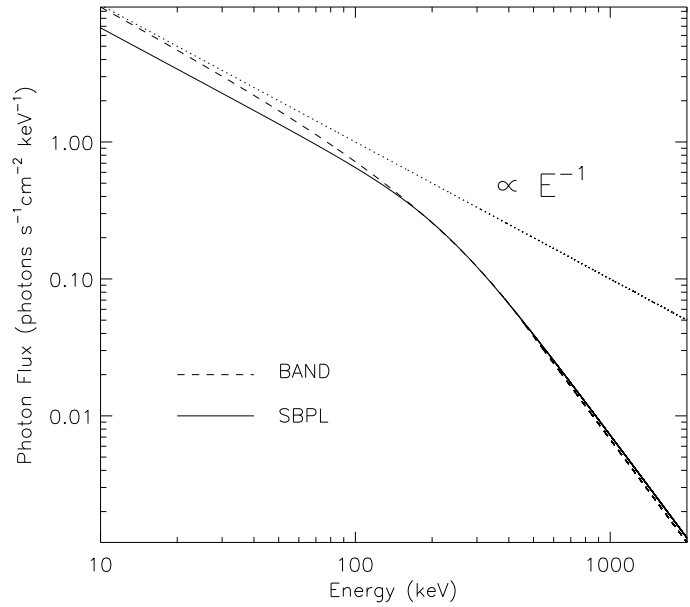}
\caption{The BAND model (dashed curve) and the SBPL model (solid curve) with
the same low-energy index values of $\alpha = \lambda_1 = -1$. 
The other parameters are also kept the same ($E_{\rm peak}$ = 300 keV and 
$\beta = \lambda_2 = -2.5$).  A dotted line is a power law with index of --1.
The difference in the low-energy behavior is evident.}
\label{fig:sbpl_band_pwrl}
\end{figure}

The discrepancy between the BAND (or COMP) $\alpha$ and the SBPL $\lambda_1$
can be especially severe when the $e$-folding energy, $E_{\rm 0} = E_{\rm peak} 
/ (2 + \alpha)$, is close to the lower energy threshold and the actual 
low-energy power law component of the BAND or COMP is assumed to lie far below 
the lower energy bound of the data.
As an example, we plot in Figure \ref{fig:sbpl_band_pwrl}, the BAND and the 
SBPL model with the same $E_{\rm peak}$, high-energy index, and $\alpha = 
\lambda_1 = -1$.
\placefigure{fig:sbpl_band_pwrl}
The difference between the --1 power law and the actual BAND model behavior is
obvious, whereas the SBPL low-energy component does follow the --1 power law.

\begin{figure} 
\epsscale{0.6}
\plotone{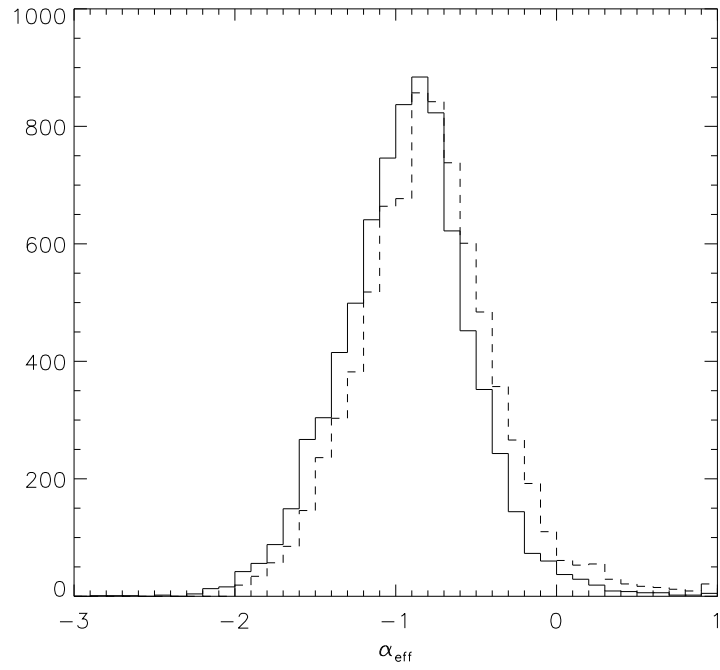}
\caption{The comparison of the effective $\alpha$ distribution (solid line) 
and the fitted $\alpha$ distribution (dashed line) of the BAND model fits to
the 7861 time-resolved spectra.}
\label{fig:band_aleff}
\end{figure}

To resolve this issue and to more accurately represent the low-energy behavior
of the LAD data, we calculate the ``effective" $\alpha$ of the BAND and COMP 
models, introduced by \citet{pre98b}.
The effective $\alpha$ is simply the spectral slope in log$f - $log$E$ 
(Equations \ref{band_model} and \ref{comp_model}) determined at some fiducial 
energy, chosen to be 25 keV, and is given by
\begin{equation} \label{eqn:aleff}
\alpha_{\rm eff} = \alpha - \frac{25 ~{\rm keV}}{E_{\rm peak}} (2 + \alpha)
 = \alpha - \frac{25 ~{\rm keV}}{E_{0}}.
\end{equation}
By defining $\alpha_{\rm eff}$, we assume that at 25 keV the low-energy power 
law has already been reached.
The energy of 25 keV was chosen because it is the lower energy bound of LADs
\citep{pre98b} and also the tangential slope at this energy seems to represent 
the data better than at other energies we tested, such as 10 keV, 50~keV, or 
certain fraction of $E_{\rm peak}$, $E_{\rm b}$, or $E_{0}$.
The correction term is larger for lower $E_0$ values.
In Figure \ref{fig:band_aleff}, we compare the BAND $\alpha$ and 
$\alpha_{\rm eff}$ of the time-resolved spectra in our sample.
As seen in Figure \ref{fig:band_aleff}, $\alpha_{\rm eff}$ correction lowers the
index values, making them more consistent with $\lambda_1$ of SBPL (Figure
\ref{fig:bpar_aleff}, bottom right panel).
\placefigure{fig:bpar_aleff}

\subsection{Error Propagation ($\sigma_{\alpha_{\rm eff}}^{}$)}
The variance of $\alpha_{\rm eff}$, without the correlated terms, can be written as
\begin{equation}
\sigma_{\alpha_{\rm eff}}^2 =
   \sigma_{E_{\rm peak}}^2 \left(\frac{\partial f}{\partial E_{\rm peak}}\right)^2
 + \sigma_{\alpha}^2 \left(\frac{\partial f}{\partial \alpha}\right)^2,
\end{equation}
where
\begin{eqnarray*}
\frac{\partial f}{\partial E_{\rm peak}}
   = \frac{25 (2 + \alpha)}{{E_{\rm peak}}^2}  \\
\frac{\partial f}{\partial \alpha}
   = 1 - \frac{25}{E_{\rm peak}}.
\end{eqnarray*}

Therefore, the $\alpha_{\rm eff}$ uncertainty is
\begin{equation}\label{eqn:aleff_var}
\sigma_{\alpha_{\rm eff}} = \sqrt{
   \sigma_{E_{\rm peak}}^2 \left(\frac{25 (2 + \alpha)}{{E_{\rm peak}}^2}\right)^2
 + \sigma_{\alpha}^2 \left(1 - \frac{25}{E_{\rm peak}}\right)^2}.
\end{equation}

The correlation term neglected above is
\begin{equation}\label{eqn:aleff_cov}
2 \left[R_{E_{\rm peak},\alpha}\sigma_{E_{\rm peak}}\sigma_{\alpha} 
   \frac{\partial f}{\partial E_{\rm peak}}\frac{\partial f}{\partial \alpha}
\right]
= 2\left[R_{E_{\rm peak},\alpha} \sigma_{E_{\rm peak}}\sigma_{\alpha}
   \frac{25 (2 + \alpha)}{E_{\rm peak}^2}
   \left(1 - \frac{25}{E_{\rm peak}}\right)
\right],
\end{equation}
where $R_{E_{\rm peak},\alpha}$ was found to be negative (see \S 
\ref{sec:band_err}).  Consequently, the error estimated from Equation 
\ref{eqn:aleff_var} may also be larger than the actual values.

\end{appendix}


\clearpage
\begin{table}
\caption{Basic properties of 350 GRBs included in the catalog.}
\label{tab:grblist}
\begin{center}
\begin{scriptsize}
\begin{tabular}{rcrccrrcrrcr}
\hline 
\hline \\[-2ex]
  GRB~~ & BATSE & Triger Time\tablenotemark{b} & Data & LAD 
   & \multicolumn{2}{c}{Time Interval\tablenotemark{c}}   &  
   & \multicolumn{2}{c}{Energy Interval}  &
   & \# of~~ \\
\cline{6-7} \cline{9-10}
  ~~Name\tablenotemark{a} & Trig \#  
   & (UT,s)~~ & Type & \# & Start (s) & End (s) & & Start (keV) & End (MeV)  &
   & Spectra  \\
   (1)~~~ & (2) & (3)~~~~~ & (4) & (5) & (6)~~~~ & (7)~~~ && (8)~~~~~~ 
   & (9)~~~~~~ && (10)~~~ \\\\[-2ex]
\hline \\[-2ex] 
 910421 & ~105 & 33243~~~ & HERB &       7 &    0.064~~ &   5.184~ && 27.109~~~~ & 1948.42~~~~ &&  12~~~~ \\
 910425 & ~109 & ~2265~~~ & MER  &     0,4 & --14.336~~ &  86.815~ && 32.506~~~~ & 1813.54~~~~ &&  26~~~~ \\
 910430 & ~130 & 61719~~~ & CONT &       6 & --17.408~~ &  64.512~ && 36.383~~~~ & 1799.34~~~~ &&  22~~~~ \\
 910503 & ~143 & 25452~~~ & HERB &       6 &    0.000~~ &  18.560~ && 32.275~~~~ & 1900.93~~~~ &&  46~~~~ \\
 910522 & ~219 & 43929~~~ & MER  &   4,5,6 &  105.947~~ & 133.147~ && 33.638~~~~ & 1821.76~~~~ &&  13~~~~ \\
 910525 & ~226 & 69987~~~ & CONT &       5 &  --3.072~~ & 171.010~ && 32.678~~~~ & 1853.06~~~~ &&  19~~~~ \\
 910601 & ~249 & 69734~~~ & MER  & 0,2,4,6 &    0.107~~ &  45.019~ && 34.665~~~~ & 1814.08~~~~ &&  68~~~~ \\
 910602 & ~257 & 82501~~~ & HERB &       0 &    0.000~~ &  17.664~ && 33.151~~~~ & 2039.36~~~~ &&   7~~~~ \\
W910609 & ~298 & ~2907~~~ & CONT &       4 &  --1.408~~ &   0.640~ && 31.936~~~~ & 1817.69~~~~ &&   1~~~~ \\
\hline \\[-2ex]
\multicolumn{12}{l}{\hspace{2ex} $^{\rm a}$~Prefix C -- Calibration burst; 
   W -- Weak burst with a single spectrum.} \\
\multicolumn{12}{l}{\hspace{2ex} $^{\rm b}$~In seconds of day.} \\
\multicolumn{12}{l}{\hspace{2ex} $^{\rm c}$~Time since trigger.} \\
\end{tabular}
\tablecomments{The complete version of this table is in the electronic edition 
of the Journal.  The printed edition contains only a sample.}
\end{scriptsize}
\end{center}
\end{table}

\begin{table}
\caption{BATSE LAD burst data types used in this work.}
\label{tab:datatypes}
\begin{center}
\begin{tabular}{lcccccc}
\hline 
\hline \\[-2ex]
Data & Number of & Number of & Time & Detector & Time \\
Type & Energy Chan & Spectra & Resolution (s) & Subset & Coverage \\ \\[-2ex]
\hline \\[-2ex] 
HERB & 128 & 128 & 0.128\tablenotemark{a} & DSELH\tablenotemark{b} & $\lesssim 500$ s \\
HER & 128 & --- & $\sim 300$ & All & Background \\
MER & 16 & 4096 & 0.016 \& 0.064\tablenotemark{c} & DSELB\tablenotemark{d} & 163.84 s \\
CONT & 16 & --- & 2.048 & All & Continuous\tablenotemark{e} \\
\hline \\[-2ex]
\end{tabular}
\begin{minipage}[c]{6.8in}
\begin{footnotesize}
$^{\rm a}$~Minimum for time-to-spill, increases by 64 ms increments. \\
$^{\rm b}$~DSELH: 4 detectors with highest count rates, determined at trigger. \\
$^{\rm c}$~The change to 64 ms resolution is after the first 32.768 s. \\
$^{\rm d}$~DSELB: The 2--4 detectors with highest count rates, as determined 
         at the time of the trigger (MER is summed over these detectors). \\
$^{\rm e}$~Can serve as background data. \\
\end{footnotesize}
\end{minipage}
\end{center}
\end{table}

\begin{table}
\begin{scriptsize}
\caption{Example fit results to simulated BAND spectra (upper two tables) and
to simulated COMP spectra (lower two tables).
The parameters are median values and the standard deviations are shown in 
parentheses.}
\label{tab:comp_band_res}
\begin{center}
\begin{tabular}{lccccccccc}
\hline 
\hline \\[-2ex]
\multicolumn{10}{c}{\bf{Simulated BAND Parameters: 
            {\boldmath$E_{\rm peak}$ = 559.5 keV, 
            $\alpha$ = --0.52, $\beta$ = --2.24}}} \\[2pt]
 \cline{2-9}\\[-5pt]
 & \multicolumn{4}{c}{BAND Fit Parameters} &
            & \multicolumn{3}{c}{COMP Fit Parameters} & \\[2pt]
 \cline{2-5}\cline{7-9}\\[-5pt]
$\langle {\rm S/N} \rangle$ & $\langle E_{\rm peak} \rangle$ 
            & $\langle \alpha \rangle$ & $\langle \beta \rangle$ 
            & $\langle \chi^2 \rangle/{\rm dof}$ && $\langle E_{\rm peak} \rangle$
            & $\langle \alpha \rangle$ & $\langle \chi^2 \rangle/{\rm dof}$
            & $\Delta \chi^2$ \\[2pt]
\hline \\[-2ex] 
285.1 & 559 (14) & --0.52 (0.02) & --2.23 (0.05) & 111.2/112 &
            & 746 (14) & --0.66 (0.01) & 252.3/113 & 141.1 \\ 
75.9  & 585 (60) & --0.53 (0.07) & --2.21 (0.32) & 109.4/112 &
            & 740 (55) & --0.65 (0.05) & 118.4/113 & 9.0 \\
12.3  & 549 (427) & --0.45 (0.72) & --1.77 (1.78) & 108.6/112 &
            & 1083 (1E4) & --0.74 (0.29) & 109.2/113 & 0.6 \\
\hline 
\end{tabular}

\vspace{12pt}
\begin{tabular}{lccccccccc}
\hline 
\hline \\[-2ex]
\multicolumn{10}{c}{\bf{Simulated BAND Parameters: 
            {\boldmath$E_{\rm peak}$ = 493.5 keV, 
            $\alpha$ = --0.97, $\beta$ = --2.36}}} \\[2pt]
 \cline{2-9}\\[-5pt]
 & \multicolumn{4}{c}{BAND Fit Parameters} &
            & \multicolumn{3}{c}{COMP Fit Parameters} & \\[2pt]
 \cline{2-5}\cline{7-9}\\[-5pt]
$\langle {\rm S/N} \rangle$ & $\langle E_{\rm peak} \rangle$ 
            & $\langle \alpha \rangle$ & $\langle \beta \rangle$ 
            & $\langle \chi^2 \rangle/{\rm dof}$ && $\langle E_{\rm peak} \rangle$
            & $\langle \alpha \rangle$ & $\langle \chi^2 \rangle/{\rm dof}$ 
            & $\Delta \chi^2$ \\[2pt]
\hline \\[-2ex] 
143.6 & 494 (38) & --0.97 (0.04) & --2.28 (0.34) & 110.1/111 &
            & 593 (36) & --1.03 (0.03) & 122.1/112 & 12.0 \\ 
27.5  & 467 (288) & --0.91 (0.25) & --1.97 (1.56) & 105.3/111&
            & 679 (386) & --1.03 (0.13) & 106.2/112 & 0.9 \\
3.3  & 299 (517) & --0.53 (2.05) & --1.46 (1.97) & 111.4/111 &
            & 1388 (5E6) & --0.85 (3.19) & 111.0/112 & --0.4 \\
\hline 
\end{tabular}

\vspace{24pt}
\begin{tabular}{lccccccccc}
\hline 
\hline \\[-2ex]
\multicolumn{10}{c}{\bf{Simulated COMP Parameters: 
            {\boldmath $E_{\rm peak}$ = 760 keV, 
            $\alpha$ = --0.70}}} \\[2pt]
 \cline{2-9}\\[-5pt]
 & \multicolumn{5}{c}{BAND Fit Parameters} &
            & \multicolumn{3}{c}{COMP Fit Parameters} \\[2pt]
 \cline{2-6}\cline{8-10}\\[-5pt]
$\langle {\rm S/N} \rangle$ & Failed\tablenotemark{a}
            & $\langle E_{\rm peak} \rangle$ 
            & $\langle \alpha \rangle$ & $\langle \beta \rangle$ 
            & $\langle \chi^2 \rangle/{\rm dof}$ && $\langle E_{\rm peak} \rangle$
            & $\langle \alpha \rangle$ 
            & $\langle \chi^2 \rangle/{\rm dof}$ \\[2pt]
\hline \\[-2ex] 
76.9 & 32 & 753 (73) & --0.66 (0.05) & --2.84 (1.89) & 113.8/112 &
          & 797 (63) & --0.69 (0.05) & 114.5/113 \\ 
286.3 & 26 & 747 (15) & --0.67 (0.01) & --3.65 (1.18) & 110.8/112 &
          & 760 (12) & --0.68 (0.01) & 112.0/113 \\ 
533.9 & 36 & 752 (~8) & --0.67 (0.01) & --3.99 (1.02) & 114.1/112 &
          & 759 (~6) & --0.68 (0.01) & 114.2/113  \\ 
\hline
\end{tabular}

\vspace{12pt}
\begin{tabular}{lccccccccc}
\hline 
\hline \\[-2ex]
\multicolumn{10}{c}{\bf{Simulated COMP Parameters: 
            {\boldmath $E_{\rm peak}$ = 300 keV, 
            $\alpha$ = --1.20}}} \\[2pt]
\cline{2-9}\\[-5pt]
 & \multicolumn{5}{c}{BAND Fit Parameters} &
            & \multicolumn{3}{c}{COMP Fit Parameters} \\[2pt]
 \cline{2-6}\cline{8-10}\\[-5pt]
$\langle {\rm S/N} \rangle$ & Failed\tablenotemark{a}
            & $\langle E_{\rm peak} \rangle$ 
            & $\langle \alpha \rangle$ & $\langle \beta \rangle$ 
            & $\langle \chi^2 \rangle/{\rm dof}$ && $\langle E_{\rm peak} \rangle$
            & $\langle \alpha \rangle$ 
            & $\langle \chi^2 \rangle/{\rm dof}$ \\[2pt]
\hline \\[-2ex] 
87.5 & 11 & 305 (26) & --1.20 (0.05) & --3.18 (2.02) & 112.8/112 &
           & 322 (25) & --1.23 (0.05) & 113.4/113 \\ 
133.5 & ~4 & 300 (15) & --1.19 (0.03) & --3.29 (1.55) & 111.0/112 &
           & 310 (12) & --1.21 (0.03) & 111.3/113 \\ 
456.3 & ~8 & 299 (~4) & --1.20 (0.01) & --4.51 (1.79) & 109.3/112 &
           & 301 (~3) & --1.20 (0.01) & 109.8/113 \\ 
\hline  \\[-2ex]
\multicolumn{10}{l}{\hspace{2ex}
$^{\rm a}$~Number of failed fits out of 100 trials,
due to poor parameter constraints.} \\
\end{tabular}
\end{center}
\end{scriptsize}
\end{table}

\begin{table}
\begin{scriptsize}
\tabcolsep=3pt
\caption{Results of BAND and COMP fits to simulated SBPL spectra (S/N $\sim$ 
100) with various $\Lambda$.
The simulated Parameters are $A$ = 0.05, $E_{\rm b}$ = 300 ${\rm keV}$, 
$\lambda_1$ = --1.0, and $\lambda_2$ = --2.5.
``$E_{\rm peak}$" is the calculated $\nu F_{\nu}$ peak energy in the SBPL 
spectra.  The parameters are median values and the standard deviations are 
shown in parentheses.}
\label{tab:sbpl_band_res}
\begin{center}
\begin{tabular}{cccccccccccc}
\hline 
\hline \\[-2ex]
\multicolumn{2}{c}{Simulated} &\multicolumn{5}{c}{BAND Fit Parameters} &
&\multicolumn{4}{c}{COMP Fit Parameters}  \\
 \cline{3-7} \cline{9-12}\\[-5pt]
$\Lambda$ & ``$E_{\rm peak}$"\tablenotemark{a} 
            & $\langle A \rangle\tablenotemark{b} \times 10^3$
            & $\langle E_{\rm peak} \rangle$\tablenotemark{a}
            & $\langle \alpha \rangle$ & $\langle \beta \rangle$ 
            & $\langle \chi^2 \rangle/{\rm dof}$ &
            & $\langle A \rangle\tablenotemark{b} \times 10^3$
            & $\langle E_{\rm peak} \rangle$\tablenotemark{a} 
            & $\langle \alpha \rangle$ 
            & $\langle \chi^2 \rangle/{\rm dof}$ \\[2pt]
\hline \\[-2ex] 
0.01 & 302 & 70 (0.3) & 457 ( 4) & --0.68 (0.01) & --2.93 (0.06) & 962.3/112 && 68 (0.3) & 483 ( 3) & --0.71 (0.01) & 1065.6/113 \\ 
0.10 & 325 & 70 (0.4) & 448 ( 4) & --0.69 (0.01) & --2.85 (0.04) & 608.4/112 && 67 (0.3) & 479 ( 3) & --0.73 (0.01) & 733.2/113 \\ 
0.20 & 352 & 69 (0.4) & 434 ( 4) & --0.74 (0.01) & --2.64 (0.04) & 295.1/112 && 66 (0.3) & 477 ( 4) & --0.79 (0.01) & 481.6/113 \\ 
0.30 & 381 & 67 (0.5) & 419 ( 5) & --0.80 (0.01) & --2.48 (0.04) & 152.4/112 && 63 (0.3) & 479 ( 4) & --0.87 (0.01) & 377.8/113 \\ 
0.40 & 413 & 66 (0.6) & 406 ( 6) & --0.86 (0.01) & --2.33 (0.03) & 119.9/112 && 61 (0.3) & 492 ( 5) & --0.95 (0.01) & 426.3/113 \\ 
0.50 & 447 & 65 (0.5) & 402 ( 7) & --0.94 (0.01) & --2.25 (0.02) & 120.7/112 && 59 (0.3) & 504 ( 5) & --1.03 (0.01) & 408.4/113 \\ 
0.60 & 484 & 64 (0.7) & 396 ( 8) & --1.00 (0.01) & --2.18 (0.02) & 127.8/112 && 58 (0.2) & 519 ( 6) & --1.10 (0.01) & 454.0/113 \\ 
0.70 & 524 & 63 (0.5) & 396 ( 7) & --1.06 (0.01) & --2.13 (0.02) & 140.2/112 && 57 (0.2) & 540 ( 7) & --1.16 (0.01) & 432.9/113 \\ 
0.80 & 568 & 62 (0.6) & 394 ( 9) & --1.11 (0.01) & --2.07 (0.02) & 141.1/112 && 56 (0.2) & 567 ( 8) & --1.22 (0.01) & 446.3/113 \\ 
0.90 & 615 & 62 (0.7) & 397 (12) & --1.15 (0.01) & --2.04 (0.02) & 142.1/112 && 55 (0.2) & 591 ( 9) & --1.27 (0.01) & 432.6/113 \\ 
1.00 & 666 & 61 (0.6) & 404 (12) & --1.20 (0.01) & --2.02 (0.02) & 135.3/112 && 54 (0.2) & 613 (10) & --1.30 (0.01) & 400.9/113 \\ 
\hline \\[-2ex]
\multicolumn{12}{l}{\hspace{2ex}
$^{\rm a}$~in keV} \\
\multicolumn{12}{l}{\hspace{2ex}
$^{\rm b}$~in ph s$^{-1}$ cm$^{-2}$ keV$^{-1}$}
\end{tabular}
\end{center}
\end{scriptsize}
\end{table}

\begin{table}
\caption{The percentage of fits resulted in $\chi^2$ with goodness of fit
of 3$\sigma$ (99.7\%) or better out of the total number of 350 time-integrated
spectra and 8459 time-resolved spectra.}
\label{tab:goodfit}
\begin{center}
\begin{tabular}{ccccc}
\hline 
\hline \\[-2ex]
                &  PWRL  &  COMP  &  BAND  &  SBPL  \\[1ex]
\hline \\[-2ex]
Time-Integrated & 11.1\% & 68.6\% & 84.6\% & 88.6\% \\[1ex]
Time-Resolved   & 23.3\% & 95.7\% & 92.3\% & 99.2\% \\[1ex]
\hline 
\end{tabular}
\end{center}
\end{table}

\begin{table}
\caption{Free parameters in each model.}
\label{tab:par}
\begin{center}
\begin{small}
\begin{tabular}{ccccccc}
\hline 
\hline
      &      & Low-E & High-E & Peak & Break 
      & Break \\[-1ex]
Model & Amplitude & Index & Index & Energy & Energy 
      & Scale \\
\hline \\[-2ex] 
PWRL & $A$ & $\lambda$  &  ---      &   ---        & --- & --- \\
COMP & $A$ & $\alpha$   &  ---      & $E_{\rm peak}$ & --- & --- \\
BAND & $A$ & $\alpha$   & $\beta$   & $E_{\rm peak}$ & --- & --- \\
SBPL & $A$ & $\lambda_1$&$\lambda_2$&   ---        & $E_{\rm b}$ & $\Lambda$ \\
\hline 
\end{tabular}
\end{small}
\end{center}
\end{table}

\begin{table}
\caption{The median {\it good} parameter values and the dispersion (quartile) 
of the distributions determined by fitting all
spectra, both time-integrated and time-resolved, separately with four models.}
\label{tab:mode_all}
\begin{center}
\begin{tabular}{crrrrr}
\hline \hline \\[-2ex]
&\multicolumn{5}{c}{Time-Integrated Parameters} \\[2pt]
\cline{2-6} \\[-2ex]
   & Low Index & High Index & $E_{\rm peak}$~ & $E_{\rm b}$~~~ & $\Delta$S~~~~~\\
   &           &            &     (keV)~      &     (keV)~ &  \\
\hline \\[-2ex]
PWRL&$-1.64~^{+0.11}_{-0.16}$&$-1.64~^{+0.11}_{-0.16}$&---~~~~~&---~~~~~&---~~~~~ \\[5pt]
COMP&$-1.18~^{+0.24}_{-0.19}$&---~~~~~&$321~^{+202}_{-105}$&---~~~~~&---~~~~~ \\[5pt]
BAND&$-1.08~^{+0.23}_{-0.16}$&$-2.33~^{+0.22}_{-0.32}$&$262~^{+119}_{-~80}$&$195~^{+~91}_{-~64}$&$1.44~^{+0.36}_{-0.28}$ \\[5pt]
SBPL&$-1.20~^{+0.27}_{-0.18}$&$-2.43~^{+0.30}_{-0.38}$&$234~^{+101}_{-~61}$&$218~^{+109}_{-~61}$&$1.42~^{+0.56}_{-0.37}$\\[2pt]
\hline \\[6pt]
\hline \hline \\[-2ex]
&\multicolumn{5}{c}{Time-Resolved Parameters} \\[2pt]
\cline{2-6} \\[-2ex]
   & Low Index & High Index & $E_{\rm peak}$~ & $E_{\rm b}$~~~ & $\Delta$S~~~~~\\
   &           &            &     (keV)~      &     (keV)~ &  \\
\hline \\[-2ex]
PWRL&$-1.74~^{+0.19}_{-0.19}$&$-1.74~^{+0.19}_{-0.19}$&---~~~~~&---~~~~~&---~~~~~ \\[5pt]
COMP&$-1.04~^{+0.24}_{-0.27}$&---~~~~~&$364~^{+204}_{-138}$&---~~~~~&---~~~~~ \\[5pt]
BAND&$-0.90~^{+0.23}_{-0.26}$&$-2.33~^{+0.26}_{-0.32}$&$268~^{+127}_{-~98}$&$190~^{+~94}_{-~60}$&$1.57~^{+0.48}_{-0.39}$ \\[5pt]
SBPL&$-1.05~^{+0.25}_{-0.27}$&$-2.42~^{+0.30}_{-0.40}$&$235~^{+117}_{-~79}$&$225~^{+102}_{-~69}$&$1.51~^{+0.48}_{-0.38}$\\[2pt]
\hline 
\end{tabular}
\end{center}
\end{table}

\clearpage
\begin{table}
\caption{Model comparison summary. The best-fit models were first determined
         statistically by $\chi^2$ probabilities (Column 1).
         In the case that several models resulted in comparable good fits, 
         Column 1 lists these by their initial letters, separated by forward 
         slashes. 
         In a second step, the model with the more constrained parameters 
         was designated as the best-fit (BEST) model. 
         Columns 3 - 6 and 8 - 11 lists the number of spectra for each BEST 
         model.
         \label{tab:models}}
\begin{center}
\begin{scriptsize}
\begin{tabular}{lrrrrrrrrrrr}
\hline 
\hline \\[-2ex]
Best-Fit & \multicolumn{5}{c}{Time Integrated} & & \multicolumn{5}{c}{Time Resolved} \\
 \cline{2-6} \cline{8-12}\\[-2ex]
~Model   &Total &{\bf PWRL}&{\bf COMP}&{\bf BAND} &{\bf SBPL} & 
         &Total &{\bf PWRL}&{\bf COMP}&{\bf BAND} &{\bf SBPL} \\
   ~~~(1) & (2)~ & (3)~~~ & (4)~~~ & (5)~~~ & (6)~~ && (7)~ & (8)~~~
   & (9)~~~ & (10)~~~ & (11)~~ \\
\hline \\[-2ex]
PWRL     &   8 &  8 & -- &  -- &  -- & &  171 & 171 &  --  &  --  &  --  \\ 
COMP     &  45 & -- & 45 &  -- &  -- & & 1811 &  -- & 1811 &  --  &  --  \\ 
BAND     &  48 & -- & -- &  48 &  -- & &  823 &  -- &  --  &  823 &  --  \\ 
SBPL     &  50 & -- & -- &  -- &  50 & &  101 &  -- &  --  &  --  &  101 \\ 
P/B/C/S  &   2 &  0 &  0 &   0 &   2 & &   24 &   2 &    3 &    0 &   19 \\ 
C/B/S    &  54 & -- &  2 &   7 &  45 & & 1535 &  -- &   88 &  192 & 1255 \\ 
P/C/B    &   4 &  0 &  4 &   0 &  -- & &  149 &  37 &  104 &    8 &  --  \\ 
P/C/S    &   1 &  1 &  0 &  -- &   0 & &   19 &   9 &    6 &  --  &    4 \\ 
P/B/S    &   1 &  1 & -- &   0 &   0 & &    9 &   5 &  --  &    0 &    4 \\ 
P/C      &   6 &  5 &  1 &  -- &  -- & &  181 &  85 &   96 &  --  &  --  \\ 
P/B      &   1 &  1 & -- &   0 &  -- & &   50 &  44 &  --  &    6 &  --  \\ 
P/S      &   2 &  1 & -- &  -- &   1 & &   26 &  13 &  --  &  --  &   13 \\ 
C/B      &  27 & -- &  3 &  24 &  -- & & 2288 &  -- &  609 & 1679 &  --  \\ 
C/S      &  25 & -- & 12 &  -- &  13 & &  688 &  -- &  239 &  --  &  449 \\ 
B/S      &  76 & -- & -- &  45 &  31 & &  584 &  -- &  --  &  144 &  440 \\
\cline{2-6} \cline{8-12}\\[-2ex]
{\bf TOTAL}&{\bf 350} &{\bf 17} &{\bf 67} &{\bf 124} &{\bf 142} & 
&{\bf 8459} &{\bf 366} &{\bf 2956} &{\bf 2852} &{\bf 2285} \\ 
~~~\%    &     &4.9 &19.1& 35.4& 40.6& &      & 4.3 & 34.9 & 33.7 & 27.0 \\ 
\hline
\end{tabular}
\end{scriptsize}
\end{center}
\end{table}

\begin{table}
\tabcolsep=4pt
\caption{Summary of time-integrated spectral fit results of 350 GRBs. 
               1$\sigma$ uncertainties are shown in parentheses.}
\label{tab:results}
\begin{center}
\begin{scriptsize}
\begin{tabular}{rccccccccrr}
\hline 
\hline \\[-2ex]
 & & & \multicolumn{8}{c}{Spectral Fit Parameters} \\
\cline{4-11}\\[-2ex]
  GRB~~ & BATSE & BEST & $A$  
& $E_{\rm peak}$\tablenotemark{b}
& $\alpha$, $\lambda_1$\tablenotemark{c}
& $\beta$, $\lambda_2$\tablenotemark{d}
& $E_{\rm b}$\tablenotemark{e}
& $\Lambda$ & $\chi^2$ \tablenotemark{f} & dof \\
  ~~Name\tablenotemark{a} & Trig \# & Model 
& (ph s$^{-1}$ cm$^{-2}$ keV$^{-1}$) 
& (keV) & & & (keV) &  &  & \\
\hline 
 910421 & ~105 & BAND & 0.0785 (0.0054) & 137 (~~4) & --0.80 (0.06) & --2.71 (0.10) & 122 (~~7) & ---  & 125.5 & 112 \\
 910425 & ~109 & COMP & 0.0081 (0.0004) & 523 (~87) & --1.24 (0.06) &      ---      &   ---     & ---  &  10.6 & 11  \\
 910430 & ~130 & BAND & 0.0110 (0.0015) & 180 (~23) & --1.23 (0.13) & --2.33 (0.22) & 141 (~31) & ---  &   6.5 & 10  \\
 910503 & ~143 & SBPL & 0.0482 (0.0001) & 586 (~28) & --1.06 (0.01) & --2.22 (0.03) & 420 (~13) & 0.20 & 190.6 & 112 \\
 910522 & ~219 & SBPL & 0.0198 (0.0001) & 240 (~12) & --1.23 (0.02) & --2.28 (0.03) & 191 (~~8) & 0.20 &   7.9 & 10  \\
 910525 & ~226 & COMP & 0.0037 (0.0003) & 404 (~63) & --1.05 (0.10) &      ---      &   ---     & ---  &   8.8 & 11  \\
 910601 & ~249 & SBPL & 0.0355 (0.0001) & 446 (~22) & --1.06 (0.01) & --3.30 (0.07) & 537 (~21) & 0.50 &   8.6 & 9   \\
 910602 & ~257 & PWRL & 0.0064 (0.0001) &   ---     & --1.54 (0.01) &      ---      &   ---     & ---  & 136.8 & 116 \\
W910609 & ~298 & SBPL & 0.0131 (0.0003) & 389 (~67) & --1.24 (0.04) & --2.40 (0.28) & 387 (~66) & 0.01 &  10.6 & 9   \\
\hline \\[-2ex]
\multicolumn{11}{l}{\hspace{2ex}
$^{\rm a}$~Prefix C -- Calibration burst; W -- Weak burst with a single spectrum} \\
\multicolumn{11}{l}{\hspace{2ex}
$^{\rm b}$~{\it fitted} $E_{\rm peak}$ for BAND or COMP, and 
                  {\it calculated} $E_{\rm peak}$ for SBPL.} \\
\multicolumn{11}{l}{\hspace{2ex}
$^{\rm c}$~$\lambda$ for PWRL, $\alpha$ for BAND, or COMP
                  and $\lambda_1$ for SBPL.} \\
\multicolumn{11}{l}{\hspace{2ex}
$^{\rm d}$~$\beta$ for BAND and $\lambda_2$ for SBPL.} \\
\multicolumn{11}{l}{\hspace{2ex}
$^{\rm e}$~{\it fitted} $E_{\rm b}$ for SBPL, and {\it calculated} 
                  $E_{\rm b}$ for BAND.} \\
\multicolumn{11}{l}{\hspace{2ex}
$^{\rm f}$~Small $\chi^2$ values are expected for calibration bursts
                  by default.} \\
\end{tabular}
\end{scriptsize}
\tablecomments{The complete version of this table is in the electronic edition 
of the Journal.  The printed edition contains only a sample.}
\end{center}
\end{table}

\clearpage
\begin{table}
\caption{The median parameter values and the dispersion (quartile) 
of the BEST model fits.  
The BEST model set consists of 17 (366) PWRL, 67 (2956) COMP,
124 (2852) BAND, and 142 (2285) SBPL time-integrated (time-resolved) spectra. 
The constituent parameter distribution values are also shown.}
\label{tab:mode_best}
\begin{center}
\begin{tabular}{cccccc}
\hline \hline \\[-2ex]
&\multicolumn{5}{c}{Time-Integrated Parameters} \\[2pt]
\cline{2-6} \\[-2ex]
   & Low Index & High Index & $E_{\rm peak}$~ & $E_{\rm b}$~~~ & $\Delta$S~~~~~\\
   &           &            &     (keV)~      &     (keV)~ &  \\\\[-2ex]
\hline \\[-2ex]
BEST&$-1.14~^{+0.20}_{-0.22}$ & $-2.33~^{+0.24}_{-0.26}$ 
      & $251~^{+122}_{-~68}$ & $204~^{+~76}_{-~56}$ & $1.28~^{+0.44}_{-0.28}$  \\[10pt]
PWRL&$-1.55~^{+0.11}_{-0.34}$&$-1.55~^{+0.11}_{-0.34}$&---~~~~~&---~~~~~&---~~~~~ \\[5pt]
COMP&$-1.20~^{+0.22}_{-0.19}$&---~~~~~&$319~^{+186}_{-~87}$&---~~~~~&---~~~~~ \\[5pt]
BAND&$-0.97~^{+0.23}_{-0.21}$&$-2.36~^{+0.17}_{-0.22}$&$224~^{+114}_{-~66}$&$160~^{+~77}_{-~41}$&$1.49~^{+0.33}_{-0.31}$ \\[5pt]
SBPL&$-1.22~^{+0.22}_{-0.14}$&$-2.35~^{+0.23}_{-0.25}$&$245~^{+~93}_{-~63}$&$226~^{+~77}_{-~46}$&$1.10~^{+0.39}_{-0.28}$\\[2pt]
\hline \\[6pt]
\hline \hline \\[-2ex]
&\multicolumn{5}{c}{Time-Resolved Parameters} \\[2pt]
\cline{2-6} \\[-2ex]
   & Low Index & High Index & $E_{\rm peak}$~ & $E_{\rm b}$~~~ & $\Delta$S~~~~~\\
   &           &            &     (keV)~      &     (keV)~ &  \\
\hline \\[-2ex]
BEST&$-1.02~^{+0.26}_{-0.28}$ & $-2.33~^{+0.26}_{-0.31}$
      & $281~^{+139}_{-~99}$ & $205~^{+~72}_{-~55}$ & $1.47~^{+0.45}_{-0.37}$  \\[10pt]
PWRL&$-1.86~^{+0.27}_{-0.26}$&$-1.86~^{+0.27}_{-0.26}$&---~~~~~&---~~~~~&---~~~~~ \\[5pt]
COMP&$-1.15~^{+0.28}_{-0.25}$&---~~~~~&$334~^{+228}_{-140}$&---~~~~~&---~~~~~ \\[5pt]
BAND&$-0.81~^{+0.25}_{-0.21}$&$-2.40~^{+0.24}_{-0.29}$&$273~^{+112}_{-~92}$&$189~^{+~70}_{-~53}$&$1.69~^{+0.43}_{-0.33}$ \\[5pt]
SBPL&$-1.07~^{+0.19}_{-0.20}$&$-2.29~^{+0.27}_{-0.31}$&$241~^{+104}_{-~68}$&$223~^{+~74}_{-~54}$&$1.17~^{+0.36}_{-0.29}$\\[2pt]

\hline 
\end{tabular}
\end{center}
\end{table}

\begin{table}
\caption{Comparison of the median values for time-integrated and 
         time-resolved spectral parameters. The dispersion (quartile) of each
         distribution is shown. K-S probabilities and the
         corresponding parameters for two distributions are also shown.
\label{tab:spec_param}}
\begin{center}
\begin{tabular}{cccccc}
\hline \hline  \\[-2ex]
Spectrum & Low Index & High Index & $E_{\rm peak}$  & $E_{\rm b}$ & $\Delta$S \\
Type     &           &            &     (keV)       &      (keV)  &  \\ \\[-2ex]
\hline \\[-2ex]
Time Integrated   &$-1.14~^{+0.20}_{-0.22}$ & $-2.33~^{+0.24}_{-0.26}$ 
      & $251~^{+122}_{-~68}$ & $204~^{+~76}_{-~56}$ & $1.28~^{+0.44}_{-0.28}$ \\[5pt]
Time Resolved     &$-1.02~^{+0.26}_{-0.28}$ & $-2.33~^{+0.26}_{-0.31}$
      & $281~^{+139}_{-~99}$ & $205~^{+~72}_{-~55}$ & $1.47~^{+0.45}_{-0.37}$ \\[5pt]
$P_{\rm{\scriptscriptstyle KS}}$ & 1.18 $\times 10^{-16}$ & 0.47 & 7.73 $\times 10^{-3}$ & 0.88 & 2.14 $\times 10^{-4}$\\
$D_{\rm{\scriptscriptstyle KS}}$ & 0.23 & 0.05 & 0.10 & 0.04 & 0.13 \\[2pt]
\hline
\end{tabular}
\end{center}
\end{table}

\begin{table}
\caption{Summary of total numbers of events considered for each correlation 
and percentages of events with high significance.}
\label{tab:corr_sum}
\begin{center}
\begin{tabular}{lccccc}
\hline\hline   \\[-2ex]
& $E_{\rm peak}-\alpha$ & $E_{\rm b}-\alpha$ & $E_{\rm peak}-\beta$ 
         & $E_{\rm b}-\beta$ & $\alpha-\beta$   \\
\hline   \\[-2ex]
Total Number of Events &  196   & 148   &  117  & 147  & 160 \\
Significant Correlation (\%) &  25.5  & 13.5  &  4.3  & 15.0 & 6.3 \\
\hline
\end{tabular}
\end{center}
\end{table}

\clearpage
\begin{deluxetable}{ccrrrrrrrrrrrrrr}
\tabcolsep=4pt
\tablewidth{0pt}
\tabletypesize{\scriptsize}
\tablecaption{List of all bursts in our sample with strong ($> 3\sigma$) 
         spectral parameter correlations in at least one parameter pair.
         Spearman rank correlation coefficients ($r_s$) and associated 
         significance probabilities ($P_{rs}$) are listed.
         $\alpha$ and $\beta$ denote low-energy and high-energy indices. 
\label{tab:corr}}
\tablehead{
\colhead{BATSE} & \colhead{Number of} & \multicolumn{2}{c}{$E_{\rm peak}$ -- $\alpha$} && 
                    \multicolumn{2}{c}{$E_{\rm b}$ -- $\alpha$} && 
                    \multicolumn{2}{c}{$E_{\rm peak}$ -- $\beta$} && 
                    \multicolumn{2}{c}{$E_{\rm b}$ -- $\beta$} && 
                    \multicolumn{2}{c}{$\alpha - \beta$} \\
\cline{3-4}\cline{6-7}\cline{9-10}\cline{12-13}\cline{15-16}\\[-2ex]
\colhead{Trig \#~} & \colhead{Spectra~~} &
\colhead{$r_s$~} & \colhead{$P_{rs}$~~} && \colhead{$r_s$~} & 
\colhead{$P_{rs}$~~} &&  \colhead{$r_s$~} & \colhead{$P_{rs}$~~} &&  
\colhead{$r_s$~} & \colhead{$P_{rs}$~~} && \colhead{$r_s$~} & 
\colhead{$P_{rs}$~~}
}

\startdata
 ~109 & ~26 &  0.70 & 8.85E--4  &&    \novalue      &&    \novalue      &&    \novalue      &&    \novalue      \\
 ~143 & ~46 &    \novalue      &&    \novalue      &&    \novalue      && --0.61 & 7.47E--5  &&    \novalue      \\
 ~219 & ~13 &  0.80 & 9.69E--4  &&    \novalue      &&    \novalue      &&    \novalue      &&    \novalue      \\
 ~249 & ~68 &  0.50 & 1.55E--5  &&    \novalue      &&    \novalue      && --0.59 & 3.60E--6  &&    \novalue      \\
 1085 & ~60 &  0.80 & 1.45E--13 &&  0.73 & 2.11E--10 && ~0.71 & 2.32E--9  &&  0.60 & 1.31E--6  && ~0.47 & 3.00E--4  \\
 1141 & ~42 &    \novalue      &&    \novalue      &&    \novalue      && --0.57 & 5.92E--4  &&    \novalue      \\
 1288 & ~28 &  0.63 & 4.73E--4  &&    \novalue      &&    \novalue      &&    \novalue      &&    \novalue      \\
 1385 & ~36 &    \novalue      &&    \novalue      &&    \novalue      && --0.73 & 9.80E--4  &&    \novalue      \\
 1541 & ~15 &  0.87 & 5.68E--5  &&    \novalue      &&    \novalue      &&    \novalue      &&    \novalue      \\
 1625 & ~37 &    \novalue      &&  0.63 & 1.19E--4  &&    \novalue      &&    \novalue      &&    \novalue      \\
 1652 & ~18 &  0.84 & 2.65E--5  &&    \novalue      &&    \novalue      &&    \novalue      &&    \novalue      \\
 1663 & ~61 &    \novalue      &&  0.59 & 5.57E--6  &&    \novalue      && --0.62 & 1.29E--6  &&    \novalue      \\
 1676 & ~30 &  0.74 & 2.65E--6  &&  0.84 & 9.14E--5  &&    \novalue      &&    \novalue      &&    \novalue      \\
 1695 & 142 &  0.44 & 9.95E--8  &&    \novalue      &&    \novalue      && --0.40 & 1.03E--5  &&    \novalue      \\
 1698 & ~32 &  0.79 & 1.54E--7  &&    \novalue      &&    \novalue      &&    \novalue      &&    \novalue      \\
 1872 & 110 &    \novalue      &&  0.44 & 5.35E--5  &&    \novalue      &&    \novalue      && --0.51 & 1.01E--6  \\
 1983 & ~54 &    \novalue      &&    \novalue      &&    \novalue      && --0.70 & 6.17E--6  &&    \novalue      \\
 2067 & ~45 &  0.64 & 2.08E--6  &&    \novalue      &&    \novalue      && --0.68 & 1.08E--6  &&    \novalue      \\
 2083 & ~45 &  0.91 & 5.99E--17 &&  0.75 & 5.18E--6  &&    \novalue      &&    \novalue      &&    \novalue      \\
 2138 & ~39 &    \novalue      &&    \novalue      &&    \novalue      && --0.77 & 5.00E--7  &&    \novalue      \\
 2156 & ~97 &  0.45 & 1.79E--5  &&  0.50 & 8.69E--6  &&    \novalue      &&    \novalue      &&    \novalue      \\
 2329 & 100 &    \novalue      &&  0.60 & 2.68E--7  &&    \novalue      &&    \novalue      &&    \novalue      \\
 2533 & ~70 &    \novalue      &&    \novalue      &&    \novalue      && --0.56 & 3.52E--5  &&    \novalue      \\
 2537 & ~26 &  0.67 & 3.21E--4  &&    \novalue      &&    \novalue      &&    \novalue      &&    \novalue      \\
 2661 & ~29 &  0.68 & 5.54E--5  &&    \novalue      &&    \novalue      &&    \novalue      &&    \novalue      \\
 2676 & 122 &  0.43 & 9.16E--7  &&  0.58 & 6.56E--10 &&    \novalue      && --0.47 & 1.66E--6  && --0.41 & 2.92E--5  \\
 2790 & ~19 &  0.85 & 9.11E--6  &&    \novalue      &&    \novalue      &&    \novalue      &&    \novalue      \\
 2798 & ~95 &  0.62 & 2.96E--10 &&  0.37 & 8.90E--4  &&    \novalue      &&    \novalue      &&    \novalue      \\
 2831 & 112 &  0.51 & 4.56E--8  &&    \novalue      &&    \novalue      &&    \novalue      &&    \novalue      \\
 2855 & ~35 &    \novalue      &&    \novalue      &&    \novalue      && --0.56 & 6.16E--4  &&    \novalue      \\
 2856 & 114 &    \novalue      &&  0.42 & 2.63E--4  &&    \novalue      &&    \novalue      &&    \novalue      \\
 3002 & ~19 &  0.74 & 3.36E--4  &&    \novalue      &&    \novalue      &&    \novalue      &&    \novalue      \\
 3035 & ~28 &  0.60 & 6.70E--4  &&    \novalue      &&    \novalue      &&    \novalue      &&    \novalue      \\
 3057 & 165 &  0.31 & 2.20E--4  &&  0.40 & 1.91E--5  &&    \novalue      &&    \novalue      &&    \novalue      \\
 3128 & ~44 &    \novalue      &&    \novalue      &&    \novalue      && --0.68 & 1.81E--5  &&    \novalue      \\
 3227 & ~50 &  0.62 & 1.25E--6  &&  0.60 & 8.65E--5  &&    \novalue      &&    \novalue      &&    \novalue      \\
 3245 & 116 &  0.70 & 8.02E--17 &&  0.60 & 9.03E--12 && ~0.40 & 6.83E--5  &&    \novalue      &&    \novalue      \\
 3253 & ~94 &  0.73 & 3.41E--17 &&  0.56 & 1.14E--5  &&    \novalue      &&    \novalue      && --0.46 & 4.95E--4  \\
 3408 & ~63 &    \novalue      &&    \novalue      &&    \novalue      && --0.49 & 5.29E--4  &&    \novalue      \\
 3492 & ~49 &  0.69 & 1.88E--7  &&  0.65 & 1.17E--4  &&    \novalue      &&    \novalue      &&    \novalue      \\
 3571 & ~22 &  0.73 & 1.03E--4  &&    \novalue      &&    \novalue      &&    \novalue      &&    \novalue      \\
 3658 & ~30 &  0.68 & 8.78E--5  &&    \novalue      &&    \novalue      &&    \novalue      &&    \novalue      \\
 3765 & ~33 &  0.69 & 1.16E--5  &&    \novalue      &&    \novalue      &&    \novalue      &&    \novalue      \\
 3767 & ~20 &  0.88 & 3.57E--7  &&    \novalue      &&    \novalue      &&    \novalue      &&    \novalue      \\
 3870 & ~17 &    \novalue      &&    \novalue      &&    \novalue      &&    \novalue      && ~0.89 & 5.15E--6  \\
 5304 & ~48 &  0.81 & 6.06E--12 &&  0.67 & 8.77E--7  &&    \novalue      &&    \novalue      &&    \novalue      \\
 5486 & ~32 &    \novalue      &&    \novalue      &&    \novalue      &&    \novalue      && --0.86 & 1.78E--5  \\
 5567 & ~31 &    \novalue      &&    \novalue      &&    \novalue      && --0.72 & 5.68E--4  &&    \novalue      \\
 5649 & 155 &  0.42 & 4.50E--8  &&    \novalue      &&    \novalue      &&    \novalue      &&    \novalue      \\
 5773 & ~57 &  0.72 & 3.29E--10 &&    \novalue      &&    \novalue      &&    \novalue      &&    \novalue      \\
 5995 & ~61 &  0.63 & 7.98E--7  &&    \novalue      &&    \novalue      &&    \novalue      && ~0.53 & 4.71E--4  \\
 6124 & ~60 &    \novalue      &&    \novalue      && --0.50 & 5.05E--4  && --0.73 & 1.13E--8  &&    \novalue      \\
 6266 & ~18 &    \novalue      &&    \novalue      && --0.88 & 3.30E--4  &&    \novalue      &&    \novalue      \\
 6329 & ~31 &  0.59 & 4.94E--4  &&    \novalue      &&    \novalue      &&    \novalue      &&    \novalue      \\
 6350 & ~44 &  0.54 & 6.94E--4  &&    \novalue      &&    \novalue      && --0.52 & 8.52E--4  &&    \novalue      \\
 6353 & ~35 &  0.79 & 4.22E--4  &&    \novalue      &&    \novalue      &&    \novalue      &&    \novalue      \\
 6576 & ~21 &  0.78 & 8.50E--5  &&    \novalue      &&    \novalue      &&    \novalue      &&    \novalue      \\
 6587 & ~60 &    \novalue      &&    \novalue      &&    \novalue      && --0.73 & 3.43E--9  &&    \novalue      \\
 6630 & ~17 &  0.85 & 5.17E--5  &&    \novalue      &&    \novalue      &&    \novalue      &&    \novalue      \\
 6665 & ~38 &    \novalue      &&    \novalue      &&    \novalue      && --0.58 & 2.85E--4  &&    \novalue      \\
 7113 & 243 &    \novalue      &&    \novalue      &&    \novalue      &&    \novalue      && --0.30 & 6.06E--4  \\
 7170 & ~30 &  0.75 & 7.79E--4  &&    \novalue      &&    \novalue      &&    \novalue      &&    \novalue      \\
 7301 & ~56 &  0.76 & 6.12E--9  &&  0.57 & 1.35E--5  && ~0.67 & 4.58E--4  &&    \novalue      && ~0.47 & 5.10E--4  \\
 7343 & 134 &  0.63 & 8.13E--15 &&  0.56 & 8.40E--9  &&    \novalue      &&    \novalue      && ~0.35 & 6.50E--4  \\
 7360 &  32 &  0.77 & 2.19E--7  &&    \novalue      &&    \novalue      &&    \novalue      &&    \novalue      \\
 7491 & 132 &    \novalue      &&    \novalue      &&    \novalue      && --0.35 & 3.22E--4  &&    \novalue      \\
 7575 &  21 &  0.85 & 1.35E--5  &&    \novalue      &&    \novalue      &&    \novalue      &&    \novalue      \\
 7678 &  37 &    \novalue      &&    \novalue      &&    \novalue      && --0.70 & 3.05E--5  &&    \novalue      \\
 7695 &  29 &  0.66 & 9.16E--5  &&    \novalue      &&    \novalue      &&    \novalue      &&    \novalue      \\
 7906 &  59 &  0.72 & 9.26E--10 &&  0.56 & 5.10E--6  &&    \novalue      &&    \novalue      &&    \novalue      \\
 7954 &  27 &  0.62 & 5.85E--4  &&    \novalue      &&    \novalue      &&    \novalue      &&    \novalue      \\
 7976 &  17 &  0.88 & 5.91E--5  &&    \novalue      &&    \novalue      &&    \novalue      &&    \novalue      \\
 7994 &  25 &  0.68 & 1.60E--4  &&    \novalue      &&    \novalue      &&    \novalue      &&    \novalue      \\ 
 8008 &  27 &  0.73 & 2.51E--5  &&    \novalue      &&    \novalue      &&    \novalue      &&    \novalue      \\ 
\enddata
\end{deluxetable}

\begin{table}
\tabcolsep=4pt
\caption{Summary of time-integrated spectral fit results for 17 short GRBs.
         1$\sigma$ uncertainties are shown in parentheses.}
\label{tab:short}
\begin{center}
\begin{scriptsize}
\begin{tabular}{ccccccccccrr}
\hline 
\hline \\[-2ex]
& & & & \multicolumn{8}{c}{Spectral Fit Parameters} \\
\cline{5-12}\\[-2ex]
  GRB & BATSE & \# of & BEST & $A$  
& $E_{\rm peak}$\tablenotemark{a}
& $\alpha$, $\lambda_1$\tablenotemark{b}
& $\beta$, $\lambda_2$\tablenotemark{c}
& $E_{\rm b}$\tablenotemark{d}
& $\Lambda$ & $\chi^2$ & dof \\
  Name & Trig \# & Spec & Model 
& (ph s$^{-1}$ cm$^{-2}$ keV$^{-1}$) 
& (keV) & & & (keV) &  &  & \\
\hline 
910609 & ~298 & 1 & SBPL & 0.0131 (0.0003) & 389 (~67) & --1.24 (0.04) & --2.40 (0.28) & 387 (~66) & 0.01 &  10.6 & 9   \\
910626 & ~444 & 1 & BAND & 0.5183 (0.5078) &  91 (~13) & ~~0.18 (0.65) & --2.11 (0.07) &  60 (~~7) & ---  & 124.1 & 114 \\
920329 & 1525 & 1 & SBPL & 0.0265 (0.0005) & 536 (119) & --0.76 (0.06) & --2.44 (0.20) & 375 (~59) & 0.30 &  12.7 & 9   \\
920414 & 1553 & 3 & SBPL & 0.0371 (0.0005) &   ---     & --0.88 (0.03) & --1.74 (0.04) & 225 (~16) & 0.01 & 119.0 & 110 \\
930905 & 2514 & 1 & SBPL & 0.0853 (0.0017) & 208 (~22) & --1.34 (0.04) & --2.76 (0.20) & 212 (~21) & 0.10 & 106.1 & 111 \\
931205 & 2679 & 1 & SBPL & 0.0241 (0.0009) & 792 (212) & --0.54 (0.05) & --2.13 (0.24) & 599 (~94) & 0.10 & 127.8 & 114 \\
940624 & 3044 & 1 & PWRL & 0.0018 (0.0001) &   ---     & --2.20 (0.14) &      ---      &   ---     & ---  & 125.8 & 114 \\
940717 & 3087 & 1 & BAND & 0.2097 (0.0431) & 182 (~20) & --0.59 (0.19) & --2.22 (0.10) & 118 (~13) & ---  & 139.3 & 114 \\
940902 & 3152 & 2 & SBPL & 0.0089 (0.0002) &   ---     & --0.99 (0.04) & --1.79 (0.16) & 490 (110) & 0.01 & 138.7 & 114 \\
950210 & 3410 & 1 & COMP & 0.0662 (0.0204) & 120 (~~9) & --0.78 (0.29) &      ---      &   ---     & ---  & 122.4 & 112 \\
950211 & 3412 & 1 & PWRL & 0.0155 (0.0005) &    ---    & --2.13 (0.06) &      ---      &   ---     & ---  & 115.5 & 112 \\
950805 & 3736 & 1 & COMP & 0.0102 (0.0004) & 920 (159) & --0.89 (0.08) &      ---      &   ---     & ---  &  10.3 & 10  \\
980228 & 6617 & 1 & SBPL & 0.1325 (0.0013) & 259 (~32) & --1.13 (0.02) & --3.46 (0.31) & 309 (~31) & 0.30 & 117.7 & 114 \\
970704 & 6293 & 3 & PWRL & 0.1817 (0.0026) &   ---     & --1.25 (0.01) &      ---      &   ---     & ---  & 137.2 & 115 \\
980330 & 6668 & 1 & SBPL & 0.0166 (0.0010) &   ---     & --1.22 (0.14) & --1.98 (0.19) & 149 (~37) & 0.01 & 123.0 & 112 \\
981226 & 7281 & 2 & BAND & 0.2143 (0.0345) & 140 (~~6) & --0.23 (0.14) & --2.54 (0.10) & 103 (~~6) & ---  & 125.2 & 113 \\
000326 & 8053 & 5 & BAND & 0.4870 (0.0423) &  90 (~~1) & --0.50 (0.07) & --3.68 (0.18) & 108 (~~6) & ---  & 122.0 & 112 \\
\hline \\[-2ex]
\multicolumn{12}{l}{\hspace{2ex}
$^{\rm a}$~{\it fitted} $E_{\rm peak}$ for BAND or COMP, and 
                  {\it calculated} $E_{\rm peak}$ for SBPL.} \\
\multicolumn{12}{l}{\hspace{2ex}
$^{\rm b}$~$\lambda$ for PWRL, $\alpha$ for BAND, or COMP
                  and $\lambda_1$ for SBPL.} \\
\multicolumn{12}{l}{\hspace{2ex}
$^{\rm c}$~$\beta$ for BAND and $\lambda_2$ for SBPL.} \\
\multicolumn{12}{l}{\hspace{2ex}
$^{\rm d}$~{\it fitted} $E_{\rm b}$ for SBPL, and {\it calculated} 
                  $E_{\rm b}$ for BAND.} \\
\end{tabular}
\end{scriptsize}
\end{center}
\end{table}

\begin{table}
\tabcolsep=2pt
\caption{Median and quartile values of the BEST spectral parameters for 
short and long GRBs.
K-S probabilities ($P_{\rm{\scriptscriptstyle KS}}$)
and the corresponding K-S parameters ($D_{\rm{\scriptscriptstyle KS}}$)
for two distributions are also shown.}
\label{tab:short_long}
\begin{center}
\begin{footnotesize}
\begin{tabular}{cccccccccccc}
\hline \hline  \\[-2ex]
 & \multicolumn{2}{c}{Low-Energy Index} && \multicolumn{2}{c}{High-Energy Index} 
 && \multicolumn{2}{c}{$E_{\rm peak}$ (keV)}  && \multicolumn{2}{c}{$E_{\rm b}$ (keV)}\\
\cline{2-3}\cline{5-6}\cline{8-9}\cline{11-12}  \\[-2ex]
 & {\small Short} & {\small Long} && {\small Short} & {\small Long} 
 && {\small Short} & {\small Long} && {\small Short} & {\small Long} \\
\hline \\[-2ex]
Time Integrated & --0.99 $^{+0.21}_{-0.24}$ & --1.15 $^{+0.20}_{-0.22}$ 
                  && --2.31 $^{+0.20}_{-0.23}$ & --2.36 $^{+0.23}_{-0.23}$ 
                  && 208 $^{+327}_{-~88}$ & 254 $^{+110}_{-~72}$ 
                  && 218 $^{+157}_{-101}$ & 202 $^{+~77}_{-~54}$ \\[4pt]
$P_{\rm{\scriptscriptstyle KS}}$ & \multicolumn{2}{c}{0.07} && \multicolumn{2}{c}{0.43} 
               && \multicolumn{2}{c}{0.43} &&\multicolumn{2}{c}{0.20} \\
$D_{\rm{\scriptscriptstyle KS}}$ & \multicolumn{2}{c}{0.31} && \multicolumn{2}{c}{0.25} 
               && \multicolumn{2}{c}{0.26} &&\multicolumn{2}{c}{0.31} \\[2pt]
\hline \\[-2ex]
Time Resolved   & --0.87 $^{+0.16}_{-0.39}$ & --1.02 $^{+0.26}_{-0.28}$ 
                  && --2.38 $^{+0.39}_{-0.51}$ & --2.35 $^{+0.25}_{-0.30}$ 
                  && 168 $^{+222}_{-~47}$ & 281 $^{+139}_{-~99}$ 
                  && 204 $^{+136}_{-~93}$ & 205 $^{+205}_{-~72}$  \\[5pt]
$P_{\rm{\scriptscriptstyle KS}}$ & \multicolumn{2}{c}{0.16} && \multicolumn{2}{c}{0.53} 
               && \multicolumn{2}{c}{0.08} &&\multicolumn{2}{c}{0.38} \\
$D_{\rm{\scriptscriptstyle KS}}$ & \multicolumn{2}{c}{0.22} && \multicolumn{2}{c}{0.19} 
               && \multicolumn{2}{c}{0.28} &&\multicolumn{2}{c}{0.22} \\[2pt]

\hline \\[-2ex]
\end{tabular}
\end{footnotesize}
\end{center}
\end{table}

\clearpage

\clearpage

\clearpage

\clearpage

\begin{thebibliography}{}
\bibitem[Amati et~al.(2002)]{ama02}Amati, L., et~al. 2002, \aap, 
         390, 81
\bibitem[Band(1997)]{ban97}Band, D.L., 1997, \apj, 486, 928
\bibitem[Band \& Preece(2005)]{ban05}Band, D.L. \& Preece, R.D., 2005, 
         \apj, 627, 319
\bibitem[Band et~al.(1993)]{ban93}Band, D.L., et~al. 1993, 
         \apj, 413, 281
\bibitem[Baring \& Braby(2004)]{bar04}Baring, M.G. \& Braby, M.L., 2004, 
         \textit{ApJ}, 613, 460
\bibitem[Bevington \& Robinson(2003)]{bev03}Bevington, P.R. \& Robinson, D.K., 
         2003, \textit{Data Reduction and Error Analysis for the Physical 
         Sciences}, 3rd Edition (New York: McGraw-Hill) 
\bibitem[Bhat et~al.(1992)]{bha92}Bhat, P.N., et~al. 1992, \nat,
         359, 217
\bibitem[Briggs(1996)]{bri96a}Briggs, M.S., 1996, in \textit{Gamma-Ray Bursts},
         3rd Huntsville Symposium, ed. C. Kouveliotou, M. Briggs 
         \& G. Fishman, AIP, 384, 133
\bibitem[Crider et~al.(1997)]{cri97}Crider, A., et~al. 1997, \apj, 
         479, L39
\bibitem[Crider et~al.(1999)]{cri99}Crider, A., et~al. 1999, \textit{A\&AS}, 
         138, 401
\bibitem[Dezalay et~al.(1992)]{dez92}Dezalay, J.-P., et~al. 1992, in 
         \textit{Gamma-Ray Bursts}, 1st Huntsville Symposium, ed. W. Paciesas 
         \& G. Fishman, AIP, 265, 304
\bibitem[Dezalay et~al.(1997)]{dez97}Dezalay, J.-P., et~al. 1997, \apj, 490, L17 
\bibitem[Fishman et~al.(1989)]{fis89}Fishman, G.J., et~al. 1989, 
         in {\it Proc. GRO Science Workshop}, GSFC, 2-39 
\bibitem[Ford et~al.(1995)]{for95} Ford, L.A., et~al. 1995, 
         \apj, 439, 307
\bibitem[Gallant(2002)]{gal02}Gallant, Y.A., 2002,  in 
         \textit{Relativistic Flows in Astrophysics}, ed. A. Guthmann et~al.,  
         Lecture Notes in Physics (Heidelberg:Springer-Verlag), 589, 24
\bibitem[Gehrels, Chipman, \& Kniffen(1994)]{geh94} Gehrels, N., Chipman, E., 
         \& Kniffen, D., 1994, \apjs, 92, 351
\bibitem[Ghirlanda et~al.(2002)]{ghi02} Ghirlanda, G., et~al. 2002, \aap, 
         393, 409 
\bibitem[Ghirlanda et~al.(2005)]{ghi05} Ghirlanda, G., et~al. 2005, \mnras, 
         361, L10
\bibitem[Ghisellini \& Celotti(1999)]{ghi99} Ghisellini, G. \& Celotti, A., 
         \apj, 511, L93
\bibitem[Harmon et~al.(2002)]{har02}Harmon, B.A., et~al. 2002, 
         \apjs, 138, 149
\bibitem[Kaneko(2005)]{kan05} Kaneko, Y., 2005, Ph.D. Dissertation, 
         University of Alabama in Huntsville
\bibitem[Katz(1994)]{kat94} Katz, J.I., 1994, \apj, 432, L107
\bibitem[Kouveliotou et~al.(1993)]{kou93} Kouveliotou, C., et~al. 1993, 
         \apj, 413, L101
\bibitem[Liang \& Kargatis(1996)]{lia96}Liang, E.P. \& Kargatis, V.E., 1996, 
         \nat, 381, 49
\bibitem[Lloyd, Petrosian \& Mallozzi(2000)]{llo+00}Lloyd, N.M., et~al. 2000, \apj, 534, 227
\bibitem[Lloyd-Ronning \& Petrosian(2002)]{llo02}Lloyd-Ronning, N.M. 
         \& Petrosian, V., \apj, 565, 182
\bibitem[Mallozzi et~al.(1995)]{mal95}Mallozzi, R.S., et~al. 1995,
         \apj, 454, 597
\bibitem[Mallozzi, Preece \& Briggs(2005)]{rmfit}
         Mallozzi, R.S., Preece, R.D. \& Briggs, M.S., 2005,
         ``RMFIT, A Lightcurve and Spectral Analysis Tool", 
         \copyright 2005 Robert D. Preece, University of Alabama in Huntsville 
\bibitem[Medvedev(2000)]{med00}Medvedev, M.V., 2000, \apj, 540, 704
\bibitem[Nemiroff et~al.(1994)]{nem94}Nemiroff, R.J., et al. 1994, \apj, 435, L133
\bibitem[Nakar \& Piran(2005a)]{nak05b}Nakar, E. \& Piran, T., 2005b, 
         \mnras, 360, L73   
\bibitem[Nakar \& Piran(2005b)]{nak05c}Nakar, E. \& Piran, T., 2005c, 
         astro-ph/0503517   
\bibitem[Paciesas et~al.(1999)]{pac99}Paciesas, W.S., et~al. 1999, \apjs, 
         122, 465
\bibitem[Paciesas et~al.(2003)]{pac03}Paciesas, W.S., et~al. 2003, in 
         \textit{Gamma-Ray Bursts and Afterglow Astronomy}, 
         ed. G. Ricker \& R. Vanderspek, AIP, 662, 248
\bibitem[Pendleton et~al.(1995)]{pen95}Pendleton, G.N., et~al. 1995, 
         Nuc.~Inst.~\& Met.~A, 364, 567
\bibitem[Pendleton et~al.(1997)]{pen97}Pendleton, G.N., et~al. 1997, 
         \apj, 489, 175
\bibitem[Preece et~al.(1994)]{wingspan}
         Preece, R.D., Briggs, M.S., Mallozzi, R.S., \& Brock, M.N., 1994,
         ``WINdows Gamma SPectral ANalysis (WINGSPAN)"
\bibitem[Preece et~al.(1998a)]{pre98a}Preece, R.D., et~al. 1998a, 
         \apj, 496, 849  
\bibitem[Preece et~al.(1998b)]{pre98b} Preece, R.D., et~al. 1998b, 
         \apj, 506, L23  
\bibitem[Preece et~al.(2000)]{pre00}Preece, R.D., et~al. 2000, 
         \apjs, 126, 19 (SP1)
\bibitem[Preece et~al.(2002)]{pre02}Preece, R.D., et~al. 2002, 
         \apj, 581, 1248
\bibitem[Press et~al.(1992)]{pre92}Press et~al., 1992,
         \textit{Numerical Recipes in FORTRAN}, 2nd Edition
         (New York: Cambridge University Press)
\bibitem[Rybicki \& Lightman(1979)]{ryb79}Rybicki, G.B. \& Lightman, A.P., 
         1979, \textit{Radiative Processes in Astrophysics}
         (New York: John Wiley \& Sons) 
\bibitem[Ryde(1999)]{ryd99}Ryde, F., 1999, \aplett, 39, 281
\bibitem[Sakamoto et~al.(2004)]{sak04a}Sakamoto, T., et~al. 2004, 
         \apj, 602, 875
\bibitem[Sakamoto et~al.(2005)]{sak05}Sakamoto, T., et~al. 2005, 
         \apj, 629, 311
\bibitem[Sari et~al.(1998)]{sar98}Sari, R., et~al. 1998, \apj, 497, L17
\bibitem[Tavani(1996)]{tav96b}Tavani, M., 1996, \apj, 466, 768
\bibitem[Zhang \& M\'esz\'aros(2002)]{zha02}Zhang, B. \& M\'esz\'aros, P.,
         2002, \apj, 581, 1236
\end{thebibliography}
\end{document}